\documentclass[prb,aps,twocolumn,floatfix,showpacs]{revtex4}
\usepackage{bm}
\usepackage{amssymb}
\usepackage{graphicx}

\begin{document}

\title{Spin-waves in triangular lattice antiferromagnet:\\
 decays,  spectrum renormalization, and singularities}

\author{A. L. Chernyshev}
\affiliation{Department of Physics, University of California, Irvine, 
California 92697, USA\\
 and Max-Planck-Institut f\"ur Physik komplexer Systeme,
01187 Dresden, Germany }

\author{M. E. Zhitomirsky} 
\affiliation{
Commissariat \`a l'Energie Atomique, DSM/INAC/SPSMS, F-38054 Grenoble,
France}

\date{\today}

\begin{abstract}
We present a comprehensive study of the dynamical properties 
of the quantum Heisenberg antiferromagnet on a triangular lattice 
within the framework of spin-wave theory. 
The distinct features of spin-wave excitations 
in the triangular-lattice antiferromagnet are
(i) finite lifetime at zero temperature due to spontaneous two-magnon decays,
(ii) strong renormalization of magnon energies $\varepsilon_{\bf k}$
with respect to the harmonic result, 
and (iii) logarithmic singularities in the decay rate $\Gamma_{\bf k}$.
Quantum corrections to the magnon spectrum 
are obtained using both the on-shell and off-shell solutions of the 
Dyson equation with the lowest-order magnon self-energy.
At low-energies magnon excitations remain well-defined albeit
with the anomalous decay rate
$\Gamma_{\bf k}\propto k^2$ at ${\bf k}\rightarrow 0$ and 
$\Gamma_{\bf k}\propto |{\bf k}-{\bf Q}_{\rm AF}|^{7/2}$ 
at ${\bf k}\rightarrow {\bf Q}_{\rm AF}$.
At high energies, magnons are heavily damped with the decay rate reaching 
$(2\Gamma_{\bf k}/\varepsilon_{\bf k})\sim 0.3$ for the case $S=1/2$. 
The on-shell
solution shows logarithmic singularities in $\Gamma_{\bf k}$ with the 
concomitant jump-like  discontinuities in Re[$\varepsilon_{\bf k}$]
along certain contours in the momentum space. Such
singularities are even more prominent in the magnon spectral function 
$A({\bf k},\omega)$.
Although the off-shell solution removes such log-singularities, 
the decay rates remain strongly enhanced. 
We also discuss the role of higher-order corrections and show 
that such singularities may lead to 
complete disappearance of the spectrum in the vicinity of certain 
${\bf k}$-points.
The kinematic conditions for two-magnon decays are analyzed for various 
generalizations of the triangular-lattice antiferromagnet as well as for the 
$XXZ$ model on a kagom\'{e} lattice. Our results suggest that 
decays and singularities in the spin-wave spectra must be 
ubiquitous in all these systems.
In addition, we give a detailed introduction in the spin-wave
formalism for noncollinear Heisenberg antiferromagnets and 
calculate several quantities for the triangular-lattice model
including the ground-state energy and the sublattice magnetization.
\end{abstract}
\pacs{75.10.Jm,   
      75.30.Ds,   
      78.70.Nx    
}

\maketitle

\section{Introduction}

Heisenberg antiferromagnet (HAF) on a triangular lattice has been a focus 
of much attention as one of the basic model systems in which geometric 
frustration and low dimensionality are expected to yield new physical
phenomena. Although available experimental realizations of the 
triangular-lattice  
antiferromagnet
\cite{Collins97,Coldea01,Svistov03,Nakatsuji05,Olariu06,Hasan08} 
are described by such a model only approximately, 
either due to anisotropies or because of additional interactions, 
the ideal nearest-neighbor Heisenberg antiferromagnet on 
a triangular lattice given by
\begin{eqnarray}
\hat{\cal H}  =  
J \sum_{\langle ij\rangle}  {\bf S}_i\cdot {\bf S}_j\ ,
\label{TLH}
\end{eqnarray}
remains the principal reference point. 

\begin{figure}[t]
\centerline{
\includegraphics[width=0.7\columnwidth]{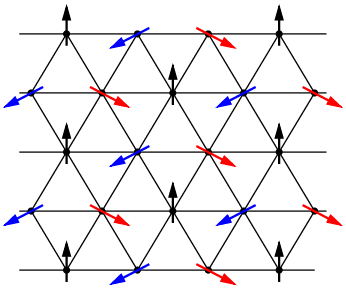}}
\caption{(Color online) The ordered 120$^\circ$ spin structure of the 
triangular-lattice HAF. 
}
\label{structure}
\end{figure}

The semiclassical $S\gg 1$ triangular-lattice HAF orders in 
the so-called 120$^\circ$ structure, see Fig.~\ref{structure}.
Historically, it was anticipated that 
enhanced quantum fluctuations destroy the long-range
antiferromagnetic order for the 
spin-$1/2$ model. \cite{Anderson74}  
However, calculations of quantum corrections 
within the spin-wave theory have 
suggested that the 120$^\circ$ magnetic structure remains stable even for
$S = 1/2$. \cite{Oguchi83,Jolicoeur89,Miyake92,Chubukov94}
The early numerical results for small clusters 
were less conclusive, some   supporting  magnetically disordered
state \cite{Leung93} and some confirming the spin-wave results.
\cite{Lhuillier94}
Since the quantum Monte Carlo suffers from the infamous sign problem
when applied to frustrated models, it
is not until the Green's function Monte Carlo work\cite{Sorella99}
that the magnetically  ordered ground state of the spin-1/2 
triangular-lattice HAF has been 
generally agreed upon. More recent series-expansion\cite{Zheng06} and
density-matrix renormalization group (DMRG) studies\cite{White07} have 
confirmed the stability of the 120$^\circ$ spin structure for 
the case of $S=1/2$ and yielded
the value of ordered moments $\langle S\rangle \approx 0.20$,
close to the previous result. \cite{Sorella99} 

Gradually, it has been recognized that the truly distinct physics of the 
quantum triangular-lattice antiferromagnet concerns its excitation
spectrum and the thermodynamic properties. 
The anomalous behavior 
of the latter has been discovered earlier by the high-temperature 
series-expansion study. \cite{Singh93} 
The temperature dependence of such quantities as entropy or
susceptibility,  
exhibits significant differences between the triangular- and 
the square-lattice models: upon lowering temperature down to about 
$J/2$ the square-lattice  antiferromagnet shows strong signs of ordering, 
while the triangular-lattice one does not. 
More recently, developments in the 
series-expansion method have allowed to calculate the 
excitation spectra of the noncollinear spin systems directly.
\cite{Zheng06,Singh06} Numerical results for the magnon band of
the spin-$1/2$ triangular-lattice HAF deviate substantially from 
the linear spin-wave theory (LSWT), with the overall band-narrowing by
$\sim\! 50$\%,  
flattening at the top of the spectrum, and extra ``roton-like'' minima 
appearing at some special ${\bf k}$-points. 
These results are so different from the well-known
square-lattice case, where the spectrum is renormalized 
only modestly and almost uniformly upward,
that Ref.~\onlinecite{Singh06} suggested that at high energies 
the elementary excitations are not spin-waves but spinons. 
This hypothesis was questioned by the subsequent
studies which have looked into the role of magnon interactions 
within the framework of the spin-wave theory. \cite{Starykh06,Chernyshev06}
It was shown that 
the first-order $1/S$ correction strongly modifies the LSWT 
spectrum in an overall qualitative agreement with the series-expansion data. 
Further detailed comparison of the series-expansion and 
the spin-wave spectra has confirmed their qualitative similarities
and outlined remaining differences. \cite{Zheng06}

Simply by the virtue of observing such
significant differences in the triangular- and square-lattice spectra, 
these works have explained the contrast in 
the thermodynamic behavior of the two systems. 
Since the spin-wave bandwidth of the triangular-lattice HAF is 
reduced to $W\approx J$ and the other features such as the 
``roton-like'' minimum are at the energies $\approx J/2$, \cite{Zheng06} 
the thermodynamic quantities must be dominated by these short-wavelength 
features down to much lower temperatures than in 
the square lattice case where the bandwidth is 
$W\approx 2J$ and the spectrum has a rather benign shape.

In the recent Letter \cite{Chernyshev06}
we have focused on another distinct feature of the spin-wave
spectrum in the triangular-lattice antiferromagnet: 
intrinsic damping due to 
spontaneous decays. 
Note, that although the spectrum renormalization in the triangular-lattice HAF
is significant, the truly dramatic {\it qualitative} difference 
from the square-lattice case
is the finite lifetime of the excitations at $T=0$.
Aside from yielding a substantial damping
for excitations in the most of the Brillouin zone, the first $1/S$ correction 
produces the logarithmic singularity in the imaginary part of
$\varepsilon_{\bf k}$ 
accompanied by the jumps in the real part along some contours in the
${\bf k}$-space. 
We have analyzed the origin of such singularities and 
have related them to the topological transitions in the decay surfaces 
of magnons, which are due to the saddle-point van Hove 
singularities in the two-magnon continuum.
Whether such logarithmic singularities  
are the artifact of the $1/S$ approximation of the theory 
or are the true features of the spectrum was
only partially addressed in our work. \cite{Chernyshev06}
For the decays into magnons that themselves acquire finite lifetimes, 
singularity will disappear in the higher $1/S$ order 
and thus is not ``real''. 
However, if the singularity is due to decays into stable 
excitations, we concluded that the singularity remains essential and 
the magnon damping should remain very strong.\cite{Chernyshev06}

We would like to mention that the numerical study 
did not observe jumps in the real part of the spectrum,
nor it reported the damping. \cite{Zheng06} 
This can be used as an argument
against any ``physical'' singularities. However, 
by design,  the series-expansion method 
finds the spectrum that is  purely real. 
Even with such restrictions, there where certain ${\bf k}$-points 
for which the numerically obtained $\varepsilon_{\bf k}$ 
had a convergence problem, shown by large error bars
in Ref.~\onlinecite{Zheng06}.

Altogether, the qualitative questions remain: 
why there is such a substantial difference of the excitation
spectrum in the triangular-lattice HAF from 
the more conventional square-lattice case? How generic 
are the decays, singularities, and anomalously large role of 
spin-wave interactions? What happens to the spectrum near the singular
points? It is the purpose of the present work to address these 
questions.
First, we would like to provide a consistent overview of the 
spin-wave formalism as applied to the  triangular-lattice model
in order to demonstrate the origin of the drastic differences between the 
spectra of the collinear and noncollinear ordered magnets. 
Second, we will elaborate on our previous findings on  magnon decays 
and demonstrate that they must exist in a wide variety 
of frustrated spin models. Finally, we would like
to extend our study beyond the on-shell approximation of the previous works
and clarify the fate of the singularities in the spectrum. 

The rest of the paper is organized as follows: 
\begin{itemize}
\item \vskip -0.2cm
Sec.~\ref{qualitative} provides
a qualitative discussion of the origin of the spectrum anomalies in 
the triangular-lattice HAF
and suggests a broader framework for the subsequent results. 
\item \vskip -0.2cm
Sec.~\ref{formalism} gives a detailed description of 
the spin-wave formalism for a noncollinear HAF. 
Here, we also address the controversy 
over two different results for the $O(1/S^2)$ correction to the 
sublattice magnetization of the triangular-lattice HAF, which exist in 
literature.\cite{Miyake92,Chubukov94} 
\item \vskip -0.2cm
The first-order $O(1/S)$ quantum corrections for the spin-wave
spectrum are considered in Sec.~\ref{onshell}.
Here we discuss the characteristic long-wavelength and short-wavelength 
features of both the spectrum renormalization and the decay rates.
\item \vskip -0.2cm
In Sec.~\ref{kinematic}, we discuss in detail 
the kinematic decay conditions for a generic single-particle spectrum 
coupled to the two-particle continuum. The relation between the 
singularities in the renormalized spectrum and the topological
transitions in the decay surfaces is established in this Section. 
\item \vskip -0.2cm
Sec.~\ref{off_shell} is devoted to the discussion of the 
off-shell solution of the Dyson equation in the complex plane. 
We show that in the strong-coupling regime, the solution may
cease to exist and the magnon pole may disappear in the vicinity of 
the logarithmic singularity. 
The results of the numerical off-shell solution of the Dyson
equation for the triangular-lattice HAF are presented and it is 
shown that the damping remains very substantial. 
In this Section we also present the results for the spectral 
function for several representative ${\bf k}$-point and 
the quasiparticle residues across the Brillouin zone. We show that 
the log-singularities become even more prominent 
in the spectral functions, complicating the conventional analysis.
\item \vskip -0.2cm
In Sec.~\ref{models} we discuss several examples of other models 
in which decays and singularities 
are ubiquitous, such as the easy-plane $XXZ$ and orthorhombically distorted 
$J$--$J'$ models on a triangular lattice as well as the $XXZ$ model on a
kagom\'{e} lattice.
\item \vskip -0.2cm
Sec.~\ref{conclusions} contains final conclusions and Appendices 
are used to provide more technical details.
\end{itemize}

\section{Qualitative discussion}
\label{qualitative}
\subsection{Cubic anharmonicities}

In the case of quantum Heisenberg antiferromagnets on bipartite 
lattices in $D\geq 2$, the spin-wave theory agrees extremely  well
with the available numerical data, even for the
``most quantum'' spin-1/2 square-lattice model. 
\cite{Zheng91,Sandvik94,Sandvik05,Zheng05}  
Therefore, not only one needs to understand the apparent anomalies
found for the Heisenberg antiferromagnet on a triangular lattice, 
but also to explain why such anomalies are not present in 
the properties of the same model on bipartite lattices. 
In the following we give
qualitative arguments that the unusual behavior 
stems from non-collinearity of spins in the ordered
state, which is, in turn, induced by geometrical frustration.

In quantum magnets with {\it collinear} spin configurations, 
the interaction between spin-wave excitations
is described by quartic and higher-order anharmonicities.
\cite{Dyson56,Oguchi60,Harris71}
In contrast, a general quantum system with non-conserved number 
of particles is {\it expected} to have anharmonicities of all orders
beginning with cubic terms, which describe interaction 
between one- and two-particle states.
The common examples are phonons in crystals \cite{Ziman} 
and excitations in superfluid bosonic systems. \cite{LL_IX}
In quantum spin-liquids, such as spin ladders and various
dimer systems, triplet excitations may also have three-particle 
interaction terms. \cite{Zhitomirsky06,Stone06} 

For the ordered spin systems, cubic anharmonicities correspond to
coupling of the transverse (one-magnon) and longitudinal 
(two-magnon) fluctuations, which would require the presence of 
mixing terms between $S^z$ and $S^{x,y}$ spin components. 
Such terms are absent in the collinear Heisenberg magnets
due to remaining $U(1)$ rotational symmetry 
about the direction of the magnetic order parameter.
On the other hand, 
in the {\it noncollinear} antiferromagnets
spin canting produces coupling of the transverse fluctuations in
one sublattice to the longitudinal ones in the others.
As a result, this yields cubic terms in the magnon-magnon interaction.
\cite{Miyake85,Miyake92,Ohyama93,Chubukov94,Nikuni98,Perkins_08}

Noncollinear spin configurations in the Heisenberg antiferromagnets
can be induced either by an external magnetic field
or by frustrating interactions. 
In the former case, spin canting and cubic anharmonicities
are small in weak fields.
At higher fields, cubic interactions dominate and lead to 
spontaneous magnon decays above threshold field
$H^*$. \cite{field,Olav08,Lauchli_09}  
In frustrated magnets, spin canting and cubic terms are substantial 
already in zero field and thus play the key role in 
the spectrum renormalization.

The above discussion creates a broader view on the ordered 
quantum magnets and their spectra. Magnets with collinear spin structures 
have excitations that are intrinsically weakly coupled. 
Not only the energy is minimized by the collinear spin orientation, but the 
interactions are also weak. Having a 
noncollinear spin structure necessarily implies 
much stronger coupling among the excitations. 
Thus, the collinear antiferromagnets
should be considered if not as an exception, but at least as a
simplified subclass of quantum antiferromagnets. Naturally, many
of their properties that are commonly assumed to be valid for all 
ordered antiferromagnets, such as the ubiquitously close agreement of the
harmonic theory with numerical results, should not be expected 
to hold in general.

Somewhat more formally, the lack of cubic anharmonicities in collinear
spin systems results in the absence of certain class of diagrams 
in the perturbative expansion. In the first order in $1/S$, 
the only non-vanishing contribution is given by
simple ``balloon'' Hartree-Fock-like terms  (see 
Fig.~\ref{vertices}). Since these are $\omega$-independent, 
the corresponding correction to  the spin-wave energy is a trivial
redefinition of  
the interaction strength. The first $\omega$-dependent correction 
appears only in the second order in $1/S$ and is already quite weak due to
small phase-space factor. In noncollinear spin
systems, cubic anharmonicity generates ``bubble'' diagrams already in 
the lowest $1/S$-order, not only inducing
a substantial spectrum correction but 
also providing a channel for the decays if the decay conditions are 
fulfilled. In the triangular-lattice HAF both effects are amplified because of
the lower dimensionality and because the tilting angle between spins
is not small.
Hence, the strong coupling of the longitudinal and transverse modes in
noncollinear magnets in general and in the triangular-lattice HAF
in particular is the key to understanding strong renormalization of
their spectra from the results of the harmonic approximation.

\subsection{Decays and singularities}

Another important property of noncollinear magnets 
is the ubiquitous propensity of the excitations 
to spontaneous decays. 
The presence of transverse-to-longitudinal coupling is
a necessary but not a sufficient condition 
for decays. In addition, the energy and momentum must be conserved
in the decay process, yielding kinematic restrictions.
Decays depend, therefore, on the {\it shape} of 
the single-particle dispersion that may, or may not, allow for
spontaneous decays.
While we will give a detailed classification of various 
kinematic conditions later (see Sec.~\ref{kinematic}), 
there are simple arguments for the
decays to exist in the triangular-lattice HAF. The complete
breaking of the $SO(3)$ rotational symmetry
by the $120^\circ$ spin structure leads to 
three Goldstone modes,\cite{Dombre89} 
one at the center of the Brillouin zone and two
at the ordering vectors $\pm{\bf Q}$. 
The former mode can be related to an infinitesimal twist of spins
about the axis which is perpendicular to the spin plane, 
while the other two correspond to twists about the axes
lying in the spin plane.
The velocity of the latter modes is smaller than that 
of the former one.
Such a difference guarantees that the energy conservation
can always be satisfied for the decay of the fast quasiparticle
into a pair of slower magnons, similar to the decay of 
the longitudinal phonon into 
a pair of the transverse ones in crystals. \cite{Ziman} Since cubic 
anharmonicities necessarily generate couplings between all magnon branches, 
this ensures finite lifetime for spin excitations in an extended 
part of the Brillouin zone at 
$T=0$. The above consideration is trivially generalized to other
frustrated antiferromagnets where symmetry guarantees existence
of more than one type of the Goldstone mode. We will elaborate
on the decays in the $XXZ$ anisotropic and orthorhombically distorted 
triangular-lattice AFs as well as on the kagom\'{e} AF in Sec.~\ref{models}.  
We would like to note that the kinematic conditions for decays 
can be completely suppressed by magnetic anisotropies,
still leaving spins in a noncollinear configuration.
In that case decays are absent but 
the renormalization of the real part of the spectrum
due to spin-wave interaction remains substantial.

In the following, we also discuss extensively the singularities 
that occur in the decay rates of single-particle spectra due to the van Hove 
singularities in the two-particle continuum. While the main conclusion is 
that the essential singularities are cut-off either by the finite lifetime 
of the decay products or by the non-singular finite lifetime of 
the decaying particle, the decay rates remain parametrically 
enhanced by such singularities. A qualitative estimate
of such a singularity-enhanced decay rate 
in the two-dimensional case is
\begin{equation}
\Gamma_{\bf k} \simeq (V_3^2/SJ) \ln (\Lambda S) \ ,
\label{qualit_gamma}
\end{equation}
where $V_3\sim \sqrt{S}J$ is the strength of a three-particle decay vertex and 
$\Lambda$ is the momentum cut-off.
Even for large values of spin $S$ the decay rate $\Gamma_{\bf k}$ in (2) is
logarithmically enhanced relative to a perturbative
result $\Gamma_{\bf k} \sim J$.

\section{Spin-Wave Formalism }
\label{formalism}

\subsection{Bosonic Hamiltonian}

\begin{figure*}[t]
\includegraphics[width=0.6\columnwidth]{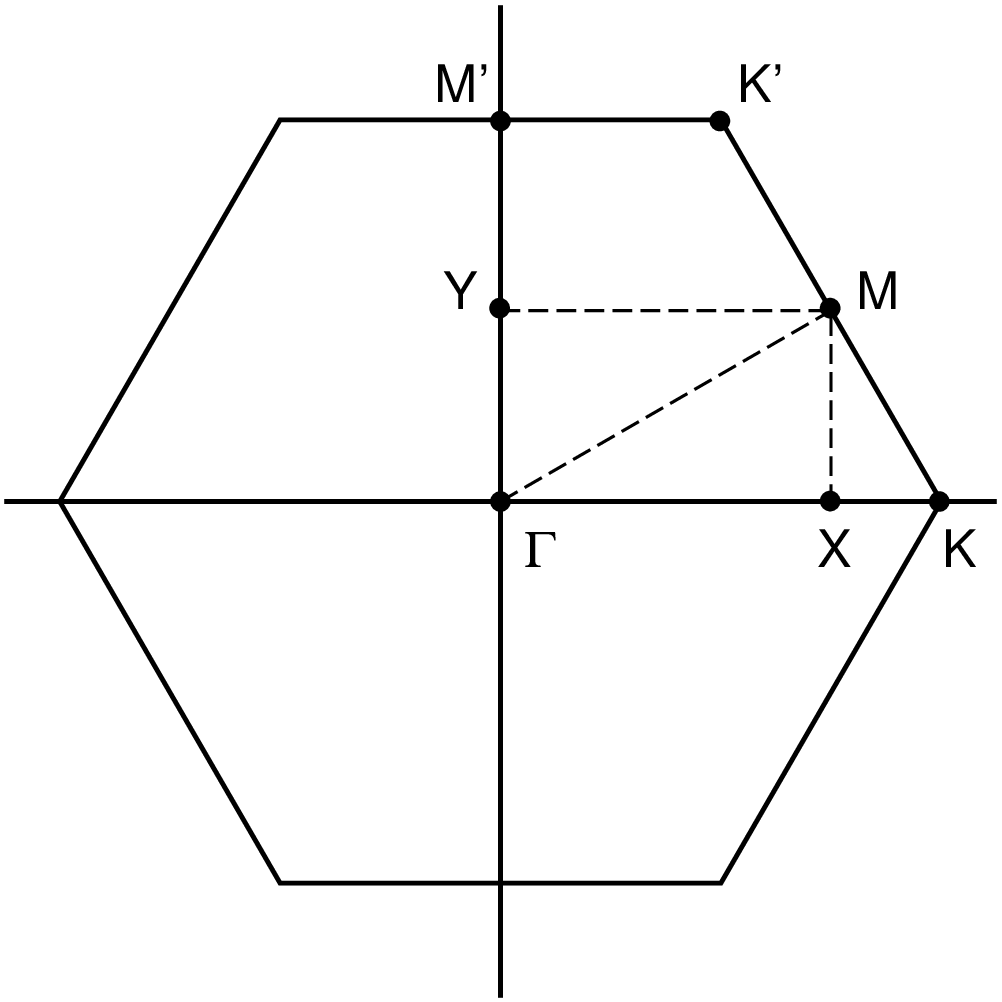}\ \hskip 1cm
\includegraphics[width=0.6\columnwidth]{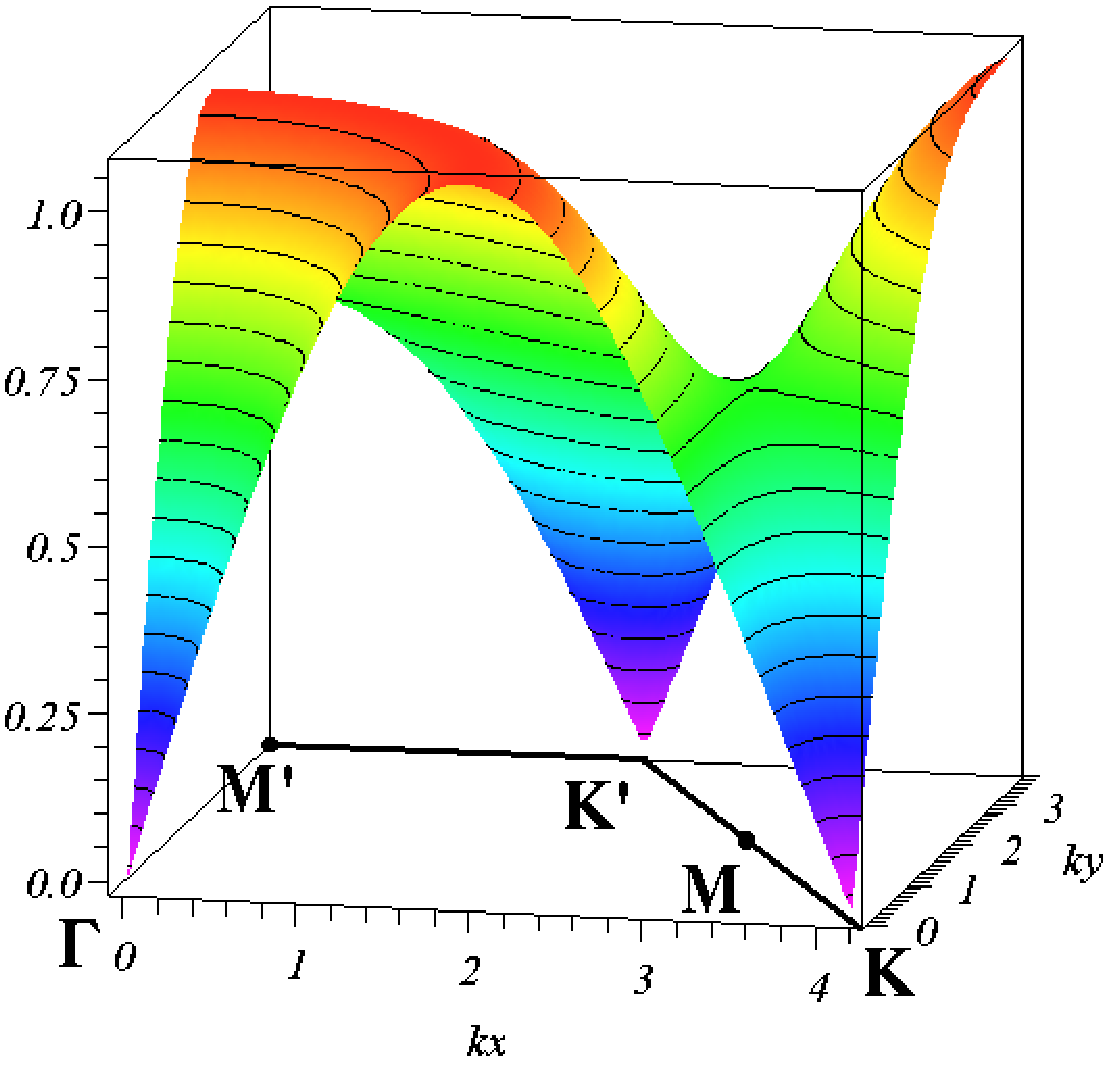}\ \hskip 1cm
\includegraphics[width=0.6\columnwidth]{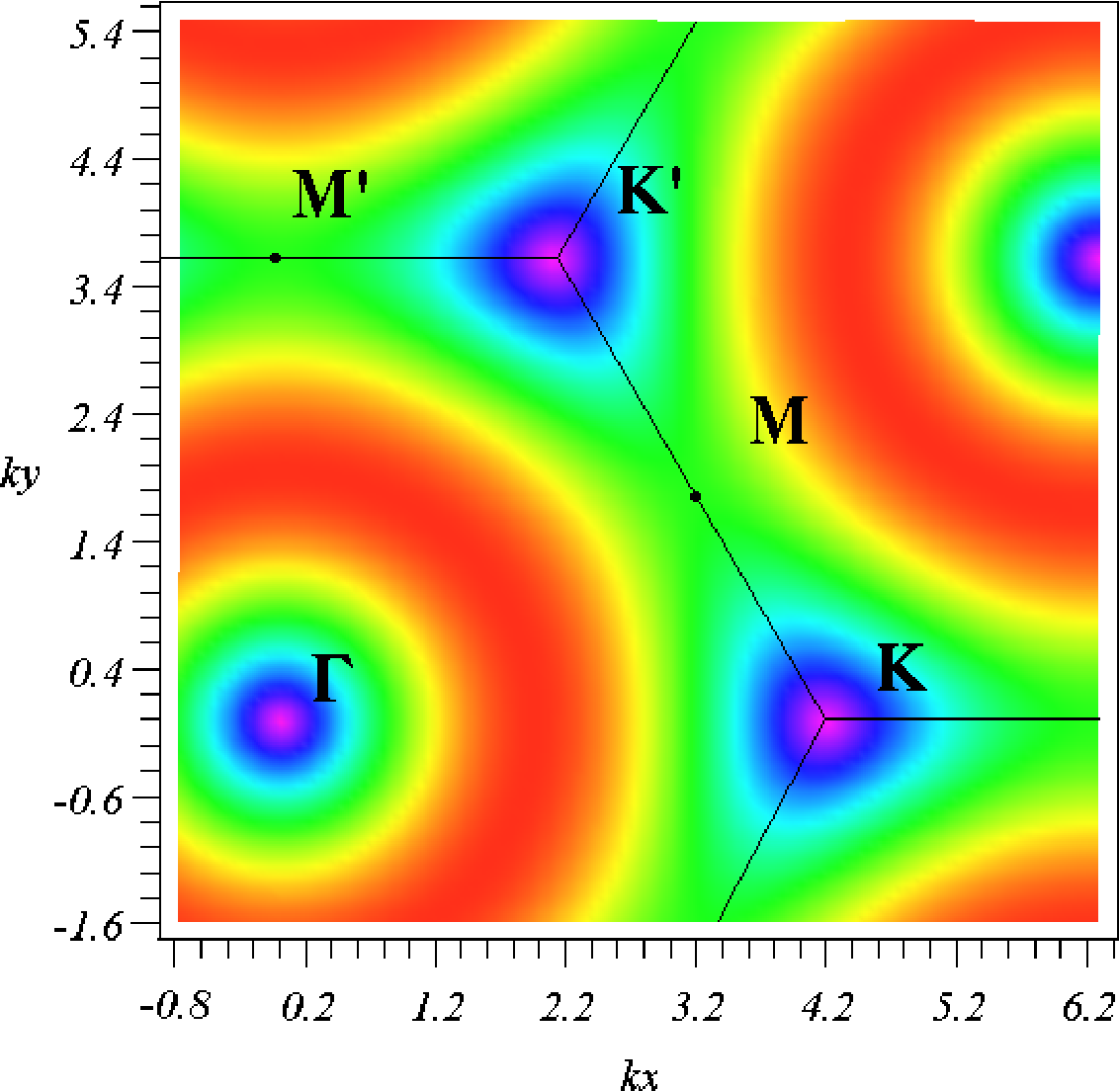}
\caption{(Color online) 
Left panel: the Brillouin zone of a triangular lattice, lines are 
representative cuts.
Central panel: 3D plot of the linear spin-wave energy 
$\omega_{\bf k}$ in the triangular lattice HAF.
Right panel: intensity plot of $\omega_{\bf k}$. 
Note different velocities of 
the Goldstone modes 
and different symmetries  of the dispersion 
near $\Gamma$ and $K$ $(K')$ points, 
and the saddle point at $M$ ($M'$).
}
\label{shape_wk}
\end{figure*}

The spin-wave theory of collinear antiferromagnets 
on bipartite lattices is commonly formulated
by introducing bosonic operators according to the number
of magnetic sublattices.\cite{Oguchi60,Harris71}
In the Heisenberg triangular-lattice antiferromagnet spins form the 
three-sublattice $120^\circ$ structure at $T=0$,
see Fig.~\ref{structure}. In order to go beyond the
linear spin-wave analysis, \cite{Oguchi83,Jolicoeur89}  
it is convenient to transform to a rotating frame
with the $z$-axis pointing along the local 
spin direction. Then, one needs to define only single species of
boson operators  with the wave-vectors that belong to the full paramagnetic 
Brillouin zone (BZ), Fig.~\ref{shape_wk}. 
We would like to note that even in collinear
antiferromagnets the above approach has substantial advantage when
calculating 
higher-order spin-wave corrections. Furthermore, such a 
``single-sublattice'' procedure has no alternatives when considering 
generic, incommensurate (spiral-like) antiferromagnetic ordering. 

As a first step, we assume that spins lie in
the $x$--$z$ plane and perform transformation from the laboratory frame 
$(x_0,z_0)$ into the rotating frame $(x,z)$:
\begin{eqnarray}
S_i^{z_0} & = & S_i^z \cos\theta_i - S_i^x \sin\theta_i  \ ,\nonumber \\
S_i^{x_0} & = & S_i^z \sin\theta_i + S_i^x \cos\theta_i  \ ,
\label{transformation}
\end{eqnarray}
where $\theta_i={\bf Q}\cdot{\bf r}_i$ and ${\bf Q}=(4\pi/3,0)$ is 
the ordering wave-vector of the $120^\circ$ spin structure. 
The spin Hamiltonian, Eq. (\ref{TLH}), 
in the new coordinate system takes the following form:
\begin{eqnarray}
\hat{\cal H} & = & 
J \sum_{\langle ij\rangle} \Bigl[ S_i^yS_j^y  + 
\cos(\theta_i-\theta_j) (S^z_iS^z_j +  S^x_iS^x_j)     
\nonumber  \\
& & \mbox{} + 
\sin(\theta_i-\theta_j)(S^z_iS^x_j-S^x_iS^z_j) \Bigr]\, ,
\label{H1}
\end{eqnarray}
where $\langle ij\rangle$ denotes, as usual, summation over 
the nearest-neighbor bonds.

The bosonization of Eq.~(\ref{H1}) is performed via 
the hermitian Holstein-Primakoff transformation:
\begin{eqnarray}
S_i^z=S- a^\dagger_ia_i\ , \ \ 
S_i^-=a^\dag\sqrt{2S- a^\dagger_ia_i} 
\label{HP}
\end{eqnarray} 
($S_i^\pm = S_i^x \pm iS_i^y$) with subsequent expansion of square roots 
to the first order in $a^\dagger_i a_i/2S$.
Such an approximation is sufficient to calculate
the $O(1/S^2)$ corrections to the ground-state energy and 
to the sublattice magnetization and to determine
the $O(1/S)$ correction to the spin-wave dispersion.
The resultant spin-wave Hamiltonian is given by:
\begin{equation}
\hat{\cal H}_{SW} = \hat{\cal H}_0 + \hat{\cal H}_2+
\hat{\cal H}_3+\hat{\cal H}_4 + O(S^{-1}), 
\label{H_SW}
\end{equation}
where $\hat{\cal H}_n$ denote terms of the
$n$th power in the original (Holstein-Primakoff) boson operators
$a^\dagger_i$ and $a_i$.  
The first term in this expansion 
is the classical energy $\hat{\cal H}_0 = -\frac{3}{2}JS^2N$. 
Terms that are linear in $a_i$ and $a^\dagger_i$ vanish automatically 
because the 120$^\circ$ spin-structure corresponds to a minimum
of the classical energy. Note, that the Hamiltonian (\ref{H_SW})
also yields a series in powers of $S$, with 
$\hat{\cal H}_0=O(S^{2})$,
$\hat{\cal H}_2=O(S^{1})$,
$\hat{\cal H}_3=O(S^{1/2})$, and
$\hat{\cal H}_4=O(S^{0})$, respectively. 
The subsequent treatment of $\hat{\cal H}_{SW}$ consists
of diagonalizing the harmonic part $\hat{\cal H}_2$ 
exactly  and, then, treating 
$\hat{\cal H}_3$ and $\hat{\cal H}_4$ as perturbations.

\subsubsection{Linear spin-wave theory}

Noninteracting magnons are described by the linear spin-wave
theory (LSWT) or the harmonic approximation.
After the Fourier transformation, 
the quadratic part of $\hat{\cal H}_{SW}$ reads  
\begin{eqnarray}
\hat{\cal H}_2 & = & \sum_{\bf k} \:
\Bigl[A_{\bf k}\,a^\dagger_{\bf k} a_{\bf k} - \frac{1}{2}\,B_{\bf k}\,
\bigl(a^\dagger_{\bf k} a^\dagger_{-\bf k} +
a_{-\bf k} a_{\bf k} \bigr) \Bigr] \ , \nonumber \\  
 & & A_{\bf k} = 3JS (1+{\textstyle \frac{1}{2}}\gamma_{\bf k}) \ , \ \ 
B_{\bf k} =   \frac{9}{2}JS \gamma_{\bf k} \ ,  
\label{H2f}
\end{eqnarray}
where $\gamma_{\bf k}$ is a sum over the nearest-neighbor sites
\begin{equation}
\gamma_{\bf k} = \frac{1}{6} \sum_{\mbox{\boldmath\scriptsize  $\delta$}}
e^{i{\bf k}\mbox{\boldmath\scriptsize   $\delta$}}
= \frac{1}{3}\,\Bigl(\cos k_x + 2 
\cos\frac{k_x}{2}\cos\frac{\sqrt{3}}{2}k_y \Bigr).
\end{equation}

Diagonalization of $\hat{\cal H}_2$ is performed
with the help of the canonical Bogolyubov transformation: 
\begin{equation}
a_{\bf k} = u_{\bf k} b_{\bf k} + v_{\bf k} b^\dagger_{-\bf k} \ 
\label{bogoliubov1}
\end{equation}
under conditions $u_{\bf k}^2 - v_{\bf k}^2 = 1$ and 
\begin{equation}
u_{\bf k}^2+v_{\bf k}^2 = \frac{A_{\bf k}}{\sqrt{A_{\bf k}^2-B_{\bf k}^2}}
\ , \ \ \ \
2u_{\bf k}v_{\bf k} = \frac{B_{\bf k}}{\sqrt{A_{\bf k}^2-B_{\bf k}^2}}\ . 
\label{bogoliubov2}
\end{equation}
As a result, the linear spin-wave Hamiltonian 
takes the following form: 
\begin{equation}
\hat{\cal H}_0+\hat{\cal H}_2 = -\frac{3}{2}\,JS(S+1)N
+ \sum_{\bf k} \varepsilon_{\bf k} \Bigl(b^\dagger_{\bf k}b_{\bf k} +
\frac{1}{2}\Bigr)\ .
\label{H2}
\end{equation}
In the harmonic approximation, spin-waves are noninteracting
bosons with the energy
\begin{equation}
\varepsilon_{\bf k} = \sqrt{A_{\bf k}^2-B_{\bf k}^2} = 
3JS \omega_{\bf k}\ ,
\label{epsilon}
\end{equation}
where we define the dimensionless frequency 
\begin{equation}
\omega_{\bf k} = 
\sqrt{(1-\gamma_{\bf k})(1+2\gamma_{\bf k})} \ .
\label{omega}
\end{equation} 
The shape of $\omega_{\bf k}$ is shown in Fig.~\ref{shape_wk}.
The harmonic spectrum of the triangular-lattice HAF has several distinct
features. The intensity map clearly shows that 
the velocities of the Goldstone modes at $\Gamma$ (${\bf k}=0$)
and $K$, $K'$ (${\bf k}=\pm {\bf Q}$) points 
are different. They are given by
\begin{equation}
v_0^{(0)} = 3JS \, \frac{\sqrt{3}}{2}\ ,\ \ \ \ \  
v_Q^{(0)} = 3JS \, \sqrt{\frac{3}{8}}\ .
\label{v0}
\end{equation}
Another notable feature is the clear three-fold symmetry of 
the modes near ${\bf Q}$ points. Instead of the usual 
convex and isotropic form, the spin-wave energy 
at small $\tilde{\bf k}={\bf k}-{\bf Q}$ is nonanalytic
with varying convexity: 
\begin{equation}
\varepsilon_{\bf k}\approx v_Q 
\tilde k(1-\alpha_\varphi\tilde k)\ , \ \ \ \mbox{where} \ \ \ 
\alpha_\varphi \sim \cos 3\varphi\ . 
\label{wQ}
\end{equation}
The overall shape of the dispersion 
is also  more complicated than in the square-lattice antiferromagnet.
At the $M$ point, the center of the BZ edge, $\omega_{\bf k}$ 
has a saddle point with the energy roughly 
half of the bandwidth. Thus, already with the 
harmonic spectrum the thermodynamic response 
at intermediate temperatures should be quite 
different from that in the square-lattice antiferromagnet.

We introduce now several non-zero Hartree-Fock averages 
that contribute to the spin-wave corrections of 
many static and dynamic quantities of the triangular-lattice HAF:
\begin{eqnarray}
n = \langle a^\dagger_i a_i\rangle , \ \ m = \langle a^\dagger_i a_j\rangle, 
\ \ \Delta = \langle a_i a_j\rangle , \ \ 
\delta = \langle a^2_i\rangle. 
\label{HF}
\end{eqnarray}
In the harmonic theory 
these are expressed as linear combinations of the 2D integrals 
\begin{equation}
c_l = \sum_{\bf k} \frac{(\gamma_{\bf k})^l}{\omega_{\bf k}}  \ ,
\label{Cs}
\end{equation}
with $l=0,1,2$, see Appendix \ref{app_A} for further details.
In particular, the linear spin-wave correction to the staggered
magnetization, $\langle S\rangle = S-\delta S$, is given by: 
\begin{equation}
\delta S \equiv n =\frac14\left(2c_0-2+c_1\right)= 0.2613032. 
\label{dS1}
\end{equation}

\subsubsection{Spin-wave interaction: cubic terms}

The cubic interaction terms $\hat{\cal H}_3$ in Eq.~(\ref{H_SW}) 
have no analog in collinear magnets. They originate from 
the coupling of local $S^z$ and $S^x$ spin components 
in Eq.~(\ref{H1}). In terms of the original  boson operators (\ref{HP}) 
the cubic interaction  is given by
\begin{equation}
\hat{\cal H}_3 =  J\sqrt{\frac{S}{2}} \sum_{\langle ij\rangle} 
 \sin(\theta_j-\theta_i)\Big[a^\dagger_i a_i (a^\dagger_j + a_j)
- a^\dagger_j a_j (a^\dagger_i + a_i)\Big] \ .
\label{H3}
\end{equation}
For the collinear spin structures, $\sin(\theta_j-\theta_i)\equiv 0$
and the cubic terms vanish identically.

Performing consecutively the Fourier and Bogolyubov transformations from
$a^\dagger_i$, $a_i$ to $b^\dag_{\bf k}$, $b_{\bf k}$ operators in 
$\hat{\cal H}_3$ we obtain the interaction terms expressed via the ``new''
bosons:
\begin{eqnarray}
\label{H3b}
\hat{\cal H}_3 & =  & \sum_{{\bf k},{\bf q}}\Bigl[ \frac{1}{2!}\:
\Gamma_1({\bf q},{\bf k-q};{\bf k})\: 
b^\dagger_{\bf q} b^\dagger_{\bf k-q} b_{\bf k}   \\ 
&& \mbox{}\ \ \ + 
\frac{1}{3!} \,
\Gamma_2({\bf q},{\bf -k-q},{\bf k})\, 
b^\dagger_{\bf q}  b^\dagger_{\bf -k-q} b^\dagger_{\bf k}
+ {\rm h.\,c.} \Bigr]. 
\nonumber
\end{eqnarray}
Generally, terms linear in 
$b^\dag_{\bf Q}$ and $b_{\bf Q}$ also appear after the above substitution.
They represent quantum correction to the pitch angle of the
spin helix \cite{Ohyama93,Veillette05,Dalidovich06}
or to the spin canting angle in an external magnetic field.
\cite{Nikuni98,field,Kopietz08} 
Such a correction vanishes in the triangular-lattice HAF 
because the ordering wave-vector ${\bf Q}$ corresponds to
a stable symmetry point in BZ.

\begin{figure}[t]
\centerline{
\includegraphics[width=0.95\columnwidth]{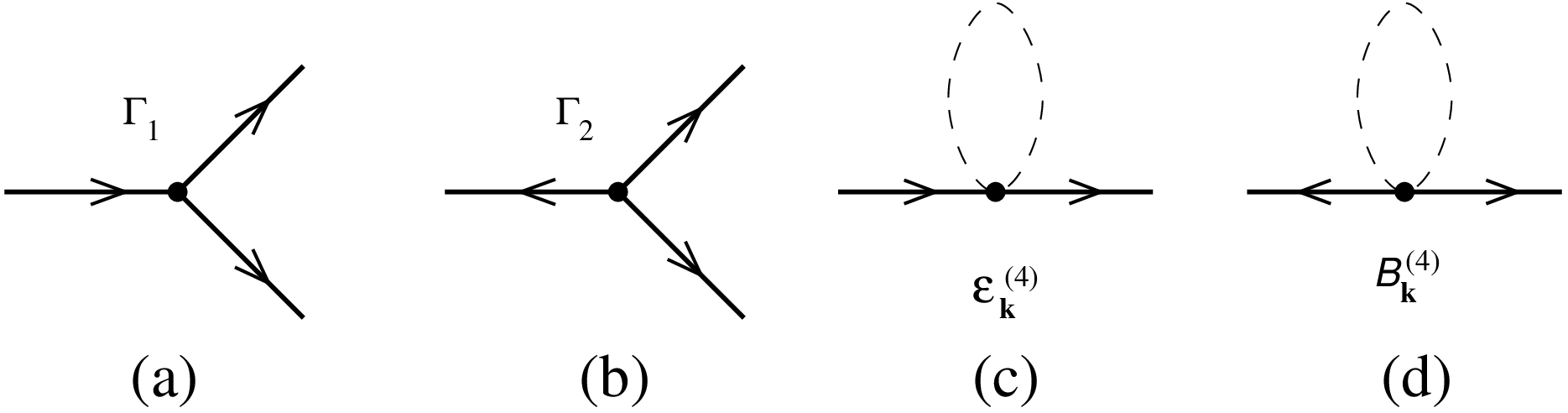}}
\caption{The lowest-order vertices that yield 
$1/S$-corrections to the spectrum and $1/S^2$
contributions to the static properties.
}
\label{vertices}
\end{figure}

The first term in Eq.~(\ref{H3b})
describes interaction between one- and two-magnon 
states and is symmetric under permutation of two outgoing momenta. 
We call it the ``decay'' term, 
although the decay processes   
may be only virtual. The second term in Eq.~(\ref{H3b}) corresponds 
to the spontaneous creation of three magnons 
and we refer to it as to the ``source'' term. 
The source vertex is symmetric under permutation
of all three momenta. Making the energy- and spin-dependence 
of the three-boson interactions explicit, we define dimensionless vertices
related to the original ones (\ref{H3b}) by
\begin{equation}
\Gamma_{1,2}(1,2;3) = 3iJ\sqrt{\frac{3S}{2}}
\;\widetilde{\Gamma}_{1,2}(1,2;3) \ ,
\end{equation}
with $\widetilde{\Gamma}_{1,2}$  given by
\begin{eqnarray}
& & \widetilde\Gamma_1(1,2;3) =  
\bar\gamma_1(u_1+v_1)(u_2u_3 + v_2v_3) + \bar\gamma_2(u_2 + v_2)  
\nonumber \\
&& \mbox{} \ \ \ 
\times (u_1u_3 + v_1v_3)-\bar\gamma_3(u_3 + v_3)(u_1v_2 + v_1u_2), 
\nonumber \\ 
&& \widetilde\Gamma_2(1,2;3) = 
\bar\gamma_1 (u_1+v_1)(u_2v_3+v_2u_3) + \bar\gamma_2(u_2 + v_2)  
\nonumber \\
&& \mbox{} \ \ \ 
\times (u_1v_3 + v_1u_3)+\bar\gamma_3 (u_3 + v_3)(u_1v_2 + v_1u_2),
\label{G12}
\end{eqnarray}
where $u_i$, $v_i$ are the Bogolyubov parameters (\ref{bogoliubov2}) 
and the function $\bar{\gamma}_{\bf k}$ is defined as 
\begin{equation}
\bar{\gamma}_{\bf k} = \frac{1}{3} \Bigl(\sin k_x - 2 
\sin\frac{k_x}{2}\cos\frac{\sqrt{3}}{2}k_y \Bigr).
\end{equation}
The diagrammatic representation of the decay and the source vertices
is shown in Figs.~\ref{vertices}(a) and \ref{vertices}(b), respectively. 
The above form of $\Gamma_{1,2}$ coincides with the expressions 
used by Miyake. \cite{Miyake85,Miyake92}

\subsubsection{Spin-wave interaction: quartic terms}

The last term in the spin-wave Hamiltonian (\ref{H_SW}) 
represents the quartic terms:
\begin{eqnarray}
\hat{\cal H}_4 & =&\frac{J}{4} \sum_{\langle ij\rangle}\: 
\Bigl[ - a^\dagger_i a_i a^\dagger_j a_j 
+ \frac{3}{4}\: \bigl(a^\dagger_i a_i a_i a_j 
+ a^\dagger_j a_j a_j a_i \bigr)  
\nonumber \\
 & & \mbox{} -\frac{1}{4}\:\bigl(a^\dagger_j a^\dagger_i a_i a_i 
+ a^\dagger_j a^\dagger_j a_j a_i  \bigr)
\Bigr]  + {\rm h.c.}
\label{H4}
\end{eqnarray}
After the Bogolyubov transformation and normal ordering
of $b$ operators, the interaction (\ref{H4}) is replaced with 
\begin{eqnarray}
\hat{\cal H}_4  =\delta E_4+\delta\widetilde H_2 +\widetilde H_4\ ,
\label{H4b}
\end{eqnarray}
where the first two terms are the Hartree-Fock 
corrections to the ground-state energy and to the magnon
self-energy, Fig.~\ref{vertices}(c) and \ref{vertices}(d), respectively.
The magnon self-energy
$\delta\widetilde H_2$ contains both the diagonal  and the
off-diagonal terms, while for the 
square-lattice HAF the anomalous off-diagonal terms vanish. 
The final term 
$\widetilde H_4$ describes two-particle scattering processes.
Similar to the case of the collinear antiferromagnets, \cite{Harris71}
this latter term yields only higher-order $1/S$ corrections compared to
$\delta E_4$ and $\delta\widetilde H_2$ and is, therefore, neglected
in the present work.

To derive the explicit form of Eq.~(\ref{H4b}), 
it is technically more straightforward 
to apply the Hartree-Fock decoupling (\ref{HF}) to (\ref{H4}) first 
and use the Bogolyubov transformation 
in $\delta\widetilde H_2$ afterward. 
This is analogous to the treatment of quartic terms 
for collinear antiferromagnets. \cite{Oguchi60} 
With the details of this derivation delineated in Appendix \ref{app_A}, we
simply list the end result for the correction to the ground-state energy:
\begin{equation}
\delta E_{4} = -\frac{3}{8}J\Bigl[(c_0\!+\!c_1\!-\!2c_2\!-\!1)^2 
- \frac{3}{2}(c_1\!-\!c_2)^2 \Bigr],
\label{E4}
\end{equation}
and to the harmonic spin-wave Hamiltonian:
\begin{equation}
\delta\widetilde{\cal H}_2 = \sum_{\bf k} 
\varepsilon^{(4)}_{\bf k} b^\dagger_{\bf k}b_{\bf k}
-\frac{1}{2}\, B^{(4)}_{\bf k}\, \bigl( b_{\bf k} b_{-\bf k} + 
b^\dagger_{-\bf k} b^\dagger_{\bf k} \bigr) \ ,
\label{H24b} 
\end{equation}
where the first-order magnon energy correction and the anomalous 
self-energy terms are expressed  as
\begin{eqnarray}
&\varepsilon^{(4)}_{\bf k} &
= \frac{3J} {4\omega_{\bf k}}\Bigr[
\gamma_{\bf k}^2\,(4c_0 +c_1 -5c_2 -4) 
\label{e4} \\
&&+ \gamma_{\bf k}(2-2c_0 + c_1 + c_2) - 2c_0 -2c_1 +4c_2 + 2\Bigr],
\nonumber \\
\mbox{and}
&B^{(4)}_{\bf k} &
 =  - \frac{9J}{8\omega_{\bf k}}(1-\gamma_{\bf k})\bigl(c_1+2c_2\gamma_{\bf
  k}\bigr)\ ,
\label{B4}
\end{eqnarray}
respectively.

\subsubsection{Effective Hamiltonian}

With the triangular-lattice model discussed in detail, 
we would like to outline the structure of the spin-wave Hamiltonian
for a generic noncollinear spin system. 
After the Holstein-Primakoff and Bogolyubov transformations
with subsequent renormalization, when necessary,
of the classical configuration,
the spin-wave Hamiltonian takes the form of 
a polynomial of terms with increasing number of
bosons, $\widetilde{\cal H}_0 + \widetilde{\cal H}_2+
\widetilde{\cal H}_3 + \ldots $, 
and decreasing power of $S$. 
Therefore, the spin-wave expansion to order $O(S^0)$
will always result in an effective Hamiltonian:
\begin{eqnarray}
\label{Heff} 
\widetilde{\cal H}_{eff} & = & \sum_{\bf k} 
\Bigl[
\tilde\varepsilon_{\bf k} b^\dagger_{\bf k}b_{\bf k}
+ V^{od}_{\bf k}( b_{\bf k} b_{-\bf k} + 
b^\dagger_{\bf -k} b^\dagger_{\bf k}) \Bigr] \\
&+&\sum_{{\bf k},{\bf q}} \Bigl[V_{{\bf k},{\bf q}} 
b^\dagger_{\bf q} b^\dagger_{\bf k-q} b_{\bf k}  
+ F_{{\bf k},{\bf q}}
b^\dagger_{\bf q}  b^\dagger_{\bf -k\!-\!q} b^\dagger_{\bf k}
+ {\rm h.c.} \Bigr]. 
\nonumber
\end{eqnarray}
The spin-wave Hamiltonian for a {\it collinear} antiferromagnet
is the restricted form of (\ref{Heff}) with 
no three-boson terms present.  
The dynamic interaction among magnons 
in such a system occurs only in the next order due to the four-boson 
terms that are substantially weaker. 
In addition, for the commonly studied case of the nearest-neighbor
square- and cubic-lattice antiferromagnets, the anomalous quadratic 
terms in Eq.~(\ref{Heff}) also vanish, leaving only a benign
energy renormalization. Altogether, the role of magnon interactions
in collinear spin states is significantly less important
than for noncollinear ones.

\begin{figure}[t]
\centerline{
\includegraphics[width=0.95\columnwidth]{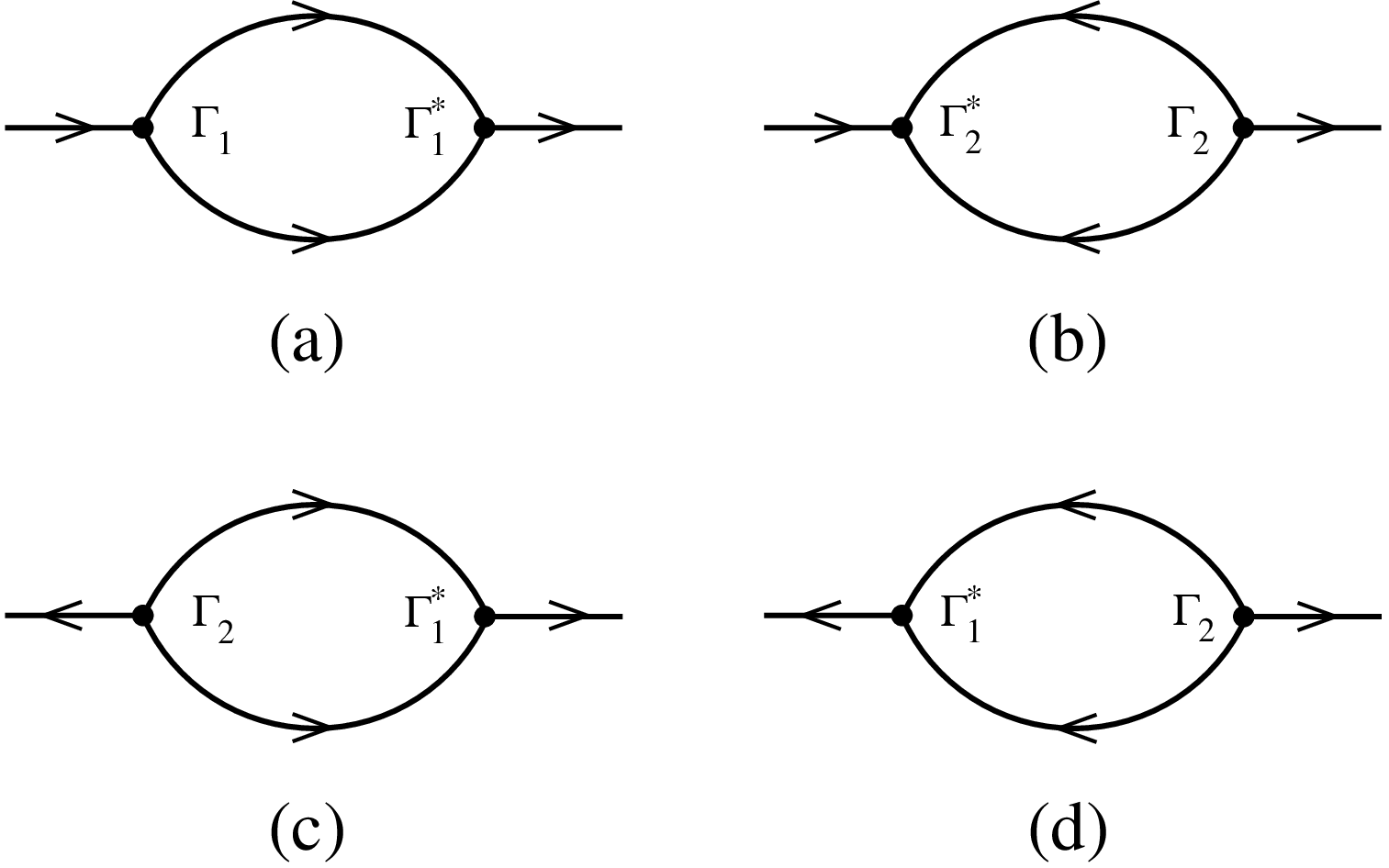}}
\caption{The lowest order normal, (a) and (b), and anomalous, (c) and (d), 
magnon self-energies generated by cubic vertices.
} 
\label{SelfE}
\end{figure}

The effective Hamiltonian (\ref{Heff}) provides 
a basis for the systematic perturbative calculations
of various properties of the triangular-lattice HAF.
Parameters of the Hamiltonian (\ref{Heff}) for the triangular-lattice case
were derived in previous subsections:
\begin{eqnarray}
\label{effTL} 
\widetilde\varepsilon_{\bf k} &=&\varepsilon_{\bf k}+
\varepsilon^{(4)}_{\bf k},\ \ \ 
V^{od}_{\bf k}=-\frac{1}{2}\, B^{(4)}_{\bf k}, \\
V_{{\bf k},{\bf q}}&=&\frac{1}{2!}\Gamma_1({\bf q},{\bf k-q};{\bf k}) ,\ \ \
F_{{\bf k},{\bf q}}=\frac{1}{3!}\Gamma_2({\bf q},{\bf -\!k\!-\!q},{\bf k}),
\nonumber
\end{eqnarray}
where $\varepsilon_{\bf k} \sim O(S^1)$,
$\varepsilon^{(4)}_{\bf k}$ and $B^{(4)}_{\bf k} \sim O(S^0)$, and
$\Gamma_{1,2} \sim  O(S^{1/2})$.

Using the standard 
diagrammatic technique for bosons at zero temperature, 
we define the bare magnon propagator as
\begin{equation}
G_0^{-1}({\bf k},\omega) = \omega - \varepsilon_{\bf k}+ i0 \ .
\end{equation}
Then, the lowest-order diagrams contributing in the order $O(S^0)$ 
to the normal and anomalous self-energies are shown
in Fig.~\ref{SelfE} and in Figs.~\ref{vertices}(c),~(d). 
The corresponding expressions for the normal self-energies are
\begin{eqnarray}
&& \Sigma_{11}^{(a)}({\bf k},\omega) = \frac{1}{2} \sum_{\bf q} 
\frac{|\Gamma_1({\bf q};{\bf k})|^2 }
{\omega - \varepsilon_{\bf q}-\varepsilon_{\bf k-q}+i0} \ , 
\label{SEa}
\\ 
&& \Sigma_{11}^{(b)}({\bf k},\omega)  = -\frac{1}{2} \sum_{\bf q} 
\frac{|\Gamma_2({\bf q};{\bf k})|^2 }
{\omega + \varepsilon_{\bf q} + \varepsilon_{\bf k+q} - i0} \ .
\label{SEb}
\end{eqnarray}
From now on we shall use, for brevity, only two momenta 
in the notations for the cubic vertices (\ref{G12}).
The anomalous self-energies are calculated as
\begin{eqnarray}
&& \Sigma_{12}^{(c)}({\bf k},\omega) = -\frac{1}{2} \sum_{\bf q} 
\frac{\Gamma_2({\bf q};{\bf k}) \Gamma_1^*({\bf q};{\bf -k})}
{\omega + \varepsilon_{\bf q} + \varepsilon_{\bf k+q} - i0} \ , 
\label{SEc}
\\ 
&& \Sigma_{12}^{(d)}({\bf k},\omega)  = \frac{1}{2} \sum_{\bf q} 
\frac{\Gamma_2({\bf q};{\bf -k}) \Gamma_1^*({\bf q};{\bf k})}
{\omega - \varepsilon_{\bf q}-\varepsilon_{\bf k-q}+i0}\ .
\label{SEd}
\end{eqnarray}
Finally, the two frequency-independent contributions to 
the normal and anomalous self-energies are given directly 
by the Hartree-Fock terms:
\begin{eqnarray}
\label{SEHF}
\Sigma^{HF}_{11}({\bf k})=\varepsilon^{(4)}_{\bf k} \ \ 
\mbox{and} \ \
\Sigma^{HF}_{12}({\bf k})=-B^{(4)}_{\bf k},
\end{eqnarray}
of Figs.~\ref{vertices}(c) and  \ref{vertices}(d), respectively. 
The structure of the lowest-order contributions 
(\ref{SEa})--(\ref{SEHF}) remain
valid, with necessary modifications of vertices, 
for an arbitrary noncollinear antiferromagnet.

\subsection{Static properties}

Previous works on the spin-wave theory for the triangular-lattice HAF
have calculated the second-order $1/S$ corrections to the ground-state
energy  and to the sublattice magnetization.  
Two approaches have been employed for calculation
of the latter: 
numerical extrapolation of the response to small staggered 
magnetic field \cite{Miyake92} and  a 
direct diagrammatic expansion.  \cite{Chubukov94} 
Surprisingly, they have 
produced two different results: $\Delta S_2 = 0.011/(2S)$
versus $0.027/(2S)$, respectively. While the latter approach has to deal
with more singular higher-dimensional integrals, the former one
relies on numerical extrapolation in a small parameter. 
Below, we resolve the controversy over the value of the staggered 
magnetization in favor of the Miyake's 
result \cite{Miyake92} by following the diagrammatic approach 
of Chubukov {\it et al.}\cite{Chubukov94}  and pointing out 
a delicate issue with numerical evaluation of 
canceling singularities under integrals. 
For completeness we also briefly discuss the ground-state 
energy correction.

\subsubsection{The ground-state energy}

The first two terms in the $1/S$ expansion of the 
ground-state energy of the triangular-lattice HAF are given
by Eq.~(\ref{H2}).
Here, we calculate the next-order correction
resulting from the magnon interactions.
The contribution from the quartic terms has already been obtained 
in the course of deriving the spin-wave Hamiltonian and  
is given by Eq.~(\ref{E4}). Another correction of the same order 
is generated by the source vertex, Fig.~\ref{vertices}b: 
\begin{equation}
\delta E_{3}  =  - \frac{1}{3!} \sum_{\bf k,q} 
\frac{|\Gamma_2({\bf q};{\bf k})|^2}{\varepsilon_{\bf k} + \varepsilon_{\bf q}
+ \varepsilon_{\bf k+q}} \ .
\label{E3}
\end{equation}
Combining all these contributions together one finds
in the second order in $1/S$:
\begin{equation}
E_{\rm g.s.}/N
 = -\frac{3}{2}J S^2\left[ 1 + \frac{I_2}{S} 
+ \frac{\left(I_4 + 2I_3\right)}{(2S)^2} \right] \ ,
\end{equation}
where $I_2 = \left (1 - \sum_{\bf k} \omega_{\bf k}\right)$ 
and the constants $I_4$ and $I_3$ are straightforwardly related 
to $\delta E_4$:
\begin{equation}
I_4 = (c_0\!+\!c_1\!-\!2c_2\!-\!1)^2 - \frac{3}{2}(c_1\!-\!c_2)^2 
\approx - 0.254293
\end{equation}
and to $\delta E_3$:
\begin{eqnarray}
I_3 = \sum_{\bf k,q} \frac{\widetilde{\Gamma}_2({\bf q},{\bf k})^2}
{\omega_{\bf k} + \omega_{\bf q} +\omega_{\bf k+q}}  =
0.13785(1) \ .
\end{eqnarray}
The above four-dimensional integral has been calculated by two 
different methods: the Monte Carlo integration and the finite-size 
extrapolation of lattice sums using clusters with different aspect
ratios, \cite{White07} both agreeing in all significant digits. 
Altogether, the ground-state energy in the second-order of
$1/S$ expansion is 
\begin{equation}
E_{\rm g.s.}/N
= -\frac{3}{2}J S^2\left[ 1 + \frac{0.436824}{2S}+ \frac{0.02141}{(2S)^2}
 \right] \ . 
\end{equation}
The above result agrees with the previous calculation \cite{Miyake85}
improving on the numerical accuracy of the last term.

\subsubsection{Sublattice magnetization}

To calculate the staggered magnetization we use the diagrammatic approach 
which is very close, aside from a few technical details, to the one used 
in Ref. \onlinecite{Chubukov94}. 
Within the spin-wave approach the sublattice magnetization is 
\begin{eqnarray}
\langle S\rangle = S - \langle a^\dagger_i a_i\rangle =
S - \delta S\ ,
\label{<S>0}
\end{eqnarray}
where the quantum correction $\delta S$ is expressed as
\begin{eqnarray}
\label{<S>}
\delta S =\sum_{\bf k}\bigl[v_{\bf k}^2+ 
\left(u_{\bf k}^2 + v_{\bf k}^2\right)
\langle b^\dagger_{\bf k}b_{\bf k}\rangle 
 + 2u_{\bf k}v_{\bf k}\langle b_{\bf k}b_{-\bf k}\rangle\bigr].
\end{eqnarray}
The first term under the sum is the LSWT
result \cite{Oguchi83,Jolicoeur89} 
already given by Eq.~(\ref{dS1}):
\begin{equation}
\delta S_1 = \sum_{\bf k} v_{\bf k}^2 \approx 0.2613032 \ .
\end{equation}
The two remaining terms in (\ref{<S>}) contain bosonic averages 
which vanish in the LSW approximation and  
contribute only to the next order in $1/S$. Therefore, 
we write:
\begin{equation}
\langle S\rangle = S - \delta S_1 - \frac{\delta S_2}{2S} \ ,
\label{deltaS}
\end{equation}
where the last correction has two contributions:
\begin{eqnarray}
&&\frac{\delta S_2}{2S}=\delta S_{2,1}+\delta S_{2,2} \ ,\\
&&\delta S_{2,1} =  \sum_{\bf k} 
\frac{1+\frac{1}{2}\gamma_{\bf k}}{\omega_{\bf k}}
\langle b^\dagger_{\bf k}b_{\bf k}\rangle, \ \ 
\delta S_{2,2} = \frac{3}{2} \sum_{\bf k} 
\frac{\gamma_{\bf k}}{\omega_{\bf k}}
\langle b_{\bf k}b_{-\bf k}\rangle .\nonumber
\end{eqnarray}

Calculation of the bosonic averages in the above expression
must be performed to the first order in $1/S$. 
As explained in Appendix \ref{app_B}, these averages are 
straightforwardly related to the normal and anomalous self-energies, 
see Figs.~\ref{vertices} and \ref{SelfE}. 
In particular, the magnon occupation number
$\langle b^\dagger_{\bf k}b_{\bf k}\rangle$
is only due to $\Sigma_{11}^{(b)}$ from Fig.~\ref{SelfE}(b), while the 
other two normal self-energy corrections, 
$\Sigma_{11}^{(a)}$ and 
$\Sigma^{HF}_{11}$, have zero contributions.
On the other hand, all three off-diagonal self-energies, 
$\Sigma^{(c)}_{12}$, $\Sigma^{(d)}_{12}$, 
and $\Sigma^{HF}_{12}$, contribute to
$\langle b_{\bf k}b_{\bf -k}\rangle$.
Leaving the details of the derivation to Appendix \ref{app_B},
we present here the final answer:
\begin{eqnarray}
\delta S_2 & = & 
 -\frac{9}{16}\,c_1c_2 + \frac{9}{16}\,(c_2-c_1) \sum_{\bf k}\, 
\frac{\gamma_{\bf k} (1-\gamma_{\bf k}) }{\omega_{\bf k}^3} \nonumber \\
& & \mbox{} + 
\frac{9}{4} \sum_{\bf k}\, \frac{\gamma_{\bf k}}{\omega_{\bf k}^2}\:
\sum_{\bf q} \frac{\widetilde{\Gamma}_1({\bf q};{\bf k})
\widetilde{\Gamma}_2({\bf q};{\bf -k})}
{\omega_{\bf q}+ \omega_{\bf k-q}+\omega_{\bf k}} 
\label{dS2}
\\
& & \mbox{} + 
\frac{3}{2} \sum_{\bf k}\, \frac{1+\frac{1}{2}\gamma_{\bf k}}{\omega_{\bf k}}\:
\sum_{\bf q} \frac{\widetilde{\Gamma}_2({\bf q};{\bf k})^2}
{(\omega_{\bf q}+ \omega_{\bf k+q}+\omega_{\bf k})^2} \ .
\nonumber 
\end{eqnarray}
This expression agrees with the formula derived previously in
Ref.~\onlinecite{Chubukov94} 
apart from the corrected sign in front of the third term.  

As is often the case with the higher-order spin-wave corrections, 
the individual contributions in Eq.~(\ref{dS2}) are divergent:
the integrands in the second and the third terms 
behave as $O(1/k^3)$ at ${\bf k}\rightarrow {\bf Q}$,
which means that not only the
leading divergences in them, but also the sub-leading ones 
$O(1/k^2)$ must cancel in order to produce finite result.
Expanding in small $\delta k= |{\bf k}-{\bf Q}|$, 
such a cancellation can be verified 
analytically. \cite{Chubukov94} 
Still, the expression given in Eq. (\ref{dS2}) 
is not well-behaved numerically. 
If one tries to evaluate $\delta S_2$ directly 
using the Monte Carlo integration, 
the outcome appears to be divergent. 
If some other methods are employed, the
result may seem to be regular. We have used the simple
finite sums in the ${\bf k}$-space that correspond to 
periodic clusters with the subsequent finite-size
extrapolation.\cite{White07}  
For any given subset of (rectangular) clusters with the fixed aspect ratio the
result of Eq.~(\ref{dS2}) converges to a finite value as the size of the
cluster $L\rightarrow\infty$. However, as an indication of the problem, 
subsets with different aspect ratio yield {\it different} values of $\delta
S_2$ in the thermodynamic limit. 

The origin of the problem is the following. The internal integrals 
in Eq.~(\ref{dS2}) over 
${\bf q}$ are not divergent and, generally, scale with the 
lattice size as\cite{White07,Ziman89}:
\begin{equation}
\label{psi}
\Psi_{\bf k}^{(L)}=\sum_{\bf q}\Big.^{(L)}\, F_{{\bf k}, {\bf q}}=
\Psi_{\bf k}^{(\infty)}+\frac{\alpha}{L} + \ldots
\end{equation}
In the thermodynamic limit, 
two such terms cancel near certain points and regularize 
the $\sim 1/k^3$ singularity
in the external integral over ${\bf k}$:
\begin{equation}
\frac{1}{k^3}\left[\Psi_{\bf k}^{(\infty)}-\Phi_{\bf k}^{(\infty)}\right] 
= \frac{1}{k^3}\left[Ak^2 + \dots\right] \ . 
\end{equation}
However, numerically such a cancellation is not complete as it 
carries a $1/L$ term as in (\ref{psi}):
\begin{equation}
\frac{1}{k^3}\left[\Psi_{\bf k}^{(L)}-\Phi_{\bf k}^{(L)}\right]
=\frac{1}{k^3}\left[Ak^2+\frac{\widetilde\alpha}{L}+\dots\right]
\label{cancel}
\end{equation}
Since the 2D integral of $1/k^3$ diverges as 
$L$, the $1/L$ correction from 
mutually canceling terms in (\ref{cancel}) will give an unphysical 
contribution to the $L\rightarrow\infty$ limit of Eq.~(\ref{dS2}).
This explains the erratic behavior of
the numerical values of $\delta S_2$ and suggest that the extra care should be
taken with Eq.~(\ref{dS2}). The way to regularize this problem 
is described in Appendix \ref{app_B}. 
After the regularization, the final answer for  Eq.~(\ref{dS2}) can be
obtained by any standard integration method, which yields:
\begin{equation}
\delta S_2 = -0.011045(5) \ .
\end{equation}
This result differs from $\delta S_2\approx 0.027$ quoted by Chubukov 
{\it et al.}\cite{Chubukov94}, which may have suffered from the above
integration problem. 
Our value agrees, though,  with  $\delta S_2\approx 0.011$
obtained by Miyake \cite{Miyake92} by a different method
that avoids highly singular integrals but deals with extrapolation in
a small auxiliary field.
 
Finally,
\begin{equation}
\langle S\rangle  = S - 0.261303 + \frac{0.011045(5)}{2S}\ .
\label{deltaS_f}
\end{equation}
Thus, the spin-1/2 Heisenberg antiferromagnet on a triangular
lattice has the following value of the ordered moments: 
$\langle S\rangle \approx 0.24974$ within the second-order spin-wave
expansion.
Note, that this spin-wave value is somewhat larger than 
the results of the Green's
function Monte Carlo,\cite{Sorella99} the series-expansion,\cite{Zheng06} and
the recent DMRG calculations\cite{White07} that give $\langle
S\rangle~\approx~0.205(15)$.

\section{Spin-wave spectrum: $1/S$ correction}
\label{onshell}

Similarly to the calculation of the static properties, 
perturbative expansion for the magnon spectrum has to be performed 
order by order in $1/S$ to ensure cancellation of all possible
divergences in the individual contributions.
This difficulty 
notwithstanding, derivation of the first $1/S$ correction 
to $\varepsilon_{\bf k}$ is straightforward. 
The anomalous terms do not contribute in that order and 
the new pole of the magnon Green's function is determined by
\begin{equation}
\varepsilon-\varepsilon_{\bf k}-\Sigma^{HF}({\bf k})-
\Sigma_{11}^{(a)}({\bf k},\varepsilon)-\Sigma_{11}^{(b)}({\bf k},\varepsilon)=
0\ , 
\label{DE_a}
\end{equation}
to which we refer to as to the Dyson equation. 
The self-energies in (\ref{DE_a}) are given by the diagrams 
in Figs.~\ref{vertices}(c) and \ref{SelfE}(a,b). 
Solving it
self-consistently  in the complex plane for a new renormalized  
spectrum $\varepsilon=\bar\varepsilon_{\bf k}-i\Gamma_{\bf k}$
constitutes the off-shell approximation discussed
in Sec.~\ref{off_shell}.

The first-order $1/S$ correction to the spectrum is obtained
within the so-called on-shell approximation.
In this approximation the self-energies are
evaluated at the bare magnon energy $\varepsilon=\varepsilon_{\bf k}$.
This leads to the following expression for 
the renormalized spectrum:
\begin{eqnarray}
\label{epsilon1}
&&\bar\varepsilon_{\bf k}-i\Gamma_{\bf k} =  \varepsilon_{\bf k} + 
\varepsilon_{\bf k}^{(4)}  \\
&& -\frac{9}{4}J \sum_{\bf q}
\biggl[ \frac{\widetilde{\Gamma}_1({\bf q};{\bf k})^2}
{\omega_{\bf q} + \omega_{\bf k-q} - \omega_{\bf k}-i0} + 
\frac{\widetilde{\Gamma}_2({\bf q};{\bf k})^2}
{\omega_{\bf q} + \omega_{\bf k+q} + \omega_{\bf k}} \biggr].
\nonumber 
\end{eqnarray}
In the right-hand side the harmonic energy is 
$O(S)$, while the rest of the terms  are $O(S^0)$.  
The finite magnon decay 
rate $\Gamma_{\bf k}$ comes only from the first term in the brackets. 
Since the Goldstone modes should be well-defined 
in the ordered AFs, 
we expect $\Gamma_{\bf k}\ll\varepsilon_{\bf k}$
in the long-wavelength limit. The details for that limit are given in 
the following subsection, while 
Sec.~IV.B is devoted to the behavior
of the renormalized spectrum in the full BZ.

\subsection{Low-energy magnons}

\subsubsection{Velocity renormalization}

The triangular lattice antiferromagnet 
has three Goldstone modes at ${\bf k}=0$ and $\pm {\bf Q}$.
The existence of these zero-energy modes follows directly
from the broken $SO(3)$ rotational symmetry in the spin space
and, therefore, should not be affected by quantum
renormalizations. \cite{Dombre89} 
As a consistency check of the $1/S$ expansion, it is important
to verify the presence of acoustic modes in the renormalized
spectrum (\ref{epsilon1}). Such a verification was first performed in 
Ref.~\onlinecite{Chubukov94} where the 
$1/S$ corrections to the velocities of the Goldstone modes were 
derived. We have reproduced these corrections with an improved 
numerical accuracy, although, technically,
a somewhat different route was followed for the derivation.
Numerical values of the spin-wave velocities are:
\begin{eqnarray}
v_0 & = & 3JS\,\frac{\sqrt{3}}{2}\, \Bigl(1-\frac{0.11488}{2S}\Bigr)\ ,
\nonumber \\
v_Q & = & 3JS\,\sqrt{\frac{3}{8}} \,
\Bigl( 1 + \frac{0.08285(2)}{2S} \Bigr)\ .
\label{v_renorm}
\end{eqnarray}

\subsubsection{Long-wavelength decays}

In the long-wavelength limit decay 
rates can be always calculated perturbatively because of 
the smallness of interaction among low-energy 
excitations and due to reduction of the phase-space volume
available for decay processes. In other terms,
presence of the well-defined Goldstone modes 
implies smallness of the damping rate 
with respect to their energy, 
$\Gamma_{\bf k}\ll\varepsilon_{\bf k}$.
As with the velocity renormalization, such a behavior 
has to be verified using Eq.~(\ref{epsilon1}).
As was remarked before, existence of the cubic interactions alone
does not immediately yield finite lifetime for the excitations since
the corresponding decays may be virtual.
The decay processes become real if both the total energy and 
the total momentum can be conserved simultaneously.
Since the two-particle excitations form a continuum
of states, the energy conservation can 
be rephrased as the requirement of an overlap
of the single-particle branch with the two-particle
continuum. This yields certain kinematic conditions
on the bare spectrum $\varepsilon_{\bf k}$, which are discussed in Section V
using the example of the triangular-lattice HAF.
Here we assume that such conditions are fulfilled and
consider several scenarios for the decays of 
the long-wavelength excitations 
relevant to the present model.

\begin{figure}[t]
\includegraphics[width = 0.9\columnwidth]{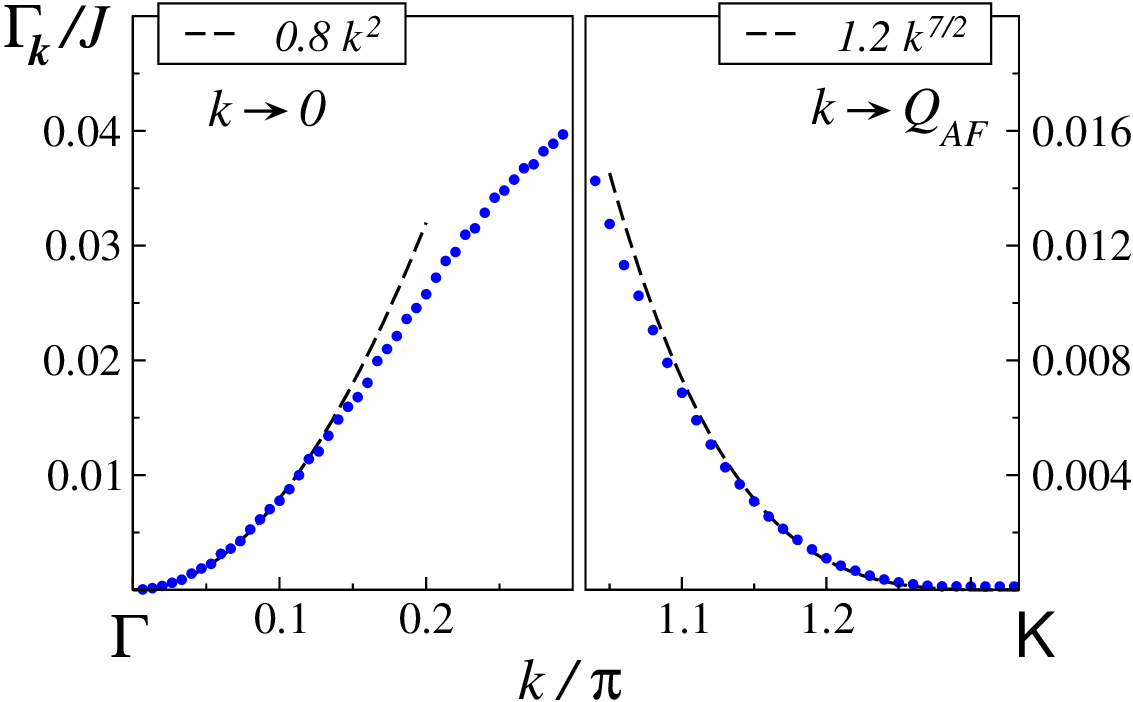}
\caption{(Color online) Spin-wave 
decay rates in the vicinity of $\Gamma$ and $K$ points
in the triangular-lattice HAF. Dots are numerical solutions
of Eq.~(\ref{epsilon1}) and dashed lines are the asymptotic results 
(see text).
}
\label{duptih}
\end{figure}

Generally, the decay rate is given by:
\begin{equation}
\Gamma_{\bf k} \sim \sum_{\bf q} |V_{{\bf k},{\bf q}}|^2
\delta(\varepsilon_{\bf k}-\varepsilon_{\bf q}-\varepsilon_{\bf k-q}) \ ,
\end{equation}
where $V_{{\bf k},{\bf q}}$ is the decay vertex.
Due to the energy conservation, the upper limit for the momentum 
of created quasiparticles should be of the order of $k$. 
Then, for the linear spectrum $\varepsilon_{\bf k}\sim k$, a naive 
power counting suggests the following answer 
\begin{equation}
\label{naive0}
\Gamma_{\bf k} \sim k^{D-1} |V_k|^2
\end{equation}
where $k^D$ comes from the $D$-dimensional phase space,  
$1/k$ is due to reduction of that space to decay surface 
from the energy conservation, and $V_k$ is the typical amplitude
of the cubic vertex on the decay surface. 
Let us assume that the cubic interaction follows the standard 
form \cite{LL_IX} 
$V_{\bf k,q}\propto \sqrt{kqq'}$, $q'=|{\bf k}-{\bf q}|$, and  
that in a typical decay process the final momenta are $q, q'\sim k$. 
This makes $|V_k|^2\propto k^3$ and 
yields a seemingly universal power law for the decay rate:
\begin{equation}
\label{naive}
\Gamma_{\bf k} \sim k^{D+2} \ .
\end{equation}
For $D=3$ this yields $\Gamma_{\bf k} \sim k^5$ 
which matches the result for the decays of the convex phonon branch,
\cite{Beliaev58} but, as we shall  
see shortly, only coincidentally.

\begin{figure*}[t]
\includegraphics[height = 0.7\columnwidth]{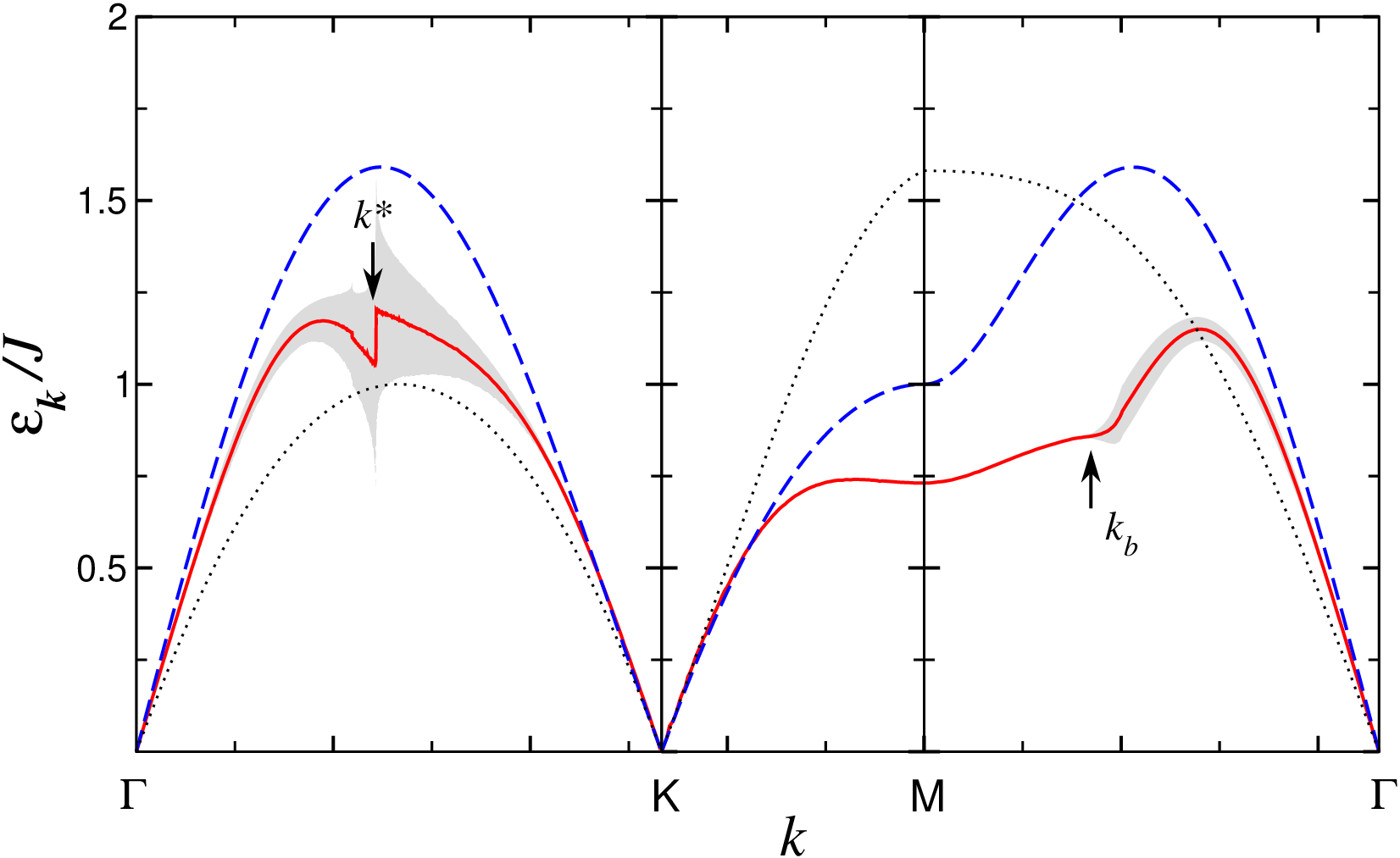} \ \hskip 1cm
\includegraphics[height = 0.7\columnwidth]{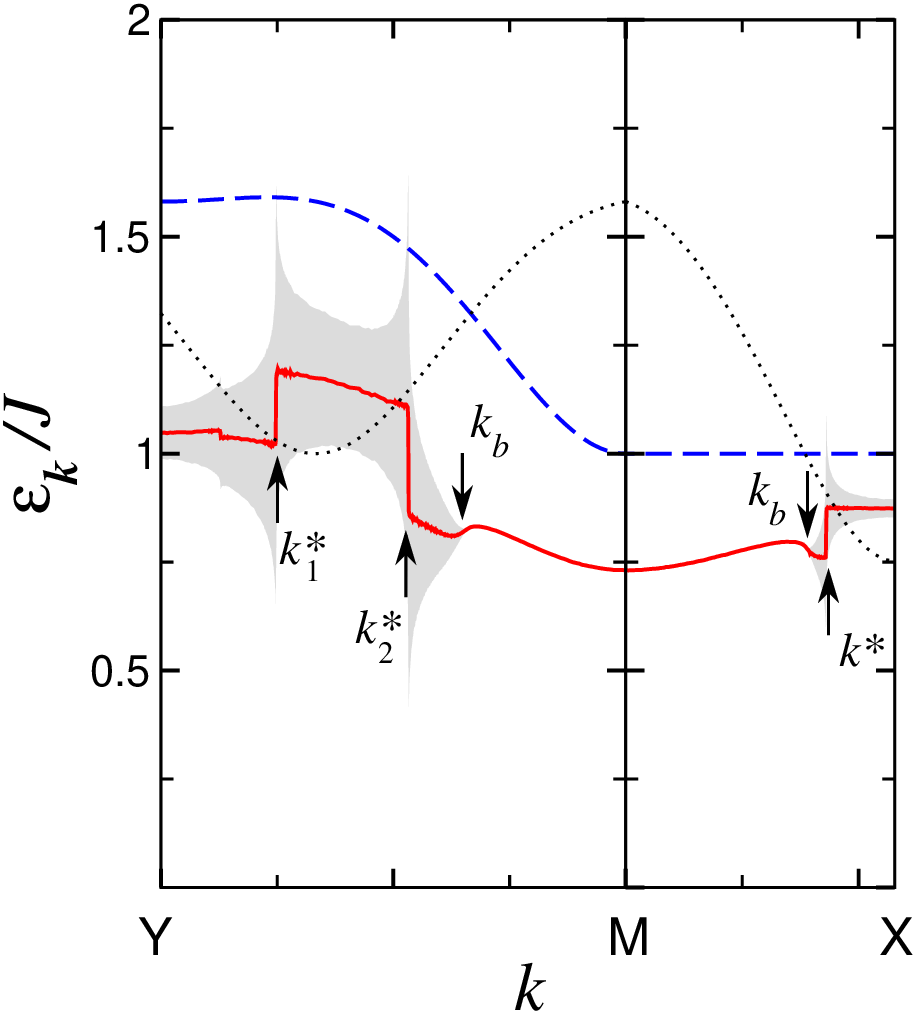}
\caption{(Color online) 
Spin-wave energy for the $S=1/2$ triangular-lattice HAF
along the symmetry directions in the BZ. 
The dashed and solid lines are the results obtained in the LSWT
approximation ($\varepsilon_{\bf k}$) 
and with the first-order $1/S$ correction ($\bar{\varepsilon}_{\bf k}$),
respectively. 
The vertical arrows indicate singularities and intersection points with 
the two-magnon continuum. Gray areas show the width of the spectral
peaks due to damping $\Gamma_{\bf k}$ from Fig.~\ref{Imw_k}.
Dotted lines represent the  minimum of the two-magnon continuum
obtained from the LSWT spectrum.  
}
\label{w_k}
\end{figure*}

In reality, the situation is more delicate and 
possible power-law exponents for the decay rate asymptote depend on the 
specifics of the problem. In the case of a single weakly nonlinear
acoustic branch, relevant to the phonon spectrum in $^4$He,
\begin{equation}
\varepsilon_{\bf k}=c k +\alpha k^3
\end{equation}
the decays are allowed only for a positive curvature 
of the spectrum,  $\alpha>0$ (convex $\varepsilon_{\bf k}$).
In this case, the unstable quasiparticle emits two
excitations in a narrow solid angle centered in 
the direction of the initial momentum.
The apex angle of the decay cone scales as 
$\theta\sim k$ such that  the phase space 
is $k^{2D-1}$ instead of $k^D$. Then the restriction from the
energy conservation gives $1/k\theta^2 \sim 1/k^3$ instead of $1/k$ 
in the previous consideration.
Altogether, for the case of the cubic 
upward curvature of the spectrum, the answer 
is universal (see also Ref.~\onlinecite{Kopietz08}):
\begin{equation}
\Gamma_{\bf k} \sim k^{2D-1},
\end{equation}
which yields $\Gamma_{\bf k} \sim k^5$ in 3D\cite{Beliaev58,LL_IX} 
and $\Gamma_{\bf k} \sim k^3$ in 2D. The 2D result
applies to the square-lattice HAF in a strong magnetic field
where the convexity of the sound-like branch  changes from 
$\alpha<0$ below the threshold field $H^*$
to $\alpha>0$ in the high-field region $H^*<H<H_s$. 
\cite{field,Kopietz08,Olav08}

If several acoustic modes with different velocities are present, 
the fast quasiparticle can always decay into two
slow ones. This situation is simpler than the previous case
since the nonlinearity of the spectra plays no role.
The phase space factor becomes $k^{D-1}$ now, in agreement with 
the above naive consideration. Physical realizations of  
this scenario include decays of the  
longitudinal phonon into two transverse ones 
in solids \cite{Ziman} as well as the decay 
of the ${\bf k}\rightarrow 0$ into two 
${\bf k}\rightarrow\pm {\bf Q}$ 
magnons in the triangular-lattice HAF. 
Clearly, such a channel of decays 
withstands quantum renormalization of the velocities and is 
pertinent to the other noncollinear AFs.

Interaction between phonons in crystals obeys the conventional scaling
asymptote $V_{\bf k,q}\propto \sqrt{kqq'}$ and, consequently, 
$|V_k|^2\propto k^3$. Therefore, the naive power 
counting of Eq.~(\ref{naive}) is valid for them. 
However, in the case of the triangular-lattice HAF, the result is yet
different from  (\ref{naive}) because the three-magnon 
vertex (\ref{H3b}) is anomalous and scales as 
$V_{\bf k,Q+q}\propto (q'-q)\sqrt{k/qq'}$ for small $k$.
For a typical decay process
$q,q' \sim k$ giving: 
$|V_{\bf k,Q+q}|^2 \sim k$ instead of $k^3$.
Altogether this yields for the noncollinear HAFs in $D$-dimensions:
\begin{equation}
\Gamma_{\bf k} \sim k^{D} \ .
\end{equation}
Direct analytic expansion of Eq.~(\ref{epsilon1}) 
gives the following decay rate of the $k\rightarrow 0$
magnons in the triangular-lattice HAF ($D=2$):
\begin{equation}
\Gamma_{\bf k} \simeq \frac{9k^2 J}{4\sqrt{2}\pi}\,
\Bigl[ 1+ \pi\Bigl(\frac{17}{32} -\frac{1}{2\sqrt{2}}
\Bigr)\Bigr]\ ,
\end{equation}
which is $\Gamma_{\bf k}\approx 0.789 J
k^2$, in agreement with the numerical data in Fig.~\ref{duptih}. 

\begin{figure*}[t]
\includegraphics[height = 0.7\columnwidth]{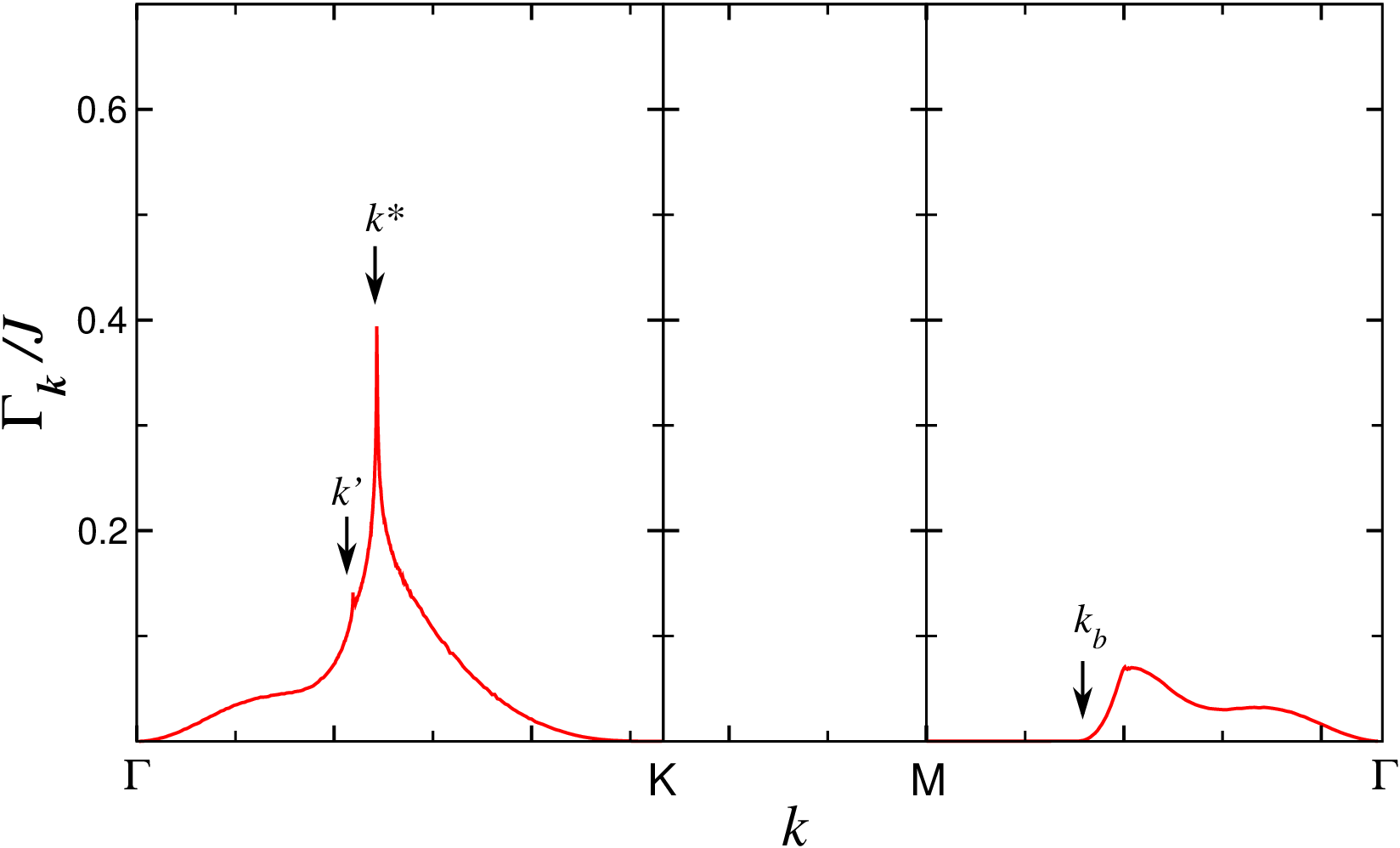} \ \hskip 1cm
\includegraphics[height = 0.7\columnwidth]{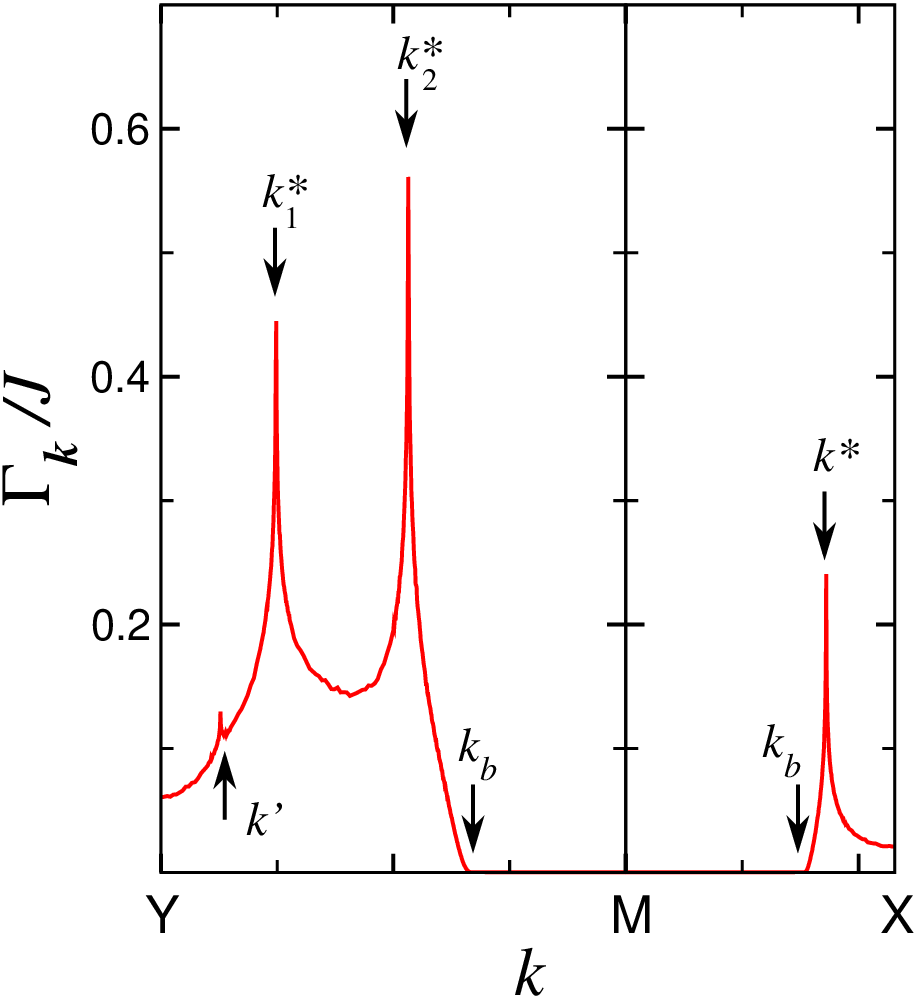}
\caption{(Color online) 
The spin-wave damping of the $S=1/2$ triangular-lattice HAF 
calculated in the first-order of the $1/S$ expansion 
along the same lines as in Fig.~\ref{w_k}. Arrows indicate
singularities and intersection points with the two-magnon 
continuum.}
\label{Imw_k}
\end{figure*}

The kinematic conditions for magnons near
$\pm{\bf Q}$ ($K$ and $K'$) points 
are more subtle. 
The magnon energy has a nonanalytic expansion in
small $\tilde{\bf k}={\bf k}-{\bf Q}$ with varying convexity: 
$\varepsilon_{\bf k}\approx c \tilde k + \alpha_\varphi\tilde k^2$, 
where the nonlinearity depends on the azimuthal
angle $\alpha_\varphi \sim \cos 3\varphi$, see Eq.~(\ref{wQ}).
This form, together with the commensurability of the AF 
ordering vector ${\bf Q}$ 
with the reciprocal lattice of the crystal,
allow for momentum and energy conservation for magnon decays 
from the steeper side of the energy cone at
${\bf k}\rightarrow{\bf Q}$ onto the flatter
sides at ${\bf q}, {\bf q}'\rightarrow-{\bf Q}$. Commensurability of
${\bf Q}$ is important here as $3{\bf Q}=0$  is the necessary
condition for the conservation of the quasi-momentum. 
Thus, magnons near the ${\bf Q}$-point are unstable only in a certain range 
of angles. 
The decay vertex for ${\bf k}\!\rightarrow\!{\bf Q}$ magnon has
the conventional scaling: $V_{\bf Q+k,-Q+q}\!\propto\! 
\sqrt{kqq'}$, so the decay
probability is $|V_k|^2\!\sim\!k^3$. However, due to  
a constraint on the angle between ${\bf k}$ and ${\bf q}$, 
the decay surface in ${\bf q}$-space is a cigar-shaped ellipse 
with length $\sim\!k$ and width $\sim\!k^{3/2}$. That 
makes the restricted phase volume of decays to scale as $k^{(3D-5)/2}$. 
This results in a nontrivial law:
\begin{equation}
\Gamma_{\bf k} \sim k^{(3D+1)/2} \ ,
\end{equation}
which gives $k^{7/2}$ for
the decay rate in 2D. Numerically, along the $K\Gamma$ line
$\Gamma_{\bf k}\approx 1.2Jk^{7/2}$, 
see Fig.~\ref{duptih}. Away from this direction the damping exhibits
an anomalous angular-dependence $\Gamma_{\bf k} \sim 1/(\cos 3\varphi)^{3/2}$.
Such a behavior is related to the saddle-point singularities 
which is discussed in the next Section.

\subsection{High-energy magnons}

The renormalized spin-wave energy and the magnon damping 
given by Eq.~(\ref{epsilon1}) have been  calculated 
using the Monte Carlo integration 
method with $10^8$--$10^9$ integration points in the full BZ. 
Numerical results for the spin-1/2 case are shown in
Figs.~\ref{w_k} and \ref{Imw_k} for representative symmetry directions
in the BZ, see Fig.~\ref{shape_wk} for notations.
Statistical errors of the calculation are comparable/smaller
than the corresponding line width.
The first prominent feature of the spectrum is that the 
renormalization is 
stronger at large momenta, in a qualitative agreement with 
the series-expansion results.
\cite{Starykh06,Chernyshev06,Zheng06} 
As we have argued before, this is a consequence of the 
considerable coupling between single-magnon excitations and  
two-magnon continuum determined by cubic anharmonicities. 
The momentum dependence of the  minimum 
of the continuum is shown in Fig.~\ref{w_k} by dotted curves. 
The intersection points of these curves 
with the magnon branch are marked as $k_b$'s.
Inside the continuum, 
magnon excitations become intrinsically damped acquiring
a non-zero $\Gamma_{\bf k}$, see Fig.~\ref{Imw_k}. 
The magnitude of the overlap 
of the continuum with the single-particle branch
gives a qualitative idea of the phase space available for decays.
The damping rate is also illustrated in Fig.~\ref{w_k} 
as the shaded area between $\bar\varepsilon_{\bf k}-\Gamma_{\bf k}$ 
and $\bar\varepsilon_{\bf k}+\Gamma_{\bf k}$. 

Another interesting property of the renormalized spectrum in Fig.~\ref{w_k} 
is its shape in the vicinity of the $M$-point (edge center of the BZ).
Quantum renormalization converts the saddle
point of $\varepsilon_{\bf k}$ at ${\bf k}_M$ 
into a local minimum surrounded by flat
parts. \cite{Singh06,Zheng06,Starykh06}  
Such local extrema must contribute significantly to  
the thermodynamic properties
of the triangular-lattice HAF. \cite{Singh06,Zheng06}   
The minimum in $\bar\varepsilon_{\bf k}$  is more pronounced 
in the numerical results\cite{Zheng06}  than in Fig.~\ref{w_k}.  
It was called a ``roton-like'' minimum and it 
was suggested that it might be a signature of spinons.\cite{Singh06}  
This has been questioned later since the same feature in 
the spin-wave results can be explained by
the enhanced density of two-magnon states near the $M$-point. 
\cite{Starykh06,Chernyshev06} 

Although the above discussion is important for benchmarking 
the spin-wave theory with the series expansion results, 
the most remarkable property of the renormalized spectrum 
is the substantial
jump-like discontinuities in $\bar{\varepsilon}_{\bf k}$,
marked as $k^*$ points in Fig.~\ref{w_k}, 
with the jump heights reaching 1/4 of the magnon bandwidth.
If considered without the concomitant behavior
of $\Gamma_{\bf k}$, such jumps  in $\bar{\varepsilon}_{\bf k}$
are especially enigmatic. 
The values of the damping at the top of the magnon band
are also quite substantial, leading to broadening of the spectral
peaks with the widths about 
$(2\Gamma_{\bf k}/\bar\varepsilon_{\bf k})\sim 1/3$. 
The most striking features of $\Gamma_{\bf k}$ are the sharp 
logarithmic singularities at several points along the selected 
cuts of the Brillouin zone. 
These are precisely the same $k^*$ points where the jumps
occur in the real part of the spectrum. Thus, the two singularities
are intrinsically related. Analytic consideration of these 
singularities in the magnon self-energy is presented in Sec.~V.B.
Here we simply state that their origin is due to the intersection 
of the single-magnon branch with the line of the van Hove saddle-point
singularities in the  two-magnon continuum.

The logarithmic singularities in the damping rate signify a
breakdown of the $1/S$-expansion in the 
vicinity of the singular $k^*$-points. 
It is remarkable that the log-singularities are obtained already in 
the lowest-order interacting spin-wave theory. Therefore, if any other 
property calculated with the $1/S$-expansion is to be compared with 
numerical methods, such singularities must be understood and 
their proper treatment within the theory outlined. 
Note, that the numerical series-expansion results
exhibit lack of convergence (large error-bars)
at certain ${\bf k}$-points.\cite{Singh06}
Although, this is not 
the same as singular jumps, such numerical features might occur 
due to similar reasons.
Analytic results for the spectrum near the singularities and 
the proper treatment of the latter will be presented in 
Secs.~\ref{kinematic} and \ref{off_shell}, respectively.

Altogether, the main effects of quantum fluctuations in the 
spectrum  of the triangular-lattice HAF are 
the substantial magnon damping in the major part of the BZ,
singular behavior of the decay rate 
along certain contours in the ${\bf k}$-space, 
and strong downward renormalization of the magnon energies  compared to the 
harmonic spin-wave theory.
All these effects underline  the 
importance of cubic anharmonicities in the noncollinear AFs and represent 
major qualitative differences from the collinear cases.

\section{Kinematics of Two-Particle Decays}
\label{kinematic}

\begin{figure}[t]
\centerline{
\includegraphics[width = 0.9\columnwidth]{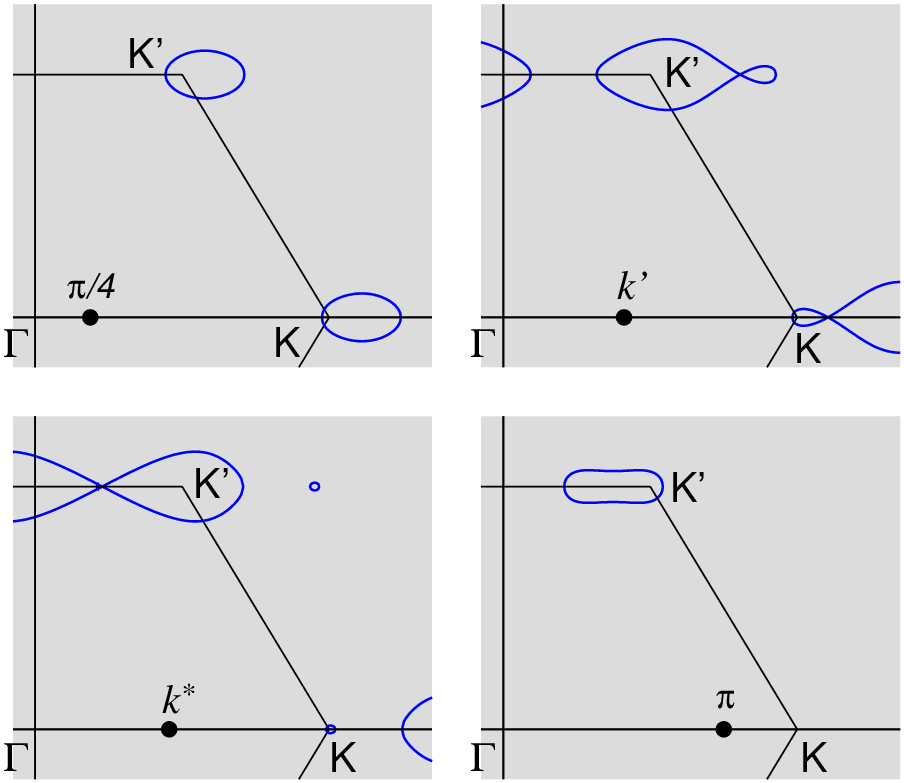}}
\caption{(Color online) 
The decay contours in ${\bf q}$-space for 
magnons with selected ${\bf k}$'s along the
$\Gamma K$ direction; $k=k'$ and $k=k^*$ are the same as in Fig.~\ref{Imw_k},
left panel. The corresponding ${\bf q}$-contours undergo topological
transitions at $k'$ and $k^*$.}
\label{q_contours}
\end{figure}

The aim of this section is to consider kinematic constraints that
follow from the energy conservation in the two-particle
decay process:
\begin{equation}
\varepsilon_{\bf k} = \varepsilon_{\bf q} + \varepsilon_{\bf k-q} \ .
\label{energy_conserv}
\end{equation}
This equation should be treated as an equation in ${\bf q}$ with the 
initial momentum ${\bf k}$ as a parameter.
The solutions of Eq.~(\ref{energy_conserv}) form
the {\it decay surface} in the ${\bf q}$-space. 
Examples of the decay surfaces (contours in 2D) 
for the triangular-lattice HAF are shown in Fig.~\ref{q_contours}
for a few representative ${\bf k}$-points along the $\Gamma K$-line.

As a function of ${\bf k}$, the decay surface changes and may disappear 
completely. In such a case, particles 
become stable with $\Gamma_{\bf  k}\equiv 0$. 
The region in the ${\bf k}$-space with stable excitations 
is separated from the decay region by the {\it decay threshold 
boundary}. Generally, two-particle excitations form a continuum of 
states in a certain energy interval 
\begin{equation}
E^{\min}_{\bf k} \leq \,E_{\bf k}({\bf q}) \equiv \varepsilon_{\bf q} + 
\varepsilon_{\bf k-q} \leq \, E^{\max}_{\bf k} \ .
\label{continuum}
\end{equation} 
Thus, the decay threshold boundary is determined  by the intersection
of the single-particle branch $\varepsilon_{\bf k}$ 
with the bottom of the continuum $E^{\min}_{\bf k}$.
For our problem, the decay region is the hexagram shown in 
Fig.~\ref{BZ}.
 
Needless to say, the two-particle decays may be prohibited 
in the entire BZ,
a situation that is common for collinear antiferromagnets
in zero field.\cite{Harris71}
In such a case, spontaneous $n$-particle decays with $n>2$ are also 
prohibited since all the energies in the $n$-particle generalization
of Eq.~(\ref{energy_conserv}) are positively defined.
Thus, the analysis of the two-particle decay conditions is crucial 
for determining  whether any spontaneous decays are possible. In the case
when the two-particle decay region is finite, the $n$-particle decays
can, in principle, be allowed in a wider region of the 
${\bf k}$-space, see, {\it e.g.},
the corresponding study for the phonon branch in 
$^4$He.\cite{Pitaevskii_76} 
However, such an expansion of the decay region 
requires rather special conditions 
and is not discussed in this work.

Aside from finding whether spontaneous decays exist or not, 
there is another important reason for considering
the decay kinematics.
As was emphasized by Pitaevskii  \cite{LL_IX,Pitaevskii59}
in the analysis of the two-roton decay threshold
in superfluid $^4$He,
the enhanced density of two-particle states
near the minimum of the continuum may produce
strong singularities in the single-particle
spectrum at the decay boundary.
This yields various unusual effects, \cite{LL_IX,Pitaevskii59,Zhitomirsky06} 
including complete disappearance of the quasiparticle
branch inside the continuum.
Apart from the singularities in the spectrum at the decay boundary, 
additional singularities may occur within
the decay region due to topological transitions of the decay surface.
\cite{Chernyshev06}
General analysis of these two effects and its application
to the triangular HAF are discussed here.

\subsection{Decay threshold boundary}
 
The boundary of the decay region corresponds to the intersection 
of the single-particle  branch $\varepsilon_{\bf k}$ with the bottom of 
the two-particle continuum (\ref{continuum}). To find 
$E_{\bf k}^{\min}$ one should analyze the extrema of the 
continuum given by
\begin{equation}
\frac{\partial E_{\bf k}({\bf q})}{\partial {\bf q}}=
\left.\frac{\partial \varepsilon_{\bf q}}{\partial {\bf q}}
\right|_{\bf q} - 
\left.\frac{\partial \varepsilon_{\bf q}}{\partial {\bf q}}
\right|_{\bf k-q}  =0 \ .
\label{extrema}
\end{equation}
Thus, the extremum condition is satisfied if the {\it velocities} of
the decay products are equal. This yields $D$ equations 
with ${\bf q}$ and  ${\bf k}$ as independent variables.
To find the decay boundary one needs to solve the decay and the
extremum conditions (\ref{energy_conserv}) and (\ref{extrema})
together. Solving them yields  a set of 
$(D-1)$-dimensional surfaces in the ${\bf k}$-space.
Since the non-trivial solution may exist not only for the minimum of
the continuum but for the other extrema as well, some parts of these
surfaces define the decay threshold boundary while the rest of the
solutions will correspond to the special surfaces within the decay
region. The latter are considered in Sec.~\ref{kinematic}.B.

For the gapless spectrum, one should also verify whether the 
emission of the acoustic (Goldstone) excitation
at ${\bf q}\rightarrow{\bf Q}_i$ corresponds to $E_{\bf k}^{\min}$.
This  condition is separate from Eq.~(\ref{extrema}) as the Goldstone 
modes are not the extrema but the end-points  of the spectrum.
In the case when one or more of these gapless modes define 
$E_{\bf k}^{\min}$, the decay boundary is given by
\begin{equation}
\varepsilon_{\bf k}=\varepsilon_{{\bf k+Q}_i} \ .
\label{Q_decay}
\end{equation}

Although the above discussion 
covers two most general cases, there exist 
a particular form of the 
solution of (\ref{energy_conserv}) and (\ref{extrema}) for which
finding the threshold boundary simplifies greatly.  
The two-particle continuum
always possesses an extremum when both decay products have equal
momenta, up to a reciprocal lattice vector ${\bf G}$:  
${\bf q},{\bf k}-{\bf q} = ({\bf k} \pm {\bf G})/2$. 
The condition of equal velocities (\ref{extrema}) is automatically
satisfied in this case.
Such an extremum 
crosses with the single-particle branch on the surface determined by
\begin{equation}
\varepsilon_{\bf k} = 2\varepsilon_{({\bf k}+{\bf G})/2} \ .
\label{decay_2eq}
\end{equation}
Among the previously studied cases, the decay boundaries
for the square- and cubic-lattice HAFs in a strong external field satisfy
the above equation. \cite{field}

\begin{figure}[t]
\centerline{
\includegraphics[width=0.7\columnwidth]{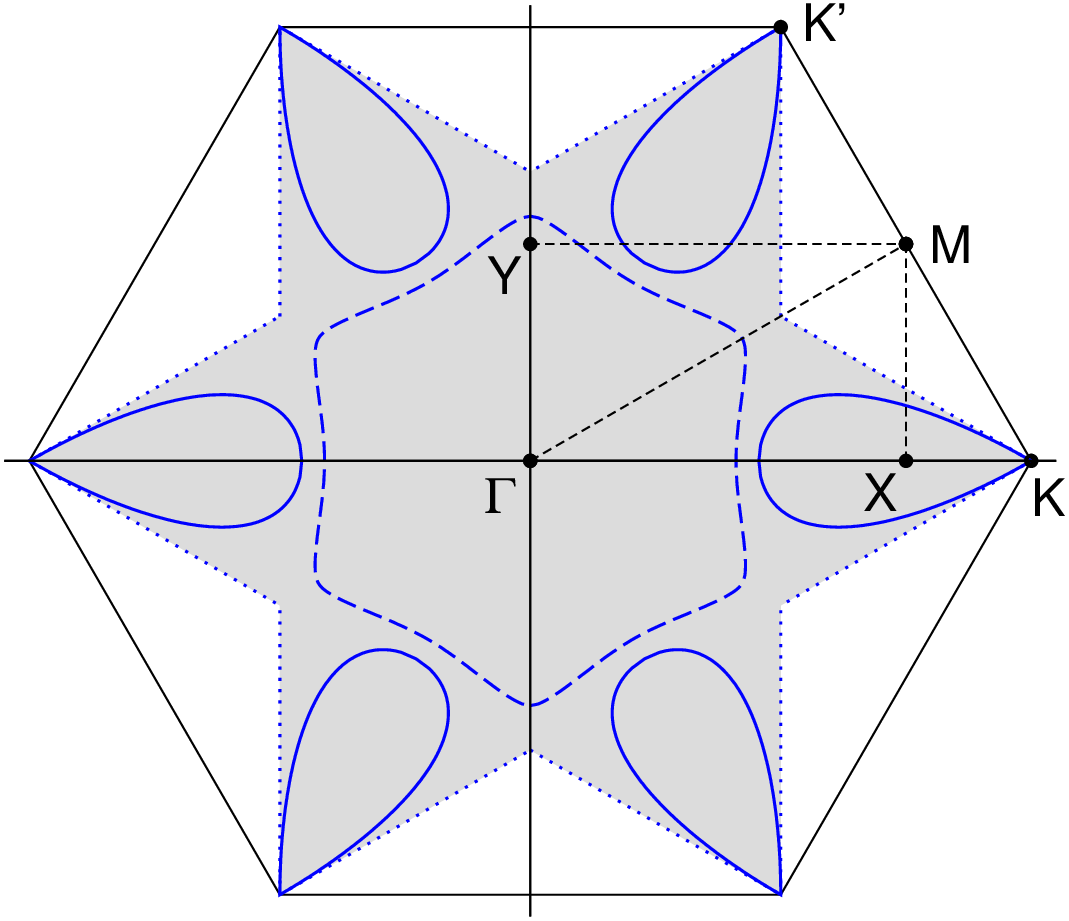}}
\caption{(Color online) The Brillouin zone of a triangular
lattice. The shaded area shows the region
where spontaneous two-magnon decays are allowed. 
The lines correspond to the extrema in the two-magnon
continuum described in text.}
\label{BZ}
\end{figure}
 
For the case of the triangular-lattice HAF, 
let us begin with the analysis of the decay involving the
Goldstone modes. 
The acoustic branch emerging from ${\bf q} = 0$
does not correspond to the crossing point of $\varepsilon_{\bf k}$
and $E_{\bf k}^{\min}$ and should be disregarded. 
The condition (\ref{Q_decay}) on the emission of Goldstone modes 
with ${\bf q} = \pm {\bf Q}$ can be rewritten,
using the expression (\ref{epsilon}) 
for the harmonic spectrum,  
as $\gamma_{\bf k} = \gamma_{{\bf k}\mp{\bf Q}}$.
This last equation is easily solved and the results are shown 
in Fig.~\ref{BZ} by the dotted lines. 
 One can show that
 ${\bf q}=\pm{\bf Q}$ points correspond to the absolute
 minimum of the continuum for all ${\bf k}$ within the 
 shaded area in Fig.~\ref{BZ}. To demonstrate that, 
$\varepsilon_{\bf k\pm Q}$ is also plotted in Fig.~\ref{w_k} 
as dotted line.

We begin the analysis of the extrema in the two-particle continuum
with the decay threshold involving magnons with equal momenta.
The solutions of Eq.~(\ref{decay_2eq}) for the
reciprocal lattice vectors 
${\bf G}_{1,2} = (\pm 2\pi,2\pi/\sqrt{3})$ and ${\bf G}_3= 
(0,4\pi/\sqrt{3})$ are readily found numerically and
are shown in Fig.~\ref{BZ} by solid lines.
No solution exists for ${\bf G}=0$.  
Solid  lines in Fig.~\ref{BZ} 
lie entirely within the shaded area 
and correspond to the saddle points of the continuum, not the minima.
The solution for the general situation of the threshold
decay into a pair of non-identical magnons
with equal velocities but different momenta and energies,
$\varepsilon_{{\bf k}-{\bf q}} 
\neq \varepsilon_{\bf q}$, is obtained by solving numerically 
the decay (\ref{energy_conserv}) and extremum 
(\ref{extrema}) conditions  simultaneously. 
The result is plotted 
in Fig.~\ref{BZ} by the dashed line. 
As in the previous case, this contour corresponds  
to the line of saddle points of the continuum within the decay region.

Altogether, the decay region is given by the union of all areas 
obtained in the above three cases. As one can see,
the decay threshold boundary in the triangular HAF 
is determined entirely by the emission of acoustic 
$\varepsilon_{\pm{\bf Q}}$ magnon. 
In accordance with the long-wavelength consideration of 
Sec.~\ref{onshell}.A, the area around ${\bf k}=0$ 
is precisely where the decay of the fast magnon 
$\varepsilon_{{\bf k}\rightarrow 0}$ into
two slow ones $\varepsilon_{{\bf q}\rightarrow \pm{\bf Q}}$ 
takes place and this area
is completely enclosed in the decay region. Similarly, 
there are only finite segments in the 
vicinity of the $\pm {\bf Q}$-points where decays are possible.
The other thresholds due to two-magnon extrema do not contribute
to the decay boundary.
As is shown in Sec.~\ref{models}, such a mutual arrangement of 
different thresholds may or may not be the case for 
other closely related systems. Already a simple
addition of the $XXZ$ anisotropy modifies
the decay boundary and switches saddle points into
minima for certain momenta.

The damping of magnons close
to the boundary of the decay region can be considered in a 
manner similar to the long-wavelength analysis of
Sec.~\ref{onshell}.A.2. Since the velocity of the decaying
magnon  near the boundary 
must be larger than the velocity of the emitted acoustic 
$\pm{\bf Q}$ mode, the phase space factor is $(k-k_b)^{D-1}$ as is 
given in Eq.~(\ref{naive0}).
The probability of the decay has a smallness only with respect
to $q\sim (k-k_b)$.
This yields the 2D decay rate near the threshold 
due to acoustic mode:
\begin{equation}
\Gamma_{\bf k}\sim (k-k_b)^2
\end{equation}
in agreement with the results in Fig.~\ref{Imw_k} 
in the vicinity of $k_b$ points.

\subsection{Topological transition in the decay surface}

A truly remarkable feature of the magnon spectrum in the 
triangular-lattice HAF is the logarithmic peaks in $\Gamma_{\bf k}$ 
obtained in the first-order of $1/S$ expansion.
After preceding discussion it is clear that the location of such
 singularities corresponds to the crossing
of single-magnon branch with the surface 
of saddle-points of two-magnon continuum. 
Indeed, the $k^*$-points 
in Figs.~\ref{w_k} and \ref{Imw_k} 
belong to the 
threshold contour for the decays into pairs of equivalent magnons, 
solid lines in Fig.~\ref{BZ}.
Much weaker anomalies, which are visible only as small peaks in 
$\Gamma_{\bf k}$ that are denoted by $k'$s in 
Fig.~\ref{Imw_k}, correspond to the threshold contour 
for the decays into non-equivalent
magnons, dashed lines in Fig.~\ref{BZ}.

In the vicinity of the crossing points of the single-particle branch 
with the saddle-point surface of the continuum,
the magnon {\it decay surface} undergoes a topological transition, 
see Fig.~\ref{q_contours}.
Expanding the energy conservation condition (\ref{energy_conserv}) in small
$\Delta {\bf k} = {\bf k}-{\bf k}^*$ and  
$\Delta {\bf q} = {\bf q}-{\bf q}^*$, where 
${\bf k}^*$ is the point on the threshold contour and 
${\bf q}^*$ is the saddle point of 
$E_{{\bf k}^*}({\bf q})$, we obtain  
\begin{equation}
\varepsilon_{\bf k} - 
\varepsilon_{\bf q} - \varepsilon_{\bf k-q} \approx
({\bf v}_1 - {\bf v}_2)\cdot \Delta {\bf k} -
\frac{\Delta q_x^2}{a^2} + \frac{\Delta q_y^2}{b^2} = 0\ ,
\label{conserv_expan}
\end{equation}
where 
${\bf v}_1$ and ${\bf v}_2$ are the velocities of the initial and final
magnons, and $a,b$ are constants. Depending on the sign of
$(v_1-v_2)\Delta k$ the solution of the above
equation are two conjugate hyperbolas, which transform into a pair of
crossing lines for $\Delta {\bf k}=0$.
Thus, ${\bf k}={\bf k}^*$ 
corresponds to the point where the decay surface splits into
two disjoint pieces, as is indeed observed in Fig.~\ref{q_contours}.

The discussion in terms of saddle points of the continuum
and topological transitions of the decay surface
complement each other.
While the van Hove singularities are always present
in the two-magnon continuum (\ref{continuum})
they do not necessarily cross with the one-magnon branch.
On the other hand, using the topological transition 
perspective one can argue that the occurrence of such a 
crossing does not depend on a precise form of 
$\varepsilon_{\bf k}$.
As was discussed in Sec.~\ref{onshell}.A, 
acoustic magnon with  small ${\bf k}$
decays into two magnons that are close to ${\bf Q}$
and $-{\bf Q}$. Its decay surface consists of two
disjoint parts near $K$ and $K'$ points, see 
$k_x=\pi/4$ in Fig.~\ref{q_contours}. Similarly, a
 magnon with  ${\bf k}\rightarrow {\bf Q}$
can emit two magnons in the vicinity of $-{\bf Q}$.
The corresponding decay surface is a single closed contour
near the $K'$ point,
 see $k_x=\pi$ in Fig.~\ref{q_contours}.
Therefore, moving along an arbitrary trajectory
in ${\bf k}$-space between $\Gamma$ and $K$ points
the decay surface must undergo at least one topological
transformation, which also 
implies crossing of at least one saddle-point threshold 
contour. The actual number of such transformations is determined
by the short-wavelength details of the spectrum. 
For the triangular-lattice HAF there are two of them 
as is demonstrated in Figs.~\ref{q_contours} and \ref{BZ}.

Let us now consider behavior of the magnon self-energy 
in the vicinity of singular points. 
The decay vertex is regular at ${\bf k}\rightarrow 
{\bf k}^*$,  ${\bf q}\rightarrow {\bf q}^*$
and gives an unimportant constant factor. 
Using the expansion 
(\ref{conserv_expan}) we obtain for the singular part of the 
magnon self-energy:
\begin{equation}
\Sigma({\bf k},\varepsilon_{\bf k}) \propto
\int \frac{d^2q} {(v_1 -v_2)\Delta k -
q_x^2/a^2 + q_y^2/b^2 + i0}.
\label{sigmaD}
\end{equation}
A straightforward integration in (\ref{sigmaD}) yields
\begin{equation}
\textrm{Re}\Sigma({\bf k},\varepsilon_{\bf k})\simeq 
\textrm{sign}(\Delta k)\ ,
\ \ \ \ \
\Gamma_{\bf k} 
\simeq  \ln\frac{\Lambda}{|\Delta k|}\ .
\label{log1}
\end{equation}
[$\Gamma_{\bf k} \equiv - \textrm{Im}\Sigma({\bf k},\varepsilon_{\bf k})$].
The cut-off parameter $\Lambda$ is determined by the characteristic 
size of the region in the ${\bf k}$-space where the expansion
(\ref{conserv_expan}) holds. 
The linear size of the smallest ``droplet'' of the decay surface 
at the topological transition can be taken as an upper bound on $\Lambda$, 
see Fig.~\ref{q_contours}.
Such an estimate explains the difference in the strength
of anomalies in $\Gamma_{\bf k}$ for $k^*$ and $k'$ points.
The topological transition at $k^*$ consists of joining/splitting 
of the two approximately equal contours of substantial size,
while the $k'$ point corresponds to the splitting off of a small piece.

To put this discussion in a broader context 
we note than in the earlier  works \cite{LL_IX,Pitaevskii59,field}
the situation was considered when 
a singularity occurs at the boundary of the decay region rather
than in the interior. In such a case, the extremum 
in the two-particle continuum that is crossed by the
single-particle branch is a minimum, not a saddle-point, 
and  the analog of Eq.~(\ref{sigmaD}) for 
$\Sigma({\bf k},\varepsilon_{\bf k})$ is given by
\begin{equation}
\Sigma({\bf k},\varepsilon_{\bf k})  \propto
\int \frac{d^2q} {(v_1 -v_2)\Delta k -
q_x^2/a^2 - q_y^2/b^2 + i0}. 
\label{sigmaB}
\end{equation}
After integration this yields
the following characteristic 
anomaly
\begin{equation}
\textrm{Re}\Sigma({\bf k},\varepsilon_{\bf k})\simeq 
\ln\frac{\Lambda}{|\Delta k|}\ ,
\ \ \ \ \
\Gamma_{\bf k} \simeq \Theta(\Delta k)\ ,
\label{log2}
\end{equation}
where $\Theta(x)$ is the Heaviside step function.
Thus, the situation is reversed in comparison to 
our case:
the log-anomaly occurs in the real and the jump 
in the imaginary parts of the spectrum. 
Since the imaginary part of $\Sigma({\bf k},\varepsilon_{\bf k})$
is related to the two-particle density of states, in 2D it is 
natural to have a jump in $\Gamma_{\bf k}$ upon 
entering the continuum and a log-singularity 
upon crossing the saddle-point line inside
the continuum. By the Kramers-Kronig relations such jumps and logs
in  $\textrm{Im}\Sigma({\bf k},\varepsilon_{\bf k})$ 
result in logs and jumps in 
$\textrm{Re}\Sigma({\bf k},\varepsilon_{\bf k})$, respectively. 
For the 3D systems, logarithmic peaks disappear and one obtains 
only square-root singularity:
$\textrm{Re}\Sigma({\bf k},\varepsilon_{\bf k})\simeq \sqrt{|\Delta k|}$.

Another important question concerns whether singularities 
in the spectrum will survive the higher-order $1/S$ treatment.
If the singularity persists, vertex corrections may 
become important, see Appendix~\ref{app_parquet}.\cite{LL_IX}
Our Sec.~\ref{off_shell} discusses this problem.

\begin{figure*}[t]
\includegraphics[angle=0,height=0.6\columnwidth]{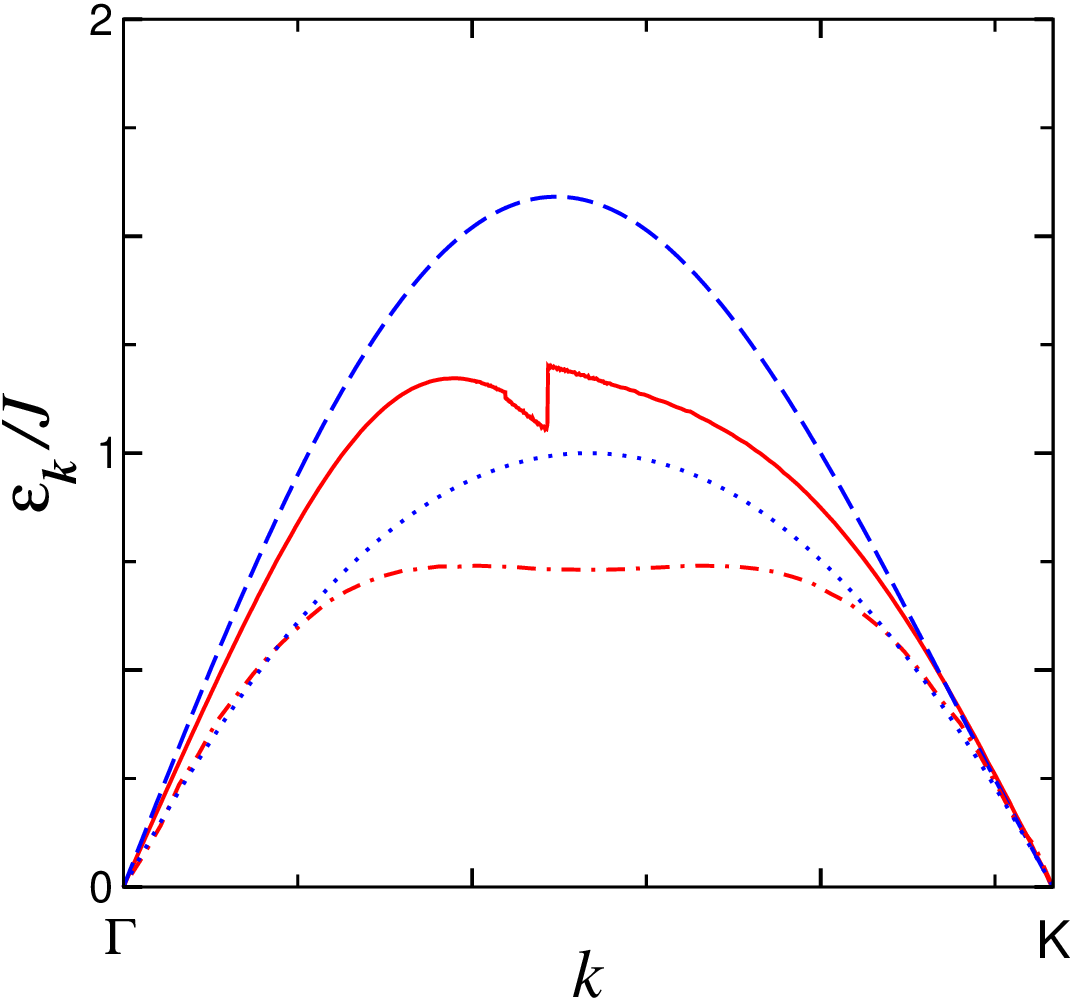} \qquad
\includegraphics[angle=0,height=0.6\columnwidth]{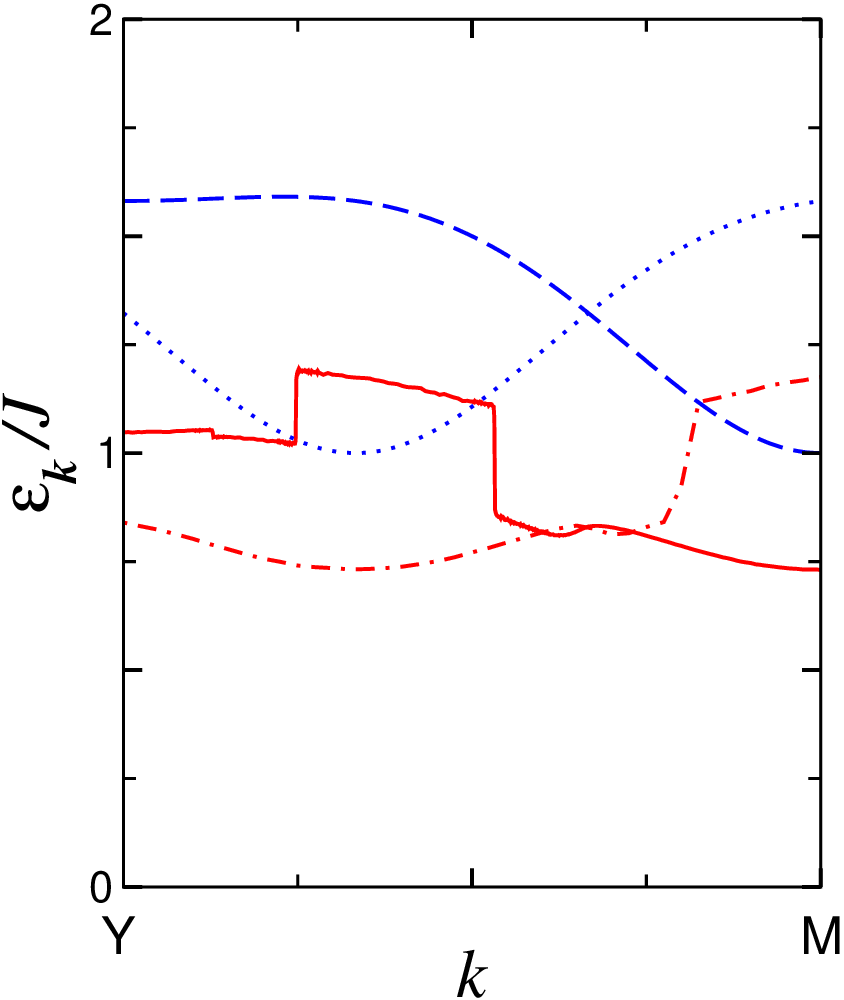} \qquad
\includegraphics[angle=0,width=0.65\columnwidth]{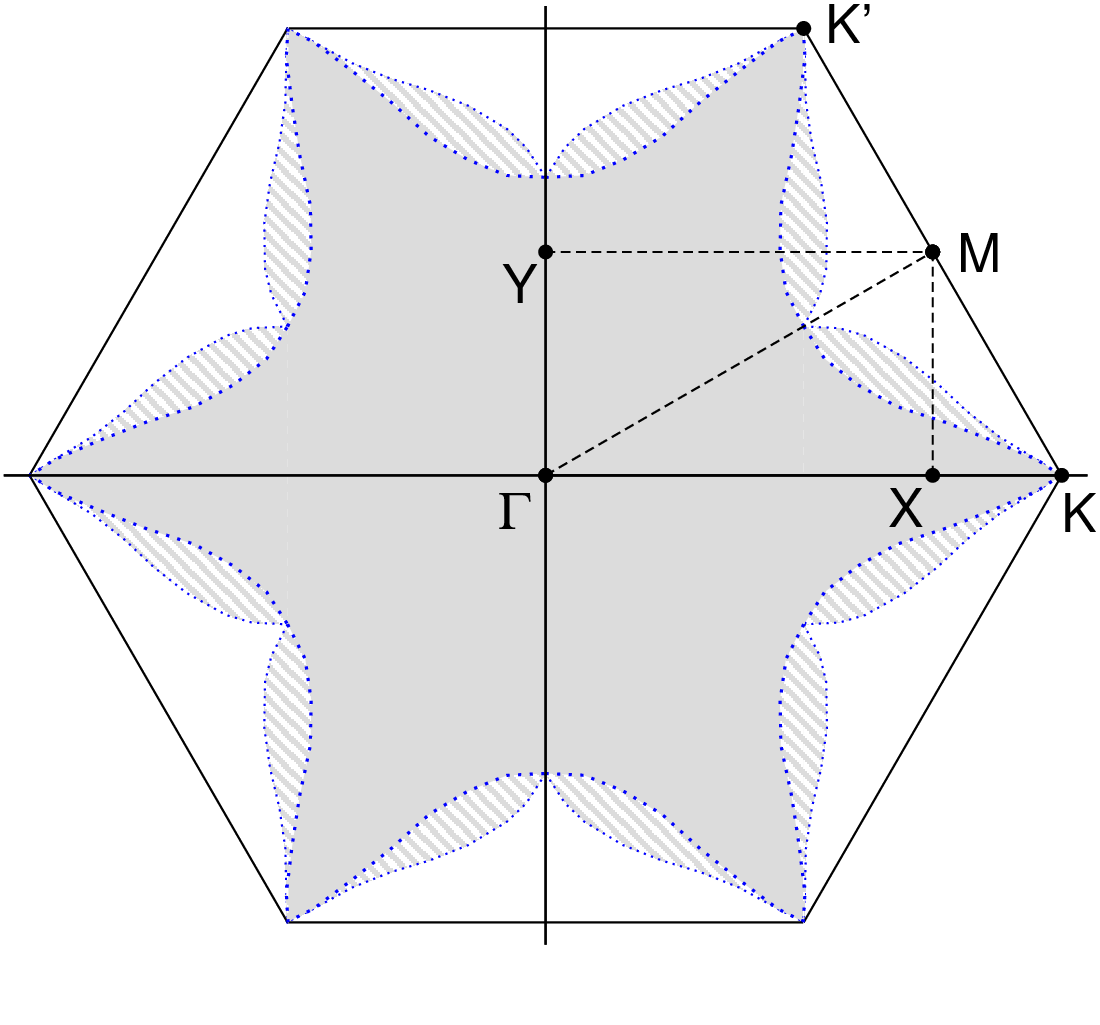}
\caption{(Color online) Renormalization of the decay region 
for the spin-1/2 triangular-lattice HAF. Left and middle panels:
Dashed and dotted lines are the LSWT predictions for the 
magnon dispersion $\varepsilon_{\bf k}$ (dashed)
and for the minimum of the two-magnon continuum (dotted);
solid and dotted-dashed lines represent same results for the  spectrum 
$\bar{\varepsilon}_{\bf k}$ with the first-order quantum correction,
respectively. 
Right panel: Modified decay region; lightly shaded regions 
highlight the area of uncertainty due to the re-entrant behavior 
seen in the middle panel.}
\label{2_mag}
\end{figure*}

\section{Off-shell Dyson equation and spectral function}
\label{off_shell}

The unusual logarithmic singularities in the magnon decay rate 
$\Gamma_{\bf k}$ found in the first-order $1/S$
corrections signify a breakdown of the standard spin-wave expansion. 
They represent an extra theoretical challenge and 
have to be renormalized in order to obtain the actual 
dynamical behavior. 
The purpose of this Section is to describe, at a technical level,
different approaches to this problem. 
In Sec.~\ref{off_shell}.A we show that the singularities are
regularized, for the most part, by allowing for the finite 
lifetime of the magnon in the {\em initial} state within the 
so-called off-shell approach. However, in the strong-coupling case,
the single-particle excitation may disappear in the vicinity of the
singularity, similar to the termination point in the quasiparticle spectrum
of superfluid $^4$He. \cite{Pitaevskii59} 
In Sec.~\ref{off_shell}.B we discuss the magnon spectral function
$A({\bf k},\omega)$ and find additional singularities in the $\omega$-space
that are directly connected to the van Hove singularities in the
two-magnon continuum.

\subsection{Singularities and off-shell solution}

\subsubsection{Modified decay region}

The first-order quantum correction to the magnon dispersion
(Sec.~\ref{onshell}) 
leads to a significant narrowing  of the magnon bandwidth
for the spin-1/2 triangular-lattice HAF. Therefore,  one may ask
how strongly the spectrum renormalization affects the decay boundary
and whether the decay condition Eq.~(\ref{energy_conserv}) 
is still satisfied for the {\it renormalized} 
$\bar{\varepsilon}_{\bf k}$. 
In the long-wavelength limit,  
the difference between the velocities of the acoustic modes $v_0>v_Q$, 
see Eq.~(\ref{v_renorm}),
guaranties that the kinematic conditions 
for decays at ${\bf k}\rightarrow 0$ are always fulfilled. 
While this argument is applicable to any system with several types
of the Goldstone mode, the decay boundary is defined by the 
short-wavelength features, specific to a particular system. 

In Fig.~\ref{2_mag} we present numerical results,
which show that the size of the decay region does not change appreciably
even for the $S=1/2$ triangular-lattice HAF.
The left and the middle panels demonstrate the renormalization
of the bottom of the two-magnon continuum for two representative
paths in the BZ.
The lowest-energy two-magnon states within the decay region
still correspond to the emission of the $\pm{\bf Q}$ Goldstone mode, 
$\bar{E}^{\min}_{\bf k}=\bar{\varepsilon}_{{\bf k}\pm{\bf Q}}$.
While the renormalizations of the spectrum and the continuum  
are significant, the intersection points of the two change only weakly. 
For the YM path in the BZ one observes a re-entrant crossing 
of the  
single-magnon branch with the bottom of the continuum, which leaves some
uncertainty in defining the new decay boundary. This behavior is related to 
the jump-like singularities in the single-magnon 
$\bar{\varepsilon}_{\bf k}$, which should become more
well-behaved once the singularities are regularized. The right panel
of Fig.~\ref{2_mag} presents
the ``new'' magnon decay boundary for the 
spectrum that includes $1/S$ renormalization. 
The light-shaded regions show the uncertainty areas where the 
re-entrant behavior of the spectrum and the continuum occur. Overall, 
the decay region does not change significantly in comparison with
the LSWT boundary in Fig.~\ref{BZ}.

\subsubsection{Higher-order diagrams perspective}

One way to regularize the decay diagram is to allow for a finite
lifetime of the decay products by dressing the inner lines in the ``bubbles''  
in Fig.~\ref{SelfE}. 
The effect of such a dressing 
depends, however, on whether the saddle-point momenta
of the decay products $\bf q^*$ and   ${\bf k^*}-{\bf q}^*$ in 
Fig.~\ref{q_contours} fall inside or outside of the decay region. 
If ${\bf q}^*$ lies inside the decay region for at least one of the
final magnons, then this magnon will acquire a finite lifetime
in the next order. 
The logarithmic singularity in $\Gamma_{\bf k}$  will be removed in this case
since the energy conservation law Eq.~(\ref{energy_conserv}) is
now satisfied only on average.
For the triangular-lattice HAF such a scenario is realized 
for a large fraction of the ``weak''
singularities ($k'$ points in Figs.~\ref{w_k}, \ref{Imw_k}). 
However, all of the ``strong'' singularities 
($k^*$ points in Figs.~\ref{w_k}, \ref{Imw_k}) and some of the ``weak'' 
ones belong to another class in which the saddle points for 
both magnons created in the 
decay process are {\it outside} of the decay 
region. Hence, at the saddle points, 
the logarithmic divergence of the one-loop diagrams will persist 
even for the renormalized spectrum and the singularities seems to remain 
essential. 

Thus, the above approach requires summation of an infinite series of 
diagrams that contain leading-order divergences, similar to 
the Pitaevskii's treatment of the spectrum termination problem.\cite{LL_IX,Pitaevskii59} 
Such a treatment is hindered within the spin-wave theory by the divergence 
of the individual terms at ${\bf k}\rightarrow{\bf Q}$ in each order 
of the $1/S$ expansion, the problem already mentioned in 
Secs.~\ref{formalism} and \ref{onshell}.
While a qualitative statement on the result of such a regularization 
can be made (see Appendix \ref{app_parquet}) 
any quantitatively reliable calculation 
are problematic in the light of this problem. 

\subsubsection{Off-shell Dyson equation}
\begin{figure*}[t]
\includegraphics[angle=0,width=1.6\columnwidth]{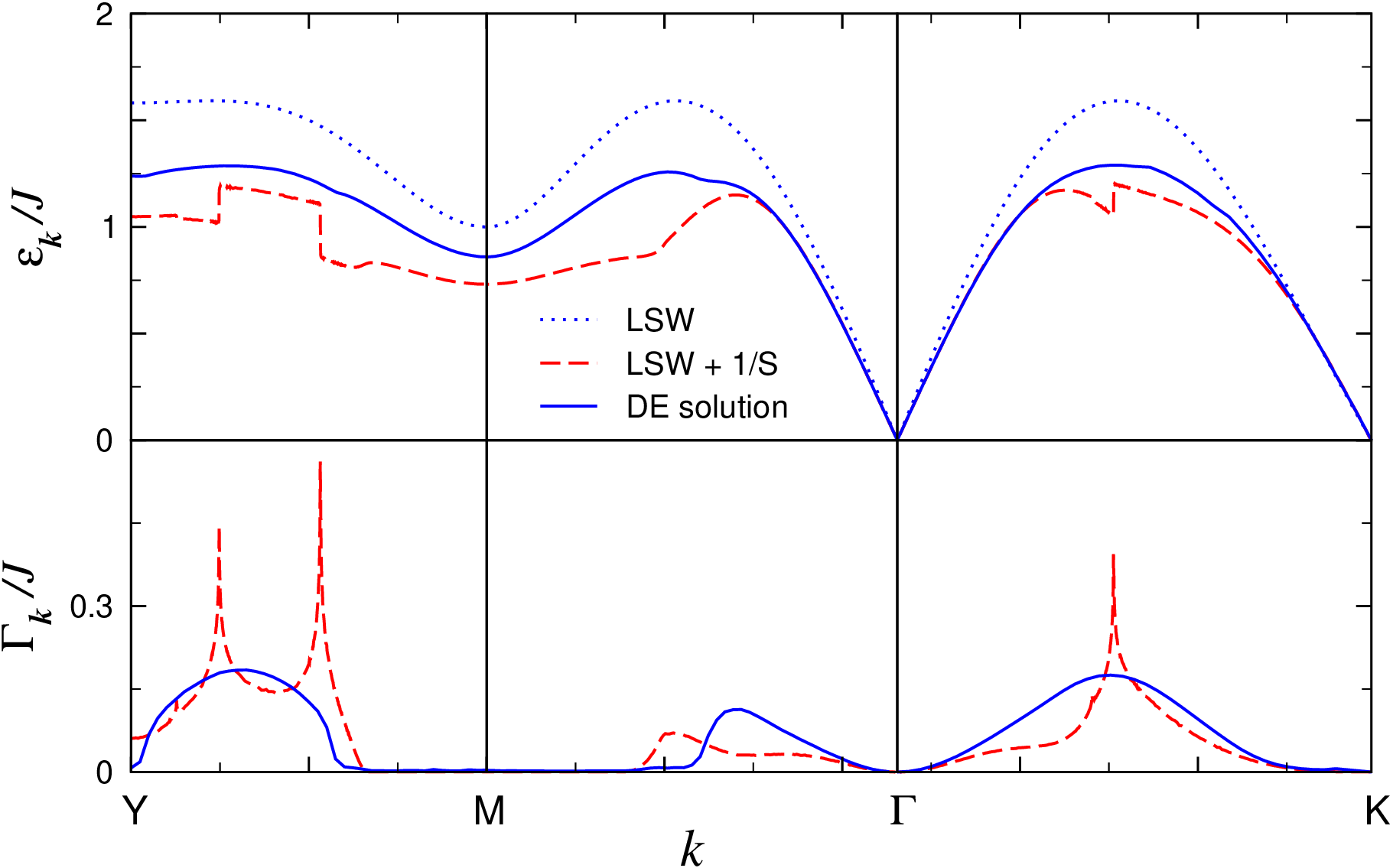}
\caption{(Color online) Comparison of the magnon spectrum (upper row)
and the decay rates (lower row) obtained in the harmonic approximation  
(dotted lines), with the first-order $1/S$ corrections 
(dashed lines), and by solving the Dyson equation (solid lines), 
respectively.}
\label{DE}
\end{figure*}

The above seemingly hopeless situation is resolved if we note that 
the logarithmic singularity occurs for a magnon which is inside the decay region. 
This means that the Dyson equation (\ref{DE_a}) should be solved ``off-shell,''
{\it i.e.}\ with the same complex energy
$\varepsilon=\bar\varepsilon_{\bf k}-i\Gamma_{\bf k}$
both outside and inside $\Sigma({\bf k},\varepsilon)$.
Physically, such a procedure allows for a finite lifetime
of the {\it initial} magnon while magnons created during the decay process 
remain stable. 
The off-shell approach avoids complications related to multi-loop
diagrams and, as we demonstrate below, 
is sufficient to regularize the singularity.
Note that, the magnon energy in the off-shell
solution contains corrections beyond the first 
$1/S$ order. Nevertheless, it can be shown that 
it is yet free from the ${\bf k}\rightarrow{\bf Q}$ divergences 
associated with the higher-order diagrams mentioned above.

After a methodological remark made in Appendix \ref{app_D}
on the proper sign of the imaginary 
part of $\varepsilon$ in the decay-like 
self-energy, the Dyson 
equation  (\ref{DE_a}) is given by
\begin{equation}
\label{DE_SC}
\varepsilon- \varepsilon_{\bf k} -\Sigma({\bf
  k},\varepsilon^*)=0 \ ,  
\end{equation}
where $\varepsilon^*$ is the complex conjugate of $\varepsilon$ and
$\Sigma({\bf k},\varepsilon)$ includes all one-loop 
contributions given by Eq.~(\ref{DE_a}). Rewriting the above equation 
explicitly  for the real and imaginary parts one obtains
the following system
\begin{eqnarray}
\bar\varepsilon_{\bf k} & = & \varepsilon_{\bf k} +
\textrm{Re}[\Sigma({\bf k},{\bar\varepsilon}_{\bf k}+i\Gamma_{\bf k})]\ ,
\nonumber \\
\Gamma_{\bf k} & = & 
-\textrm{Im}[\Sigma({\bf k},\bar\varepsilon_{\bf k}+i\Gamma_{\bf k})]>0\ .
\label{DE_SC1}
\end{eqnarray}

Let us first demonstrate how a finite $\Gamma_{\bf k}$ can regularize 
the singularity. Replacing $\varepsilon_{\bf k}$ in Eq.~(\ref{conserv_expan})
with complex $\varepsilon$ and using parametrization $\varepsilon - \varepsilon_c -
{\bf v}_2\cdot \Delta {\bf k} = \bar\rho =|\bar\rho|e^{-i\varphi}$,
where $\varepsilon_c$ is the position of the saddle point in the continuum,
we obtain after integration
\begin{equation}
\label{singular_sigma}
\Sigma({\bf k},\varepsilon)\simeq
-\frac{V_3^2}{JS}\, \biggl[\Bigl(\frac{\pi}{2}-\varphi\Bigr) +
i \ln \biggl|\frac{\Lambda JS}{\bar\rho}\biggr|\biggr] \ .
\end{equation}
Here, $V_3^2\propto\Gamma_1^2({\bf k^*},{\bf q^*})\sim J^2S$ and 
$\Lambda$ is the momentum cutoff.
Omitting the non-singular contributions, 
Eqs.~(\ref{DE_SC1}) are rewritten as
\begin{eqnarray}
&&\bar\varepsilon_{\bf k}\simeq\varepsilon_{\bf k} 
-\frac{V_3^2}{JS}\left(\frac{\pi}{2}-\varphi\right),
\nonumber \\
\label{DE_sing}
&&\Gamma_{\bf k}\simeq\frac{V_3^2}{JS}
\ln \Bigl|\frac{\Lambda JS}{\bar\varepsilon_{\bf k}
-\varepsilon_c-{\bf v}_2\cdot\Delta{\bf k}+i\Gamma_{\bf k}}\Bigr|\ .
\end{eqnarray}
The on-shell solution is recovered by substituting
$\varepsilon_{\bf k}=\varepsilon_c+
{\bf v}_1\cdot\Delta{\bf k}$ instead of 
$\varepsilon=\bar\varepsilon_{\bf k}+i\Gamma_{\bf k}$ in (\ref{DE_sing}). 
This yields $\bar\rho = ({\bf v}_1-{\bf v}_2)\cdot \Delta {\bf k}$ 
and also implies that $\cos\varphi=\pm 1$ (or $\varphi=0,\pi$)
depending on the sign of $\Delta k$. As a result, one finds the
jump in the real part of the spectrum and the log-singularity in the
decay rate in agreement with Eq.~(\ref{log1}).

Any renormalization should shift the ``bare''
singularity into a new crossing point 
of the single-magnon branch with the
surface of the saddle points in the two-magnon continuum.
Assuming that the real part of the energy renormalization is already
included in the definition of $\varepsilon_c$ and ${\bf k}^*$ and that $\bar\varepsilon_{\bf k}$ 
can be still expanded as 
$\bar\varepsilon_{\bf k}\simeq\varepsilon_c +{\bf v}_1\cdot\Delta{\bf k}$
we obtain 
\begin{equation}
\Gamma_{\bf k}\simeq\frac{V_3^2}{JS}
\ln \Bigl| \frac{\Lambda JS}
{\Delta{\bf v}\cdot\Delta{\bf k}+i\Gamma_{\bf k}}\Bigr| \ .
\label{DE_offshel_Im}
\end{equation}
As a result, the imaginary part of the solution at the singular point
$\Delta{\bf k}$ is now regular and can be determined by solving 
the transcendental equation:
\begin{equation}
\label{Gamma_offshell}
\gamma^*= \ln \Bigl(\frac{\Lambda (JS)^2}{V_3^2}\,\frac{1}{\gamma^*}\Bigr)
\end{equation}
with $\Gamma_{\bf k} = V_3^2 \gamma^*/JS$.
Depending on the relative strength of the three-magnon coupling
$(V_3/JS)^2$ to the size of the dimensionless momentum cutoff $\Lambda$ 
there are two different regimes:
\begin{eqnarray}
\label{Gamma_offshell_1}
&&\Gamma_{\bf k}\simeq\frac{V_3^2}{JS}
\ln \Bigl[\frac{\Lambda (JS)^2}{V_3^2}\Bigr] \ ,
\ \ V_3^2/\Lambda(JS)^2\ll 1 \ , \nonumber \\
&&\Gamma_{\bf k}\simeq \Lambda JS \ ,
\ \ \ \ \ \ \ \ \ \ \ \ \ \ \ \ \ 
 V_3^2/\Lambda(JS)^2\gg 1 \ .
\end{eqnarray}
Thus, at large $ V_3^2/\Lambda(JS)^2$ the decay rate is independent of
the coupling and is defined by the phase volume factor. Recalling
that $V_3^2\propto J^2 S$ in the case of the triangular-lattice HAF, we
obtain estimates:
\begin{eqnarray}
\label{Gamma_offshell_2}
&&\frac{\Gamma_{\bf k}}{JS}\propto\frac{1}{S}\ln (S\Lambda) \ ,
\ \ \ \ S\Lambda \gg 1 \ ,\nonumber \\
&&\frac{\Gamma_{\bf k}}{JS}\propto \Lambda \ ,
\ \ \ \ \ \ \  \ \ \ \ \ \  \ 
 S\Lambda \ll 1 \ ,
\end{eqnarray}
where the first expression is relevant to the ``strong'' singularities
with large phase space volume for decays ($k^*$-points), and the second is
for the ``weak'' ones ($k'$-points).

For ${\bf k}\rightarrow{\bf k^*}$, one finds
for the off-shell solution
\begin{eqnarray}
\label{Re_offshell_1}
\cos\varphi =
\frac{{\rm Re}(\Delta\varepsilon-{\bf v}_2\cdot\Delta{\bf k})}
{|\Delta\varepsilon-{\bf v}_2\cdot\Delta{\bf k}|}
\approx \frac{\Delta{\bf v}\cdot\Delta{\bf k}}{\Gamma_{\bf k}}
\rightarrow 0 \ .
\end{eqnarray}
Hence, $\varphi \rightarrow\pi/2$ and the singular jump in 
$\bar\varepsilon_{\bf k}$, see Eq.~(\ref{DE_sing}), also disappears
in agreement with the above assumption.

In a hypothetical case of the strong cubic term ($V_3\gg JS$)
one needs to consider vertex renormalizations discussed in 
Appendix~\ref{app_parquet}. Briefly, this replaces the log-singularity
in the self-energy with  
${\rm Im}\Sigma({\bf k},\varepsilon)\propto 1/\ln|\Lambda/\rho|$. 
Solving the Dyson equation yield the same answer as in 
(\ref{Gamma_offshell_1}) for the $V_3^2/\Lambda(JS)^2\ll 1$ limit, 
while in the opposite case solution for the single-particle 
spectrum near $k^*$ does not exist. This is similar to the
complete disappearance of the spectrum at the termination 
point of the spectrum in the superfluid $^4$He. \cite{Pitaevskii59}.

Finally, we present in Fig.~\ref{DE} the numerical solution 
of the Dyson equations (\ref{DE_SC1}) for 
the spin-1/2 triangular-lattice HAF. 
While the jumps and the logarithmic peaks disappear, the damping 
rate remains substantial throughout BZ. 
Note, that the overall shape of the off-shell 
$\bar\varepsilon_{\bf k}$ 
is in a better agreement with the series expansion 
data \cite{Zheng06} than the on-shell results. In particular, the
``roton'' minimum is well-pronounced and the ``flat regions'' 
are much less significant than in the latter case. 
However, there is an overall upward
energy scale offset of our results relative to the numerical ones.
This may be due to both the remaining higher-order $1/S$-corrections 
to the spectrum in the spin-wave theory approach and the neglect
of the imaginary part of the spectrum in the series-expansion
calculations. The upward energy renormalization of the off-shell versus 
on-shell results is natural as the latter tends to overestimate the 
energy shifts.

Altogether, the main result of the off-shell consideration is that this
approach naturally resolves the singularity problem, 
regularizing the log-singularities in the decay rates and removing the
concomitant jump-like discontinuities in the real part of the spectrum.
As the result, the decay rates remain significant  and are logarithmically
enhanced relative to the perturbative results.

\subsection{Spectral function}

Generally, a detailed information about both the
pole-like and the incoherent parts of the single-particle
spectrum is obtained from the diagonal component 
of the spectral function 
\begin{eqnarray}
\label{Akw_def}
A({\bf k},\omega)=-\frac{1}{\pi}\, 
{\rm Im}\bigl[G_{11}({\bf k},\omega)\bigr] \ .
\end{eqnarray}
In the leading one-loop approximation the diagonal Green's function
for the triangular-lattice HAF is given by
\begin{equation}
\label{Gkw_def}
G^{-1}_{11}({\bf k},\omega) = \omega-\varepsilon_{\bf k}-
\Sigma({\bf k},\omega) \ ,
\end{equation}
with  
$\Sigma({\bf k},\omega)=\Sigma^{HF}({\bf k})+
\Sigma_{11}^{(a)}({\bf k},\omega)
+\Sigma_{11}^{(b)}({\bf k},\omega)$ expressed  by Eq.~(\ref{DE_a}).
In quantum antiferromagnets the spectral function
 $A({\bf k},\omega)$ is also
related to the dynamical  structure factor
$S({\bf k},\omega)$ which is directly measured 
in inelastic neutron experiments. Generally, $S({\bf k},\omega)$ 
also has contributions from the
off-diagonal and two-particle correlations,
\cite{Ohyama93,Veillette05,Dalidovich06} but
the spectral function (\ref{Akw_def})  still provides the major component. 

In the absence of intrinsic damping, the quasiparticle peak in 
$A({\bf k},\omega)$ occurs precisely at $\omega=\bar\varepsilon_{\bf k}$ 
found from the solution of the Dyson equation.
In the presence of spontaneous decays the solution of 
Eq.~(\ref{DE_SC1}) differs from the position and the width of a
quasiparticle peak in the spectral function because the latter
is defined on the real $\omega$-axis.
Another characteristic feature of $A({\bf k},\omega)$ for all
non-collinear AFs is the  contribution from the two-magnon continuum 
due to  a non-orthogonality of the one- and two-particle
excitations. In particular, for the 
momenta $\bf k$ inside  the decay region, the  
spectral weight in $A({\bf k},\omega)$ should become non-zero above the 
bottom of the continuum. On the same ground, 
one should also expect singular behavior due to the
van Hove singularities of the continuum 
to be prominent in the spectral function at any ${\bf k}$, not only at
special contours of ${\bf k^*}$. 
This is because frequency scans through all possible energies and is 
not limited to the ``mass-surface'' $\omega=\varepsilon_{\bf k}$. 
Since we are restricted 
to the one-loop approximation for 
$\Sigma({\bf k},\omega)$  due to difficulties with the higher-order
diagrams discussed above, such van Hove singularities will appear as sharp
features in $A({\bf k},\omega)$. This is due to both the one-loop
approximation for the self-energy 
and because $\omega$ at which the
system is probed is purely real. 
Qualitatively, all singularities are expected to be regularized
by the higher-order contributions. The complication is, of course,
that a quantitative calculation of such a regularization can be
difficult if not impossible. 
\begin{figure}[b]
\includegraphics[width=0.99\columnwidth]{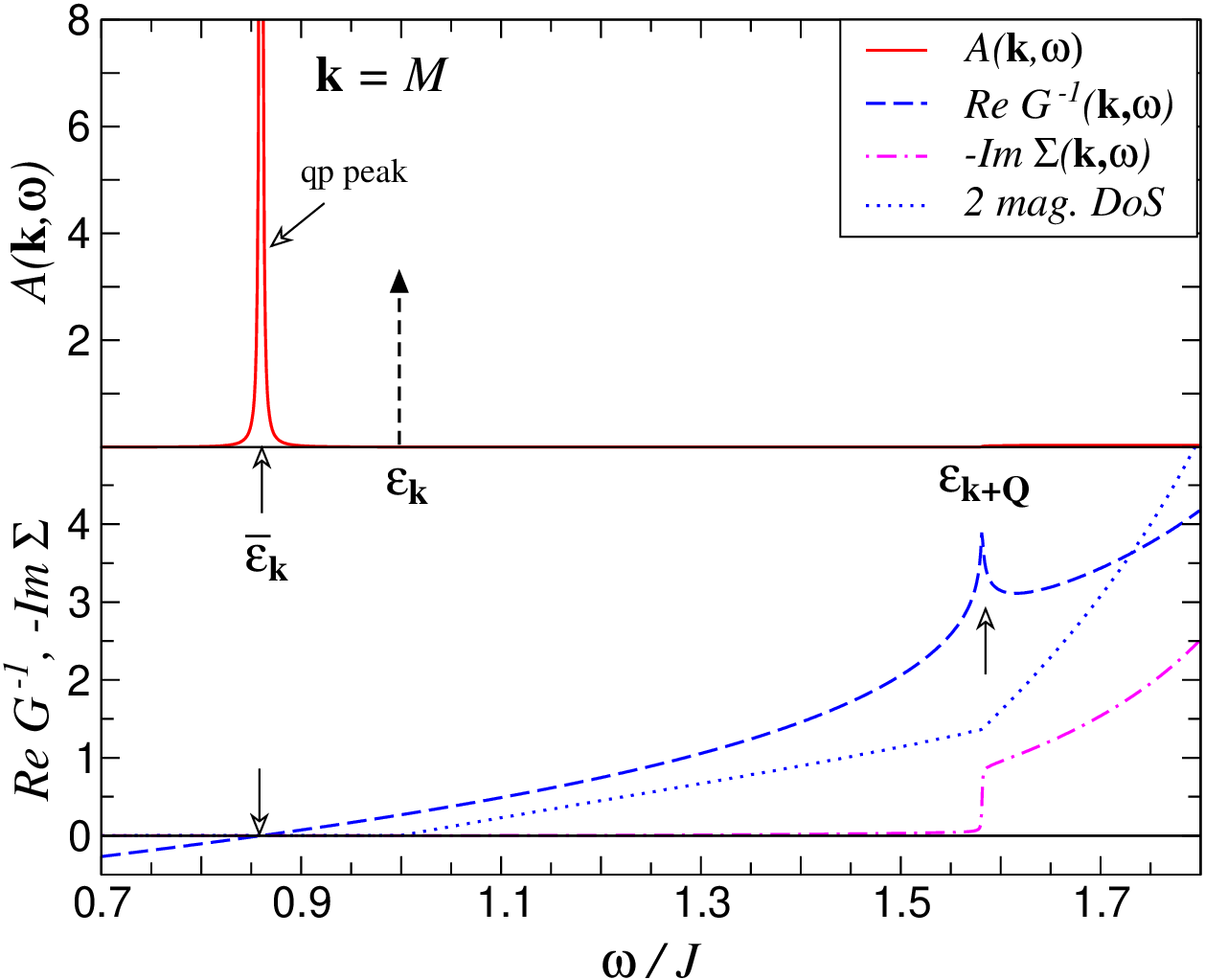}
\caption{(Color online) Magnon spectral function $A({\bf k},\omega)$, 
upper panel; Real and imaginary parts of one-loop 
$G^{-1}_{11}({\bf k},\omega)$ together with the two-magnon DoS (dotted line),
lower panel, all for  ${\bf k}=(\pi,\pi/\sqrt{3})$ ($M$-point).
Dashed arrow in the upper panel indicates position of 
the ``bare'' magnon peak. Solid  arrows denote position of the 
renormalized quasiparticle peak, 
zero of ${\rm Re}[G_{11}({\bf k},\omega)^{-1}]$, and 
the van Hove singularities.}
\label{Akw1}
\end{figure}

\begin{figure*}[t]
\includegraphics[width=0.99\columnwidth]{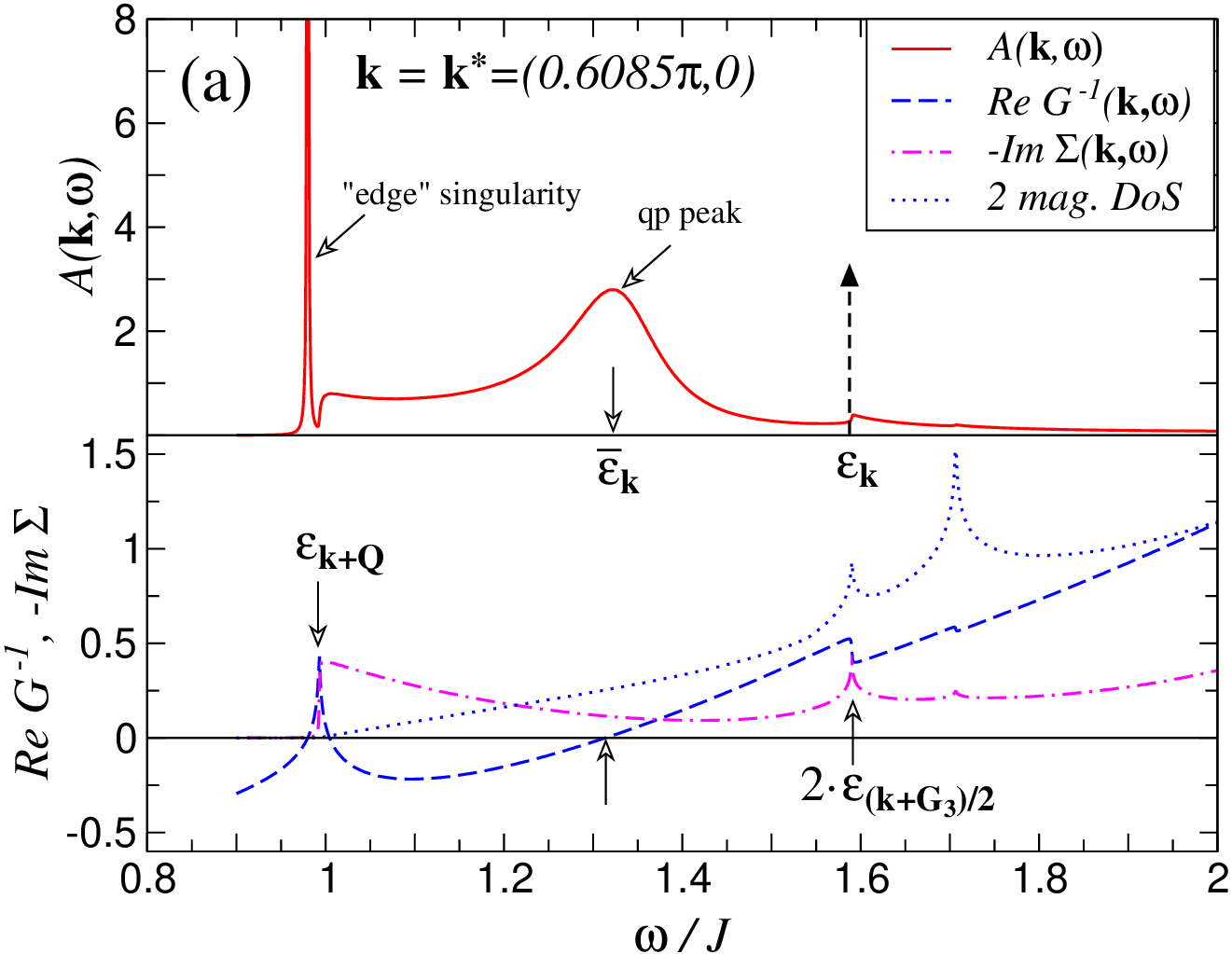}\ \hskip 0.5cm
\includegraphics[width=0.99\columnwidth]{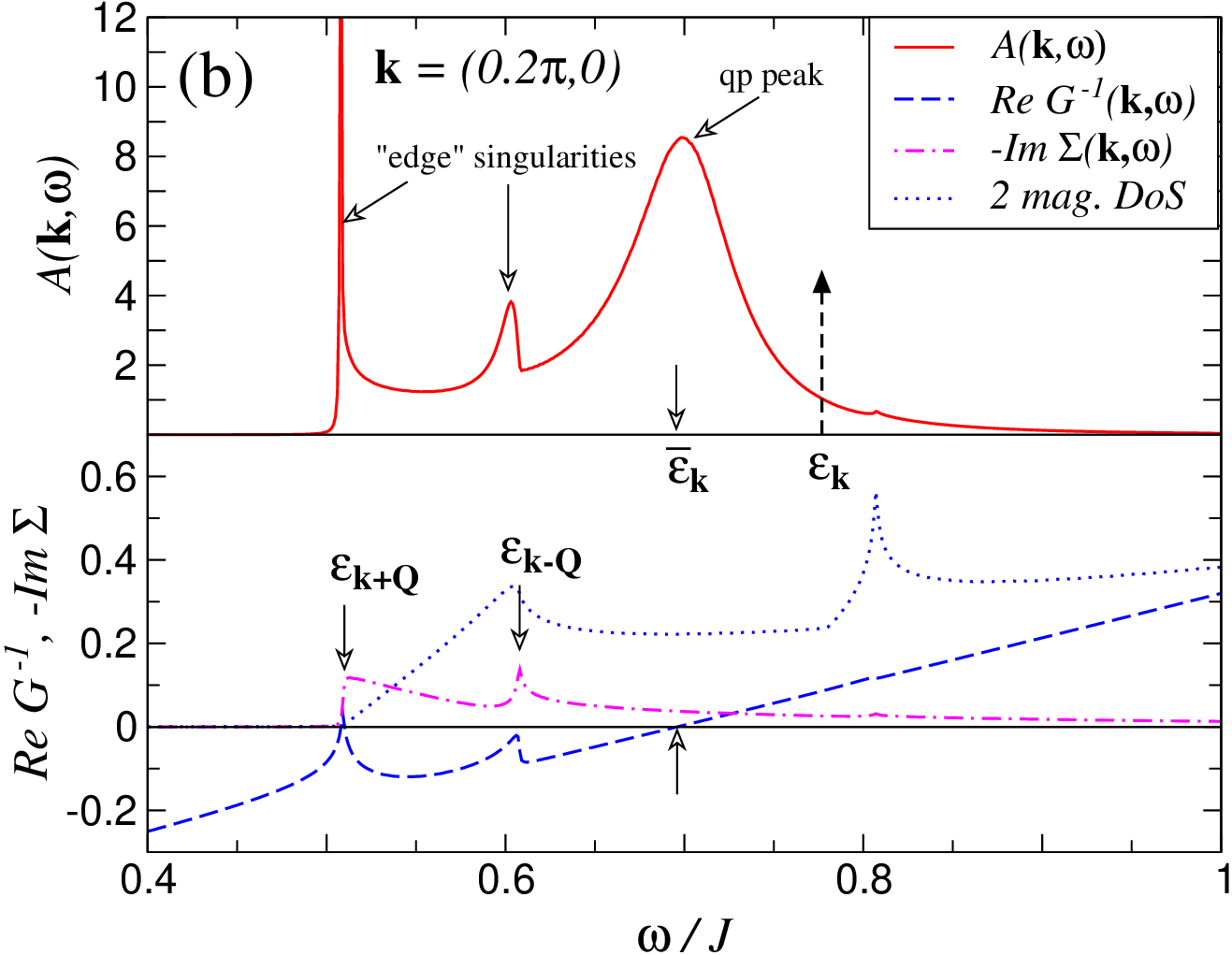}
\caption{(Color online) Same as in Fig.~\ref{Akw1}, 
for momenta on the $\Gamma K$ line, (a)
${\bf k}={\bf k^*}=\pi(0.6085,0)$ and (b) ${\bf k}=\pi(0.2,0)$. 
Solid vertical arrows indicate position of the 
quasiparticle peak,  zero of ${\rm Re}[G^{-1}_{11}({\bf k},\omega)]$, 
van Hove and ``edge'' singularities  (see text).}
\label{Akw2}
\end{figure*}

Figures \ref{Akw1} and \ref{Akw2} show the spectral
function (\ref{Akw_def}) for three different momenta:
the $M$-point (face-center of the BZ) outside of the decay region,
 and the two points on the $\Gamma K$ line, 
${\bf k}=(0.6085\pi,0)$ and ${\bf k}=(0.2\pi,0)$, both inside of 
the decay region.
The momentum for Fig.~\ref{Akw2}(a) is
${\bf k^*}$, which corresponds to the 
logarithmic peak in the on-shell $\Gamma_{\bf k}$. 
The upper panels in Figs.~\ref{Akw1} and \ref{Akw2} show the spectral
function $A({\bf k},\omega)$, while the lower panels
contain ${\rm Re}\,G^{-1}({\bf k},\omega)$, 
${\rm Im}\,G^{-1}({\bf k},\omega)$, and the two-magnon 
density of states (DoS): 
\begin{eqnarray}
\label{2mag_DOS}
D_{\bf k}^{\rm 2mag}(\omega)=\sum_{\bf q}\delta\left(
\omega -\varepsilon_{\bf q}-\varepsilon_{\bf k-q}\right). 
\end{eqnarray}
The energies of the non-interacting spin-waves $\varepsilon_{\bf k}$ 
are indicated by dashed arrows. The quasiparticle peaks in 
$A({\bf k},\omega)$ plots  are  marked by solid arrows. 
The lower panels shows that ${\rm Re}[G_{11}({\bf k},\omega)^{-1}]$ 
vanishes at the same $\omega=\bar\varepsilon_{\bf k}$, as is expected for 
the pole behavior. For momenta inside the decay region, 
these peaks are also significantly broadened, see Figs.~\ref{Akw2}(a),~(b).
While $\bar\varepsilon_{\bf k}$ differs very little from the one found 
in the numerical solution of the Dyson equation
(\ref{DE_SC1}),  the damping in 
Figs.~\ref{Akw2}(a),~(b) is somewhat 
smaller than the one found self-consistently
from (\ref{DE_SC1}). As we discuss in Appendix~\ref{app_residue}, 
one can properly define the quasiparticle residue even for the case 
of the non-zero damping. The issue of defining it at a singularity 
point is also discussed there.

Confirming our previous discussion, the spectral weight in the decay 
region, Fig.~\ref{Akw2}, 
is non-zero above the two-magnon continuum $\omega$-boundary.
The lower panels in Figs.~\ref{Akw2}(a),~(b) demonstrate
that every singularity in the two-magnon DoS 
is reflected in both the real and the imaginary parts of the self-energy 
and, as a result, in the spectral function.
For ${\bf k}={\bf k^*}$ in Fig.~\ref{Akw2}(a) the bare magnon energy 
$\varepsilon_{\bf k}$ coincides with one such singularity, which 
corresponds to the decay into two magnons with energies 
$\varepsilon_{({\bf k}+{\bf G_3})/2}$. This intersection with the 
singularity surface causes the anomaly in the on-shell spectrum, 
see Sec.~\ref{onshell}. For a non-singular momentum in Fig.~\ref{Akw2}(b), 
which is not on the ${\bf k^*}$ contour, similar singularity is 
above the magnon energy.

Perhaps the most spectacular and also unexpected feature of all the data
in Figs.~\ref{Akw2}(a),~(b) are the sharp peaks at the bottom of 
the spectrum that {\it are not} associated with a concomitant peak
in the two-magnon DoS. At the first glance, it may even be concluded that
these peaks are the ``true'', well-defined quasiparticle peaks with 
zero damping. A close inspection of ${\rm Re}[\Sigma({\bf k},\omega)]$
and ${\rm Im}[\Sigma({\bf k},\omega)]$ in the lower panels clearly
connects the peaks in  $A({\bf k},\omega)$ to the log- and jump-like 
singularities in the one-loop self-energy. The origin of them
is slightly more delicate than just a two-magnon DoS feature.
The two-magnon DoS at the bottom of the continuum $\varepsilon_b$ 
in the decay region corresponds to the boundary to the emission
of the $\varepsilon_{\bf \pm Q}$ magnon. Thus, with 
$\varepsilon_b=\varepsilon_{\bf k\pm Q}$, 
$\Delta\omega=\varepsilon-\varepsilon_b$, and ${\bf q}$ in the
vicinity of $\pm{\bf Q}$, the threshold behavior of it is:
\begin{eqnarray}
\label{2mag_DOS1}
D_{\bf k}^{\rm 2mag}(\omega)\propto\int q\, dq\, \delta\left(
\Delta\omega -v_Q|{\bf q}|\right)\propto
\Theta(\Delta\omega)\cdot
\left(\Delta\omega\right).
\end{eqnarray}
Within the self-energy, the decay vertex exhibits anomalous behavior 
at small $\delta {\bf q}=\pm{\bf Q}+{\bf q}$ for 
${\bf k}$ away from ${\bf k}=0$ and from the decay boundary: 
${\widetilde\Gamma}_1({\bf k},{\bf q})\propto 1/\sqrt{|\delta {\bf q}|}$.
This yields a jump-like threshold behavior in the decay rate 
\begin{eqnarray}
\label{Sigma_thresh}
{\rm Im}[\Sigma({\bf k},\omega)]\propto\int dq\, \delta\left(
\Delta\omega-v_Q|{\bf q}|\right)\propto \Theta(\Delta\omega), 
\end{eqnarray}
and the concomitant log-singularity in 
${\rm Re}[\Sigma({\bf k},\omega)]$, as in Sec.~\ref{kinematic}.B. 
Thus, the weak threshold singularity of the type 
$\Theta(\Delta\omega)\cdot\Delta\omega$ in the two-magnon DoS is enhanced by 
the singular decay vertex. This leads to a zero in 
${\rm  Re}[G_{11}({\bf k},\omega)^{-1}]$ and  
a ``pseudo''-quasiparticle peak in $A({\bf k},\omega)$.
 We refer to these anomalies in Figs.~\ref{Akw2}(a),~(b) 
as to the ``edge'' singularities. While in Fig.~\ref{Akw2}(a)
this singularity is due to $\varepsilon_{\bf - Q}$ magnon,
in Fig.~\ref{Akw2}(b) there are two such singularities, one 
associated with $\varepsilon_{\bf -Q}$ and the other 
$\varepsilon_{\bf + Q}$ magnon emission boundaries.
\begin{figure*}[t]
\includegraphics[height = 0.7\columnwidth]{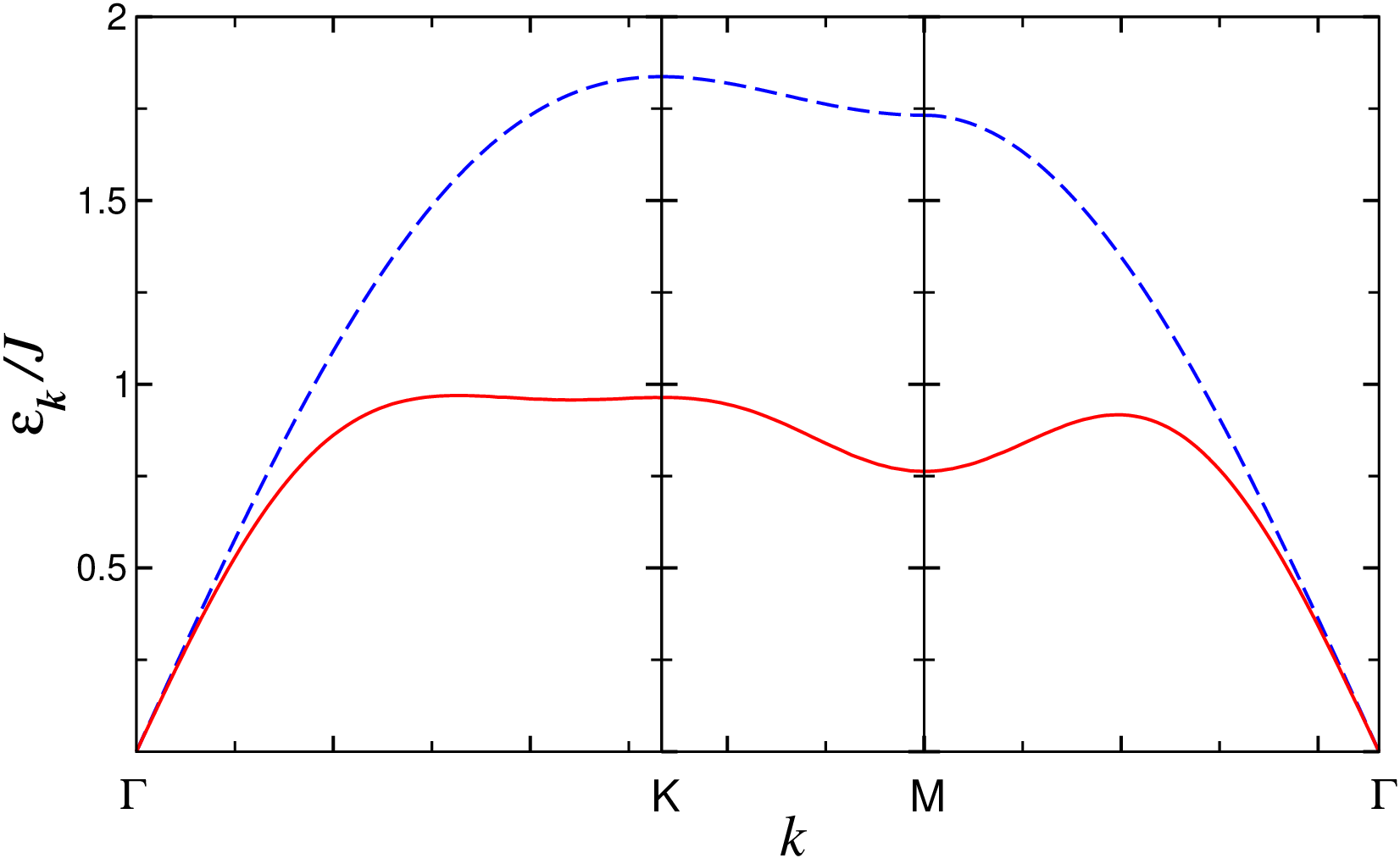} \ \hskip 1cm
\includegraphics[height = 0.7\columnwidth]{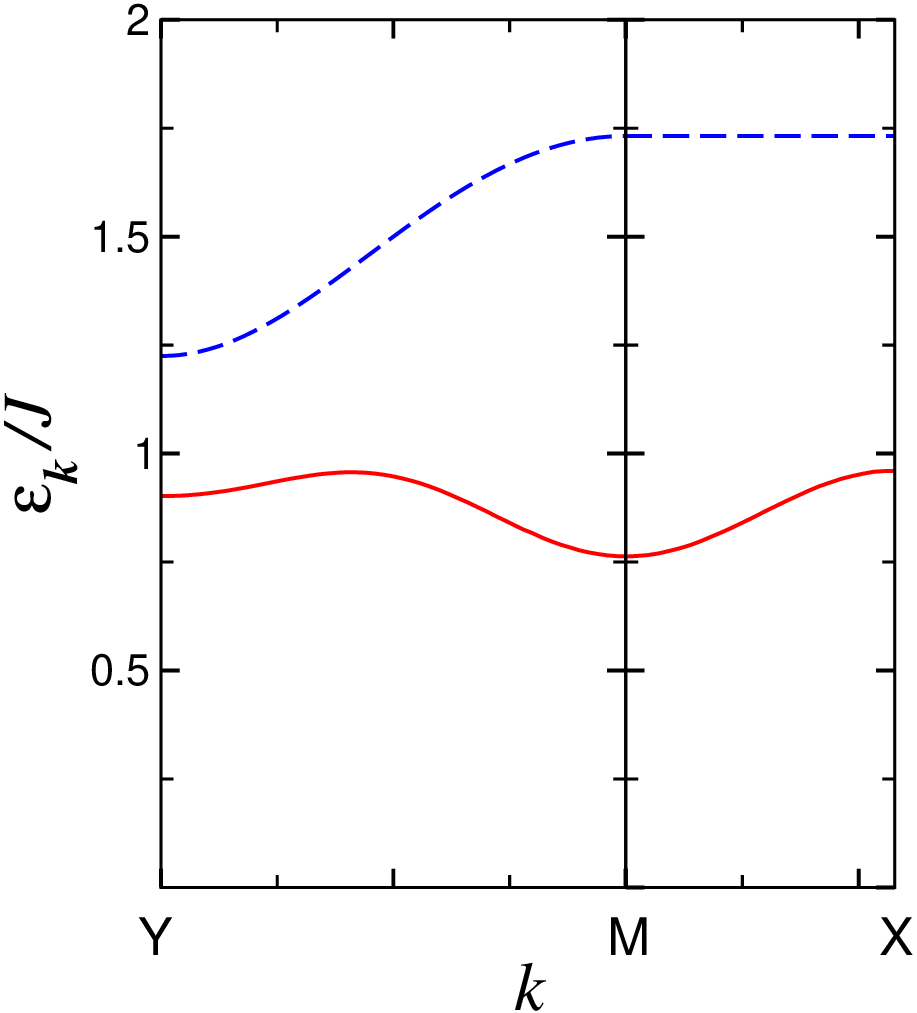}
\caption{(Color online) 
Magnon energy for the $XY$ spin-1/2 
triangular-lattice AF along representative directions. 
Dashed line is the linear spin-wave dispersion and 
solid line is including the first $1/S$ correction.
}
\label{w_k_XY}
\end{figure*}

As is discussed earlier, all such singularities
should be regularized by the higher-order treatment. For the
``edge''-singularities, the regularization requires 
summation of the Pitaevskii's sequence, see Appendix~\ref{app_parquet},
since the external energy is purely real in $A({\bf k},\omega)$.

To summarize, the conventional consideration of the single-particle
spectral function in the triangular-lattice HAF is complicated by the
logarithmic singularities associated with the van Hove singularities in the
two-magnon continuum and with the singular behavior of the
three-magnon coupling. Despite the contamination with these spurious
features, a sensible analysis of the spectrum in terms of the
broadened quasiparticle peak on the background of the two-magnon
continuum is still possible.

\section{Other models }
\label{models}

In this Section we demonstrate that anomalous 
features found in the spectrum of 
the triangular-lattice Heisenberg 
antiferromagnet are generic and appear in
a wide variety of noncollinear antiferromagnets. 
One straightforward generalization from the considered case
is to the easy-plane $XXZ$ model.
Another modification consists of reducing lattice symmetry 
while keeping isotropic interaction between spins.  
This gives the so-called $J$--$J'$ model on an 
orthorhombically distorted triangular lattice: 
the Heisenberg exchange along horizontal chains $J$ 
is stronger than the interchain coupling $J'$. Another 
system discussed here is the kagom\'e-lattice 
antiferromagnet with the $XXZ$
anisotropy.  In the following we outline basic 
kinematic conditions for two-magnon decays in these models
and emphasize the singularities in the magnon spectrum.

\subsection{$XXZ$ model}
The $XXZ$ antiferromagnet 
on a triangular lattice is defined by the following Hamiltonian
\begin{eqnarray}
\hat{\cal H}  =  J
\sum_{\langle ij\rangle}  
\bigg[S^x_i S^x_j + S^y_i S^y_j +\alpha S^z_i S^z_j\bigg]\ ,
\label{H_XXZ}
\end{eqnarray}
For the easy plane system with the anisotropy parameter
$\alpha=J_z/J<1$ spins form the same 120$^\circ$ structure as 
in the Heisenberg case. In the harmonic approximation
the spin-wave energy is given by a
simple modification of the Heisenberg formula (\ref{omega}):
\begin{equation}
\varepsilon_{\bf k} = 3JS \sqrt{(1-\gamma_{\bf k})
(1+2\alpha\gamma_{\bf k})}\ .
\label{epsilonXXZ}
\end{equation}
In addition, the three-boson interaction terms retain the same functional 
form as in Eqs. (\ref{H3})--(\ref{G12}).
Therefore, for the $1/S$ consideration of the $XXZ$ model the
 changes concern only
Bogolyubov parameters and quartic terms. While a detailed
consideration of this model
is beyond the scope of the present work, we would like
to highlight two ubiquitous features of its spin-wave spectrum 
determined by  the three-boson interactions: strong renormalizations 
and decays.

Let us first focus on the strongly anisotropic case and put $\alpha=0$.
The ground-state energy  for such an $XY$ antiferromagnet is
\begin{equation}
E_{\rm g.s.}/N
= -\frac{3}{2}J S^2\biggl[ 1 + \frac{0.064515}{2S}+ \frac{0.013326}{(2S)^2}
\biggr]
\end{equation}
in agreement with Ref.~\onlinecite{Miyake85}.
The harmonic spin-wave spectrum for this case 
is shown in Fig.~\ref{w_k_XY} by dashed lines. 
Because of the reduced spin-rotational symmetry
there is only one acoustic branch near the $\Gamma$-point.
The zero-point fluctuations reduce
the sublattice magnetization  to
$\langle S\rangle = S -0.051467$ in 
the linear spin-wave approximation.
Even for the spin-1/2 case this amounts only to a 10\% renormalization.
Yet, the $1/S$ corrections
to the spectrum shown by solid lines in Fig.~\ref{w_k_XY} 
are even larger than in the isotropic case!
Magnon bandwidth narrowing in  $\alpha=0$ case is almost 50\% of
its bare value. One can also observe that the much-discussed 
``roton-like'' minimum at the $M$-point is much more pronounced
here than in the spin-wave results for the Heisenberg limit.
The origin of this minimum can be traced to the 1D-like van Hove 
singularity  in the 
two-magnon density of states at the $M$-point, similarly
to the isotropic case, Sec.~\ref{onshell}.B. In addition, one can 
notice that the top of the renormalized magnon band exhibits much more 
extended flat regions than in the Heisenberg limit as well as some 
weaker minima (between $\Gamma$ and $K$) and other extrema.
These features must affect the thermodynamics of this model substantially.
Thus, interpolating between $\alpha=0$ and $\alpha=1$ cases, one can 
conclude that the anharmonic three-boson terms lead to 
very strong spectrum renormalization throughout the BZ for the 
$XXZ$ model on the triangular lattice for all values of $\alpha$. 

\begin{figure}[b]
\includegraphics[width=0.65\columnwidth]{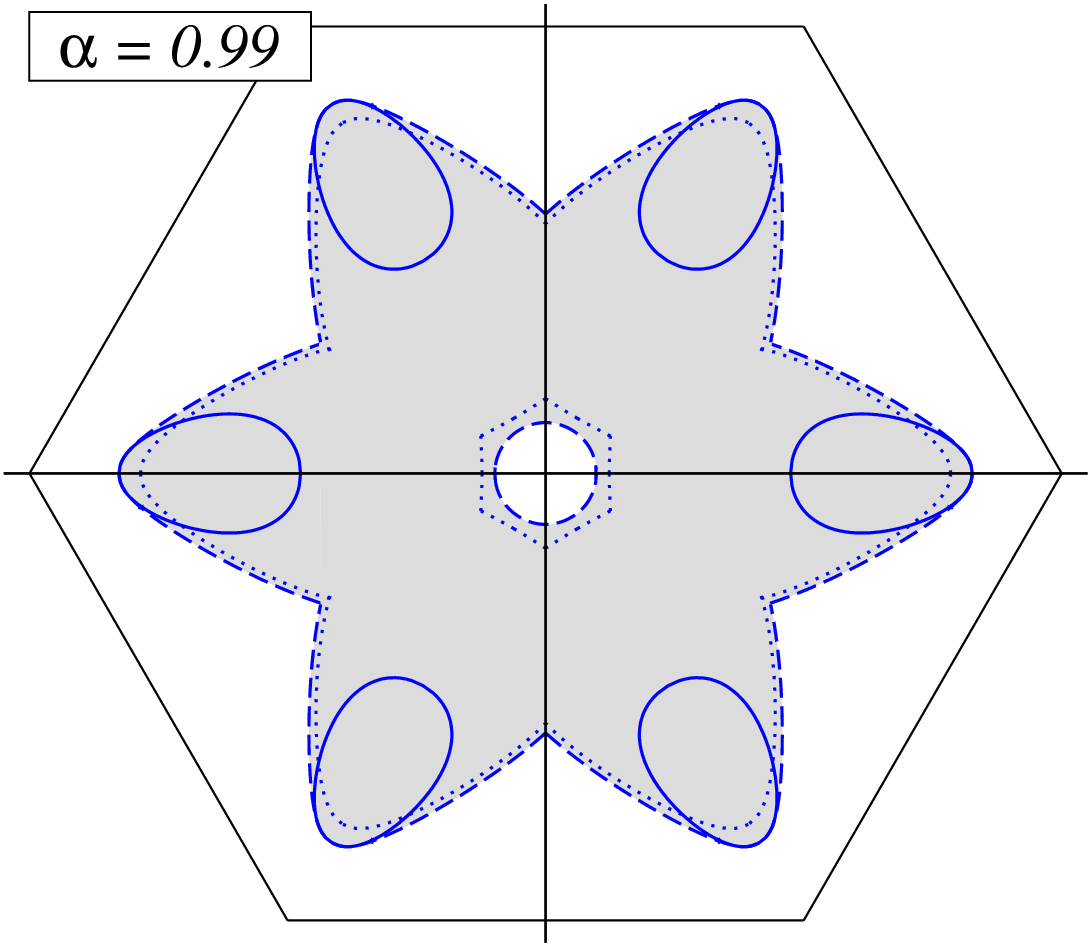}
\vskip 1cm
\includegraphics[width=0.65\columnwidth]{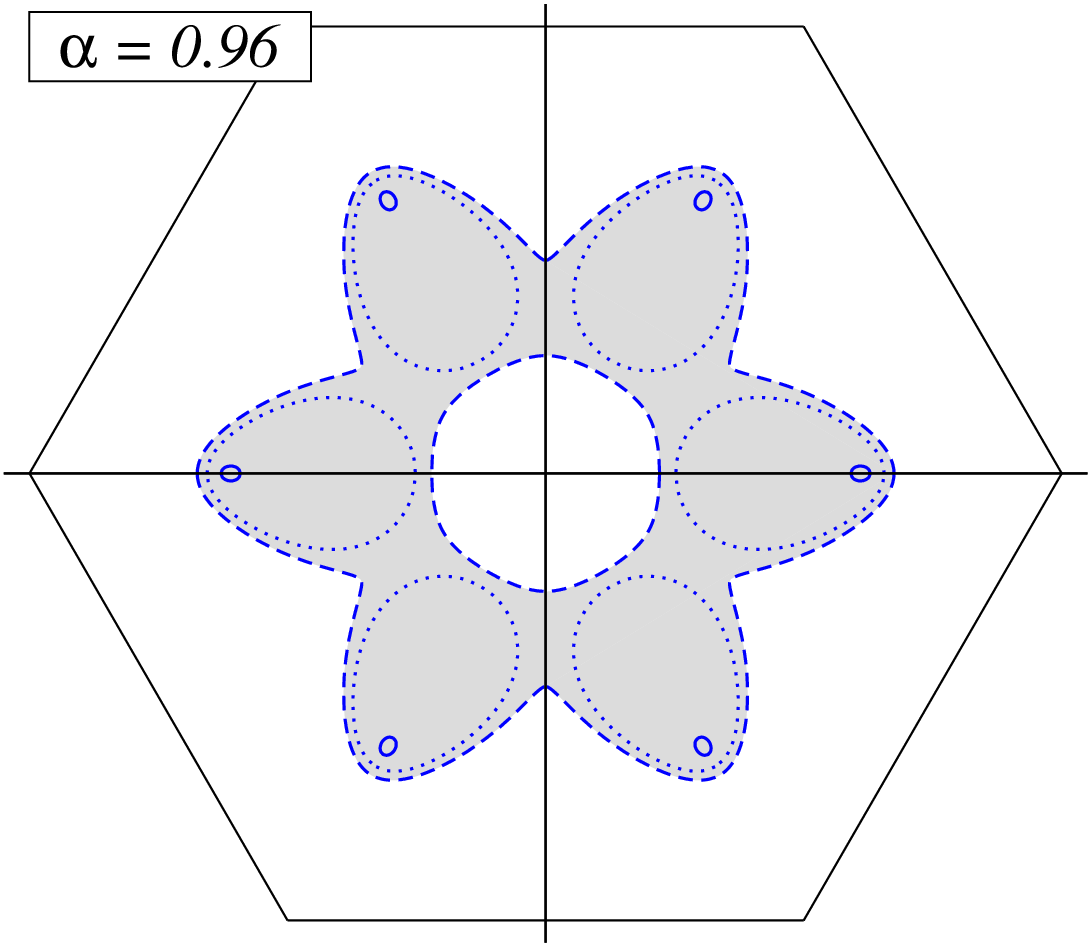}
\caption{(Color online) Decay region and singularity lines for 
the $XXZ$ triangular-lattice antiferromagnet
 with $\alpha=0.99$ (top) and $\alpha=0.96$ (bottom). 
Definition of lines is the same as 
in Fig.~\ref{BZ}.}
\label{alpha}
\end{figure}

\begin{figure*}[t]
\includegraphics[width=0.6\columnwidth]{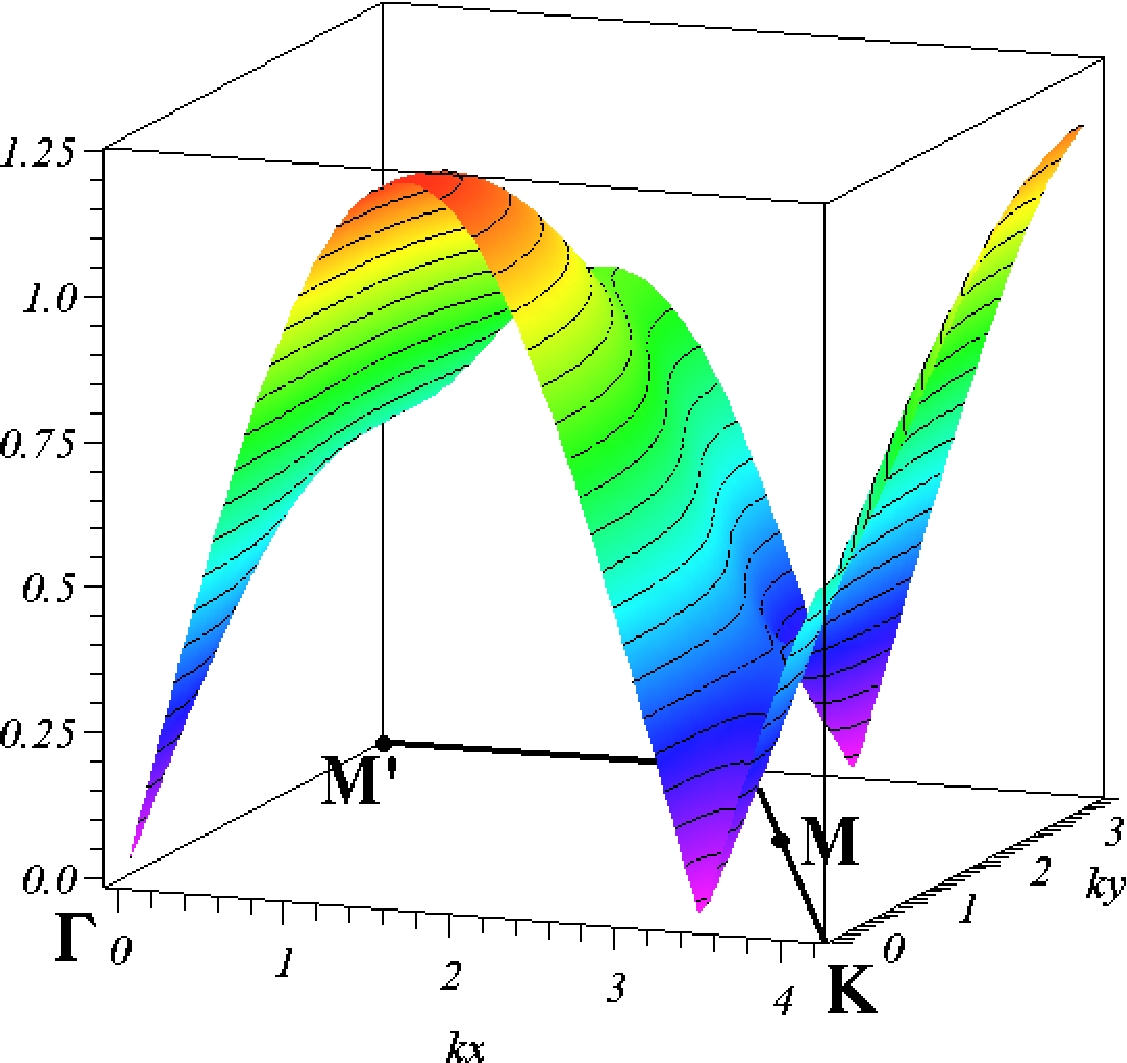}\ \hskip 1cm
\includegraphics[width=0.6\columnwidth]{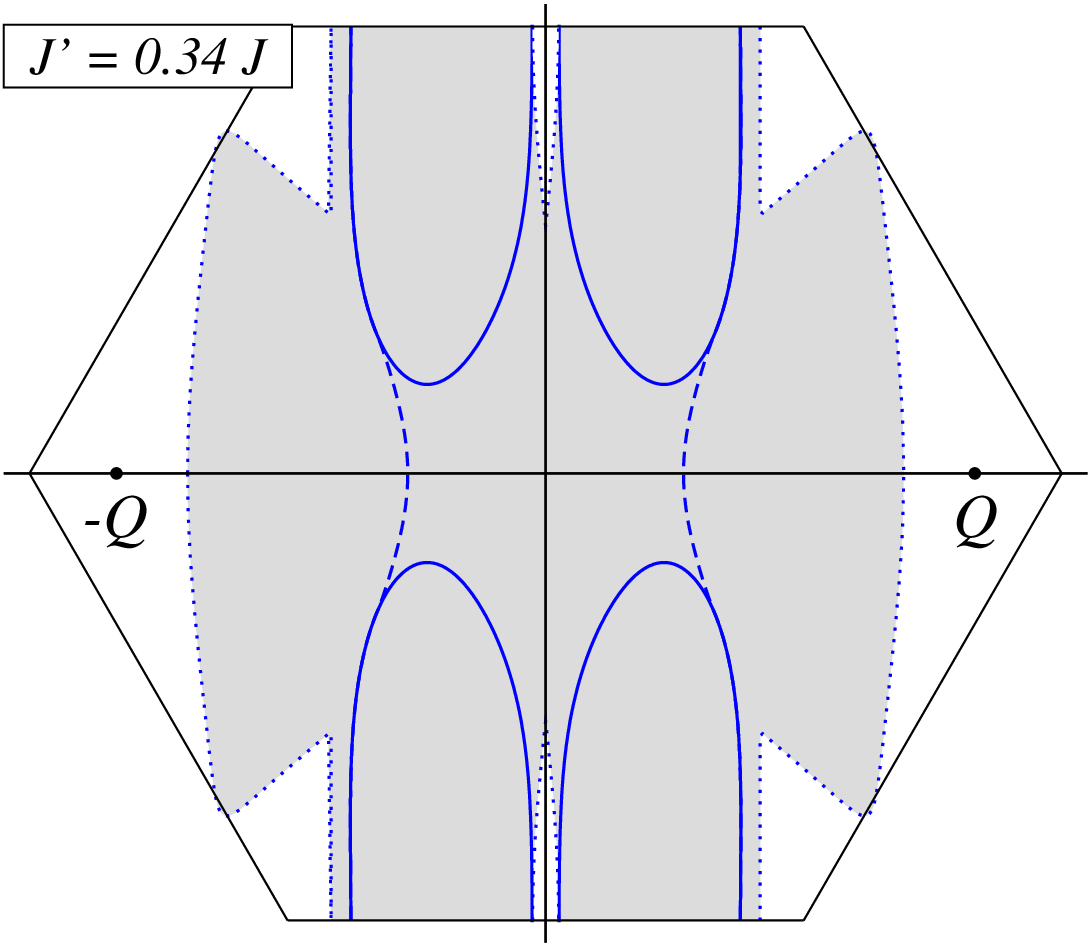}\ \hskip 1cm
\includegraphics[width=0.6\columnwidth]{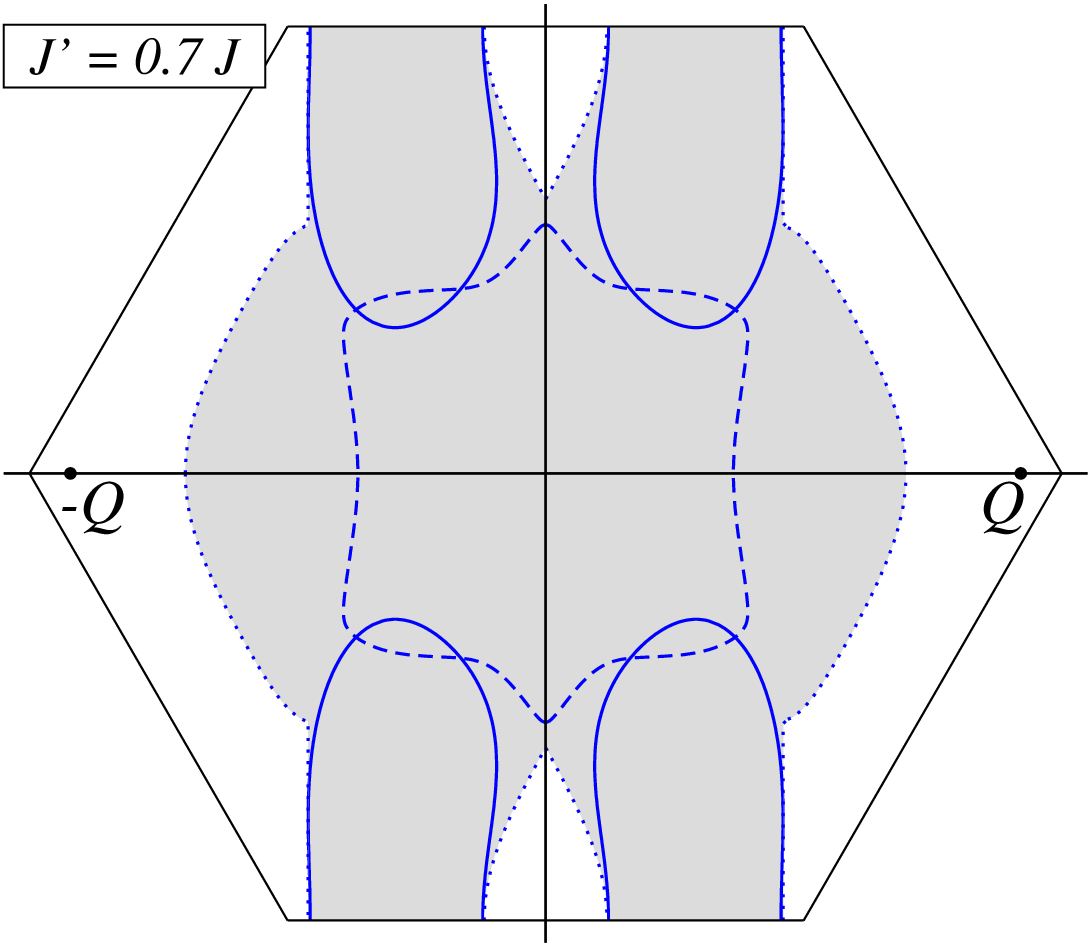}
\caption{(Color online) 
Left panel: the 3D shape of the 
linear spin-wave frequency 
$\omega_{\bf k}=\varepsilon_{\bf k}/2JS$ 
for the $J-J'$ antiferromagnet with $J'/J=0.34$.
Central panel: the decay region and the singularity lines 
at $J'/J=0.34$ (definitions are the same as 
in Fig.~\ref{BZ}). Right panel: same for $J'/J=0.7$.
}
\label{shape_wk_J_J1}
\end{figure*}

On the other hand, 
magnons in the $\alpha=0$ $XXZ$ model are stable at $T=0$
and decays are absent.
To study the evolution of the two-magnon decays
we begin with the nearly Heisenberg limit $1-\alpha\ll 1$. 
Magnons with ${\bf q}= \pm {\bf Q}$ are gapped with 
$\varepsilon_{\bf Q}\!=\!\sqrt{3(1-\alpha)/2}$. This 
has two immediate consequences in comparison with the $\alpha=1$ limit: 
(a) magnons at ${\bf k}\rightarrow 0$ 
cannot decay into two $\varepsilon_{\bf Q}$ magnons
up to ${\bf k}$-values such that 
$\varepsilon_{\bf k}=2\varepsilon_{\bf Q}$ and
(b) ${\bf Q}$-magnons become stable themselves for the same reason.
As a result, the star-shaped decay region of Fig.~\ref{BZ} develops a hole in 
the middle and has vertices shrunk and rounded, see Fig.~\ref{alpha}. 
The evolution of the character of the decay boundary with $\alpha$ is
non-trivial. Initially, the emission of a ${\bf Q}$ magnon remains 
an absolute minimum of the two-magnon continuum for most 
of the decay region except the tips of the star vertices. 
At $\alpha_1\!\approx\!0.993$ 
the decay into a pair of non-equivalent magnons 
switches from being a line of saddle 
points into the absolute minima of the continuum and takes over the
decay boundary, compare dotted and dashed lines in Figs.~\ref{BZ} and
\ref{alpha}. 
Without going into details, we simply note that 
in such a case
the real part of the spectrum is singular, 
$\textrm{Re}\,\Sigma(k_b,\omega)\propto \ln|k-k_b|$,  see Eq.~(\ref{log2}),
which is similar to the case of 
excitations in $^4$He near the threshold of decay into a pair of 
rotons.\cite{LL_IX,Pitaevskii59}
Similarly to the latter case, finding the spectrum near the boundary
will require summation of the higher-order diagrams.
Fig.~\ref{alpha}  shows the evolution of  the decay region 
and singularity lines between $\alpha=0.99$ and 
$\alpha=0.96$. 
Further decrease of $\alpha$ completely eliminates the decay 
region at around $\alpha\approx 0.92$. 
Therefore, magnon decays are present in an anisotropic 
triangular-lattice AFs, but only for not very strong anisotropies.

\subsection{$J$--$J'$ model}

Another variety of generalizations of the nearest-neighbor
Heisenberg model can be generated by  spatial anisotropies, keeping
spin-space intact. A particular model of this type corresponds to 
an orthorhombically distorted triangular lattice:
\begin{eqnarray}
\hat{\cal H}  =  
J\sum_{\langle ij\rangle}^{x}  {\bf S}_i\cdot {\bf S}_j + 
J'\,\sum_{\langle ij\rangle}^{z.-z.}  {\bf S}_i\cdot {\bf S}_j \ ,
\label{H_JJ1}
\end{eqnarray}
in which interactions along the ``1D chains'' running parallel
to the $x$-axis is $J$, while 
zig-zag interaction between the chains is $J'$. This model has attracted
a lot of attention due to experimentally available systems, Cs$_2$CuCl$_4$
\cite{Coldea01} and  
Cs$_2$CuBr$_4$ \cite{Tsujii07} with $J'/J$ is $\approx 0.34$ 
and $\approx 0.7$, respectively.
For the former system, a comprehensive 
experimental neutron scattering analysis of the spin-excitation 
spectrum has been performed\cite{Coldea01} and an extensive theoretical 
analysis using both  the spin-wave theory and the 1D spinon-based approach
have been carried out.\cite{Dalidovich06,Veillette05,Essler,Starykh_Balents}
We do not intend to repeat any of these calculations here, but
would like to emphasize that a substantial broadening of the spin-waves 
in a major part of the BZ must persist throughout the phase
diagram of the $J$--$J'$ model.

The ground state of the classical 
$J$--$J'$ model is an incommensurate spin spiral.
In the harmonic approximation, the energy of the 
spin-waves in the $J$--$J'$ model with $J'/J<2$ is:\cite{Dalidovich06}
\begin{equation}
\varepsilon_{\bf k} = 6SJ
\sqrt{\left(\gamma_{\bf k}-\gamma_{\bf Q}\right)
\left(\left(\gamma_{\bf Q + k}+
\gamma_{\bf Q -\bf k}\right)/2 - \gamma_{\bf Q} \right) },
\label{epsilonJJ1}
\end{equation}
where
\begin{equation}
 \gamma_{\bf k}=\frac{1}{3} \Bigl(\cos{k_x}+2\frac{J'}{J}
\cos{\frac{k_x}{2}} \cos{\frac{\sqrt{3} k_y}{2}}\Bigr) \ ,
\label{Jk}
\end{equation}
and the ordering vector ${\bf Q}=(Q_x,0)$ is given by 
\begin{equation}
Q_x=\pi + 2 \arcsin (J'/2J)\ .
\label{Q_JJ1}
\end{equation}
Since the $SO(3)$ symmetry is preserved in the  $J$--$J'$ model,
the Goldstone modes remain at ${\bf k}=0$ and ${\bf k}=\pm{\bf Q}$ points
with the incommensurate ordering wave-vector ${\bf Q}$, Eq. (\ref{Q_JJ1}).
Thus, because the velocities of these modes must generally stay different, 
magnon decays will always be allowed. In Fig.~\ref{shape_wk_J_J1} we show
the 3D shape of the linear
spin-wave energy for $J'/J=0.34$. All notations
and symmetry points are the same as in Fig.~\ref{shape_wk}, $k_x$ 
is in units of $1/a$ and $k_y$ is in units of $1/b$, where
$a$ and $b$ are the lattice constants in the chain and interchain
directions, respectively. One can
observe a more sophisticated shape of the dispersion with three saddle points
at different energies ($M$, $M'$, and between $M'$ and $K'$ points). 
Comparing this Figure with the $J=J'$ case of  Fig.~\ref{shape_wk} one
also finds that with the decrease of $J'$ the spin-wave 
spectrum develops a low-energy branch in the $y$-direction 
(between the chains with strong $J$). That should make the rest of the
spectrum  
prone to decays into it.

To analyze the decays on a more qualitative level we show the decay 
regions and the singularity lines for two representative values of $J'/J$ 
in Fig.~\ref{shape_wk_J_J1}. First, the boundary of the decay region
is determined by the emission of the gapless $\pm {\bf Q}$ magnon for
any value of $J'/J$ as in the $J'=J$ case. Second, the incommensurability
of the ordering wave-vector ${\bf Q}$ does not change the kinematics of the 
decays near ${\bf k}\rightarrow 0$ where the spin-waves can always decay 
into two modes near $+{\bf Q}$ and $-{\bf Q}$. However, 
the incommensurability forbids the decays from near the ${\bf Q}$-point 
into the vicinity of $-{\bf Q}$ point since the $2{\bf Q}$ wave-vector 
is not equal to any reciprocal lattice vector anymore and quasi-momentum
cannot be conserved in such a decay. This leads to shrinking of the vertices 
of the star-shaped decay region in the $k_x$-direction, 
see Fig.~\ref{shape_wk_J_J1}.
However, the vertices expand considerably in the $k_y$ direction. The 
singularity lines due to the saddle points in the 
two-magnon continuum also
expand and stretch in the $k_y$ direction as $J'$ is lowered, in agreement 
with the expectations from the shape of the dispersion.
Overall, the decay region grows with the decrease of $J'$. At 
$J'\approx 0.34J$, relevant to Cs$_2$CuCl$_4$,\cite{Coldea01} 
the decay region covers most of 
the BZ, see Fig.~\ref{shape_wk_J_J1}.

Within the spin-wave theory, 
a significant phase space volume for the decays found at $J'/J=0.34$
can be somewhat compensated by smaller noncollinearity 
of the spins which reduces the decay amplitudes. 
In real Cs$_2$CuCl$_4$, additional 
Dzyaloshinskii-Moriya interactions make this angle large 
(close to 90$^\circ$) between the spins in the nearest-neighbor
1D chains,\cite{Coldea01} 
but almost antiparallel within the individual chains
with strong $J$. This means that, effectively, the 
decay vertices are proportional to the weaker coupling $J'$. 
However, the decays of the spin-waves still result in a 
substantial damping, see Refs.~\onlinecite{Dalidovich06,Veillette05}. 
Altogether, the $J$--$J'$ model should exhibit magnon decays and 
singularities in their spin-wave spectrum throughout all
the ranges of $J'$ where the spin-wave theory is applicable.

\subsection{$XXZ$ model on the kagom\'{e} lattice}

The Heisenberg antiferromagnet on the kagom\'{e} lattice is magnetically disordered
at $T=0$ in both the classical ($S=\infty$) and the $S=1/2$ 
limits.\cite{Chalker92,Harris92,Chubukov92}
The degeneracy of the classical ground state reveals itself in the presence of
the dispersionless zero-energy branch of magnons in the harmonic spectrum.
The easy-plane anisotropy does not lift such a degeneracy, while opening up a constant gap 
for the zero-energy mode. A somewhat different behavior is realized in
the recently discovered kagom\'{e}-lattice compound  potassium jarosite.\cite{Matan06} 
In this antiferromagnet, the Dzyaloshinskii-Moriya  interactions lift the zero-energy 
mode to finite energies removing simultaneously the classical degeneracy in favor of 
a so-called ${\bf q }=0$ spin configuration.
Still, the easy-plane $XXZ$ model,
\begin{eqnarray}
\hat{\cal H}  = J 
\sum_{\langle ij\rangle}  
\bigg[S^x_i S^x_j + S^y_i S^y_j +\alpha S^z_i S^z_j\bigg]\ ,
\label{H_kagXXZ}
\end{eqnarray}
is more advantageous for a qualitative consideration of the magnon spectrum, 
because analytic expressions for the spin-wave energies can
be easily derived.

\begin{figure}[t]
\includegraphics[width=0.95\columnwidth]{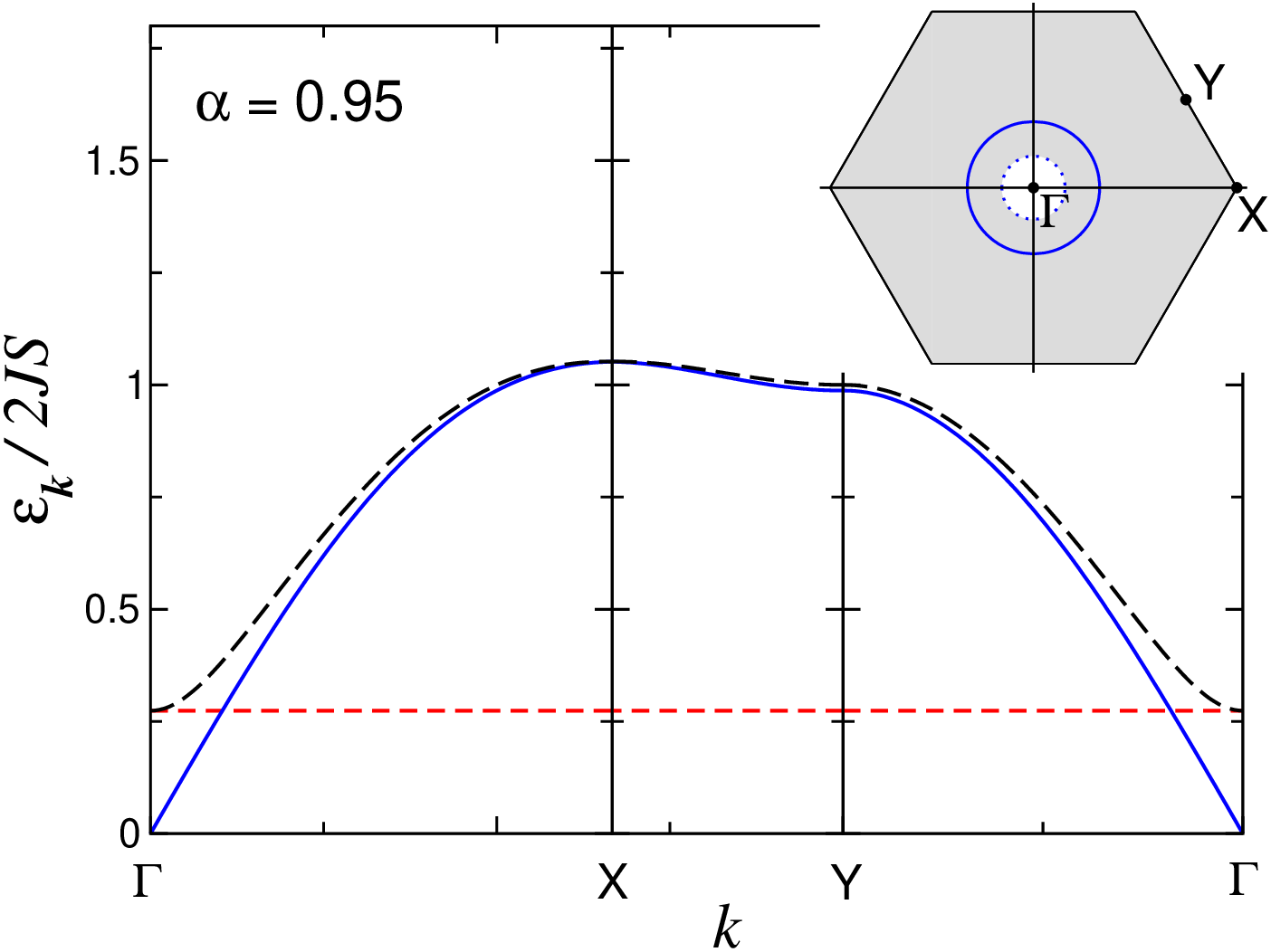}
\caption{(Color online) Linear spin-wave frequencies
of the excitations in the $XXZ$ model on the  kagom\'{e} lattice, 
Eq.~(\ref{omega1_kag}) for
$\alpha=0.95$, along $\Gamma XY$ cuts. Notations for the BZ 
are the same as in Ref.~\onlinecite{Harris92}. Lines and shaded area 
are the decay boundary (dotted), strong singularity line 
(solid), and the decay region, respectively, see text.}
\label{kag_disp}
\end{figure}

Since the unit cell of the kagom\'{e} lattice consists of three atoms, there
are three branches of magnetic excitations. For the 120$^\circ$ structure,
the energies of these branches in the harmonic approximation are given by:
\begin{equation}
\varepsilon^{(i)}_{\bf k} =2JS \omega^{(i)}_{\bf k}\ ,
\label{epsilon_kag}
\end{equation}
where
\begin{eqnarray}
\label{kag_wk}
&&\omega^{(1)}_{\bf k} = \sqrt{\frac32}\sqrt{1-\alpha} =\omega_0\ ,
\label{omega1_kag}\\
&&\omega^{(2,3)}_{\bf k} = 
\sqrt{1-\alpha\gamma_{\bf k}-
\frac{(1-\alpha)}{4} \left(1\pm\sqrt{1+8\gamma_{\bf k}}\right)} \ ,\nonumber
\end{eqnarray} 
with
\begin{equation}
\gamma_{\bf k}=\cos k_1 \cos k_2 \cos k_3 \ ,
\label{gamma_kag}
\end{equation} 
and $k_1=k_x$, $k_{2,3}=\pm k_x/2+k_y\sqrt{3}/2$, respectively.  
Thus, there is a gapped dispersionless mode, $\omega_0$, gapped
dispersive one, $\omega^{(2)}_{\bf k}$, and gapless one, 
$\omega^{(3)}_{\bf k}$, see Fig.~\ref{kag_disp} where $\omega^{(i)}_{\bf k}$ 
are shown for $\alpha=0.95$. In the real system, the lowest branch 
is weakly dispersive.\cite{Matan06} Using symmetry consideration, 
the BZ for the kagom\'{e} lattice HAF can be reduced to the one smaller 
than the triangular-lattice BZ, see Ref.~\onlinecite{Harris92}.
In Fig.~\ref{kag_disp}, $\Gamma XY$ cuts are according to the notations
of this work, with $X=(0,2\pi/3)$ and $Y=(\pi/2,\pi/2\sqrt{3})$.

Since the three modes in (\ref{kag_wk}) 
are some linear combinations of local spin-flips, 
the three-boson interaction due to non-collinearity of the spin structure
will necessarily facilitate couplings of all three branches with each other.
Such a coupling should not have any apparent smallness aside from the
$1/\sqrt{S}$ 
factor, same as in Eq.~(\ref{H3b}). Therefore, some qualitative 
conclusions about magnon decays in this system can be made without the
detailed  
technical analysis.

A distinct feature of the spin-waves in the  kagom\'{e} AFs is an (almost) 
dispersionless mode. For the decays this means that in the decays 
involving one of such magnons the momentum conservation can be ignored.
This immediately implies that any spin-wave with the energy exceeding 
$\omega_0$ will have a finite life-time due to decays into one $\omega_0$
magnon and a Goldstone mode of the gapless branch. Thus, all the
excitations  above the horizontal dashed line in Fig.~\ref{kag_disp} will
be damped due to such decays.
Another observation is that there must exist  a particularly strong singularity
in both dispersive branches  $\omega^{(2)}_{\bf k}$ and $\omega^{(3)}_{\bf k}$
at the energy twice the energy of the dispersionless mode. This is because
the self-energy near $2\omega_0$ will have a resonant-like form:
$\Sigma_{\bf k}(\omega)\propto |V|^2/(\omega-2\omega_0)$.
Even for the case when the lower branch has some 
residual dispersion, the damping of $\omega^{(2,3)}_{\bf k}$ must be
anomalously  
large near $2\omega_0$ in comparison with the rest of the spectrum. 
This should be valid even for large values of spin and in both 2D and
3D systems. 
Such a singularity is much stronger than the logarithmic singularities 
discussed in the rest of this paper and should be readily observed in
experiment.  

\subsection{Summary of Sec.~\ref{models}.}
Summarizing the above examples, we conclude
that  strong renormalization, significant damping, and
singularities must be common features of the excitation spectra 
of a wide variety  of noncollinear AFs. 
The enhancement of damping along certain contours due to singularities 
should be considered as fingerprints of magnon decays. This should
help to distinguish   
the decay-induced spin-wave broadening from the other scenarios that
yield broad  
spectra of the spin excitations.

\section{Conclusions}
\label{conclusions}
To summarize, the triangular-lattice AF is a prominent example
of a geometrically frustrated magnetic system. Present study has 
demonstrated that the frustration-induced non-collinearity of 
the spin structure in such systems is both the source and 
the key to understanding of their anomalous spin-wave spectra.
The highlights of the anomalous features of the spectrum that should
be observable in experiments are the substantial
broadening of magnon peaks due to spontaneous decays
in a large part of the Brillouin zone and strong deviations of the
spectrum from the LSWT expectations for $S=1/2$ systems.
The broadening should also be enhanced  by the $\ln S$ factor in the 
vicinity of certain
contours in the momentum space due to the van Hove singularities
in the two-magnon continuum. Such an enhancement might be more visible
in the systems with larger spins, such as the spin-5/2 triangular-lattice
RbFe(MoO$_4$)$_2$, \cite{Svistov03}
although the overall damping will be smaller due to
the $1/S$ effect. 
As the measurements of the lifetimes of spin excitations in the 
neutron-scattering experiments are expected to improve 
drastically in the future,\cite{Keimer} this will 
allow for the detailed analysis of the damping. Therefore, 
the enhancement of the damping along specific contours will be able
to  serve as a fingerprint
of the spin-wave decays and could help to distinguish the spin-wave decay 
mechanism from the other 
scenarios such as fractionalization of spin excitations into
spinons. \cite{Lee_Syljuasen,Starykh_Balents} 
We have also demonstrated that the strong renormalization, decays, and
singularities in the spectrum are prominent in a wide variety of
frustrated AFs of current interest.

The strong cubic anharmonic coupling of the spin-waves
induced by the non-collinearity is also the source of various
technical challenges within the spin-wave approach. 
Such challenges are largerly unfamiliar 
in the well-studied bipartite (collinear) AFs
where interactions are weaker and are of higher order.
In this work, we have demonstrated that the spin-wave theory encounters
some new problems in treating magnon interactions in the non-collinear AFs. 
Generally speaking, the standard $1/S$ expansion for the spectrum becomes 
non-analytic, which is manifested by the singularities in the 
first-order $1/S$ self-energy.
The origin of such singularities is related to the crossing
of the single-particle branch with the surface of van Hove singularities
in the two-magnon continuum.
Regularization of such singularities also requires extra technical
effort of going beyond the lowest-order $1/S$ approximation.


\begin{acknowledgments}
We are grateful to O. Starykh and A. Chubukov for illuminating
discussions. We also thank H. Tsunetsugu for bringing 
Ref.~\onlinecite{Oguchi83} to our attention.
Part of this work has been done at 
the Max Planck Institute for the Physics of Complex Systems,
which we would like to thank for hospitality.
This work was supported by DOE under grant DE-FG02-04ER46174
(A. L. C.). The research at KITP was supported by the NSF under Grant
No. PHY05-51164 (A. L. C.). 

\end{acknowledgments}


\appendix

\section{Hartree-Fock corrections}
\label{app_A}

In the harmonic approximation the following Hartree-Fock averages are 
nonzero for the triangular-lattice HAF
\begin{eqnarray}
&& n = \langle a^\dagger_i a_i\rangle = \sum_{\bf k} v_{\bf k}^2 = 
\frac{1}{2}\sum_{\bf k} \frac{1+ \frac{1}{2}\,\gamma_{\bf k}} 
{\omega_{\bf k}} - \frac{1}{2} \ , 
\nonumber \\
&& m = \langle a^\dagger_i a_j\rangle = \sum_{\bf k} \gamma_{\bf k} 
v_{\bf k}^2 = \frac{1}{2}\sum_{\bf k} \frac{\gamma_{\bf k} + 
\frac{1}{2}\,\gamma^2_{\bf k}}{\omega_{\bf k}} \ , 
\nonumber \\
&& \Delta = \langle a_i a_j\rangle =  \sum_{\bf k}  \gamma_{\bf k} u_{\bf k}
v_{\bf k} = \frac{3}{4}\sum_{\bf k}  
\frac{\gamma^2_{\bf k}} {\omega_{\bf k}} \ ,
 \nonumber  \\
&& \delta = \langle a^2_i\rangle =  \sum_{\bf k} u_{\bf k} v_{\bf k} =
\frac{3}{4}\sum_{\bf k}  
\frac{\gamma_{\bf k}} {\omega_{\bf k}} \ . \label{AHF} 
\end{eqnarray}
These four constants can be expressed through combinations of three 
two-dimensional integrals 
\begin{eqnarray}
\label{ACs}
&& c_l = \sum_{\bf k} \frac{(\gamma_{\bf k})^l}{\omega_{\bf k}}\ , \ \ \ \ \
c_0 = 1.5747334\ , \nonumber \\ 
&& c_1 = -0.1042539\ , \ \ \ c_2 = 0.3444458 \ .
\end{eqnarray}

As is discussed in Sec.~\ref{formalism}.A.3, the quartic terms 
(\ref{H4}) yield a correction to the
ground-state energy that is given by the 4-boson averages. In the leading
order, they can be decoupled into the 
bilinear combinations (\ref{AHF}) using the Wick's theorem. The
corresponding terms in Eq.~(\ref{H4}) are given by:
\begin{eqnarray}
&& \langle a^\dagger_i a_i a^\dagger_j a_j\rangle = n^2+m^2+\Delta^2\ ,   \\
&& \langle a^\dagger_i a_i a_i a_j\rangle = 2n\Delta +  m\delta, \ \
\langle a^\dagger_j a^\dagger_j a_j a_i\rangle = 2nm + \Delta\delta.\nonumber 
\end{eqnarray}
This yields the ground-state energy correction from the quartic terms:
\begin{equation}
\delta E_{4} \!=\! -\frac{3}{2}J\Bigl[ n^2\! +  m^2 \!+ \Delta^2 \!- 3n\Delta 
- {\textstyle \frac{3}{2}}m\delta + nm + {\textstyle
  \frac{1}{2}}\Delta\delta\Bigr]. 
\label{AE4}
\end{equation}
After some further algebra the above expression is 
converted into a combination of $c_l$ constants given by Eq.~(\ref{E4}).

A similar mean-field decoupling procedure yields the following correction 
to the harmonic part of the Hamiltonian in terms of the original bosons:
\begin{eqnarray}
&& \delta
\widetilde{\cal H}_2 = \sum_{\bf k} \delta A_{\bf k} a^\dagger_{\bf k}
  a_{\bf k} 
-\frac{1}{2}\, \delta B_{\bf k} \Bigl( a_{\bf k} a_{-\bf k} + 
a^\dagger_{-\bf k} a^\dagger_{\bf k} \Bigr)\ ,
 \nonumber \\
&& \delta A_{\bf k} = -\frac{3}{4}J
\bigl[2 (c_0\!+\! c_1\! -\! 2c_2 \!-\! 1) \!+\! \gamma_{\bf k}\! 
(c_0 \!+\! c_2 \!-\! 1 \!+ \!{\textstyle\frac{1}{4}}c_1) \bigr],
\nonumber \\
&& \delta B_{\bf k} = \frac{9}{4}J \left[ 
\gamma_{\bf k}\left(1+c_2-c_0-{\textstyle \frac{1}{4}}c_1\right) - 
{\textstyle \frac{1}{2}}c_1\right]\ .  \label{AH24a}
\end{eqnarray}
After that one uses the Bogolyubov transformation 
(\ref{bogoliubov1}) and obtains the following quadratic form of magnon
operators: 
\begin{equation}
\delta
\widetilde{\cal H}_2 = \sum_{\bf k} 
\varepsilon^{(4)}_{\bf k} b^\dagger_{\bf k}b_{\bf k}
-\frac{1}{2}\, B^{(4)}_{\bf k}\, \left( b_{\bf k} b_{-\bf k} + 
b^\dagger_{-\bf k} b^\dagger_{\bf k} \right) \ .
\label{AH24b} 
\end{equation}

The coefficients in this expression are related
to $\delta A_{\bf k}$ and  $\delta B_{\bf k}$ by
\begin{eqnarray}
 \varepsilon^{(4)}_{\bf k} & = & \left(u_{\bf k}^2
+v_{\bf k}^2\right)\delta A_{\bf k} 
-2u_{\bf k}v_{\bf k}\delta B_{\bf k}\ , \nonumber \\
 B^{(4)}_{\bf k} & = & \left(u_{\bf k}^2 + v_{\bf k}^2\right)\delta B_{\bf k}  
-2u_{\bf k}v_{\bf k}\delta A_{\bf k} \ ,
\label{AeB4}
\end{eqnarray}
which leads finally to Eqs.~(\ref{e4}) and (\ref{B4}).

\section{Sublattice magnetization}
\label{app_B}

Calculation of the second-order correction to the sublattice 
magnetization (\ref{deltaS}) requires evaluation of 
the lowest-order contributions to 
$\langle b^\dagger_{\bf k}b_{\bf k}\rangle$ and 
$\langle b_{\bf k}b_{-\bf k}\rangle$, which are expressed 
in terms of the normal and the anomalous Green's functions, respectively:
\begin{eqnarray}
&& \langle b^\dagger_{\bf k}b_{\bf k}\rangle =  
i\int\frac{d\omega}{2\pi}\,G_{11}({\bf k},\omega)e^{i\omega\delta}  \ , \nonumber \\
&& \langle b_{\bf k}b_{-\bf k}\rangle = 
i\int\frac{d\omega}{2\pi}\,G_{12}({\bf k},\omega) \ .
\end{eqnarray}
One needs to keep only the first-order terms 
in the perturbative expansion of the Green's functions.
The diagonal average 
$\langle b^\dagger_{\bf k}b_{\bf k}\rangle$
has a single nonzero contribution determined by
$\delta G_{11}({\bf k},\omega) 
= G^2_0({\bf k},\omega)\Sigma_{11}^{(b)}({\bf k},\omega)$:
\begin{equation}
\langle b_{\bf k}^\dagger b_{\bf k} \rangle = 
\frac{1}{2}\, \sum_{\bf q}
\frac{|\Gamma_2({\bf q;k})|^2}{(\varepsilon_{\bf k}+\varepsilon_{\bf q}
+\varepsilon_{\bf k+q})^2} \ ,
\end{equation}
which  yields
\begin{equation}
\delta S_{2,1} = \frac{3}{4S} \sum_{\bf k,q} \frac{1+\frac{1}{2}\gamma_{\bf k}}
{\omega_{\bf k}} \frac{\widetilde{\Gamma}_2({\bf q;k})^2}
{(\omega_{\bf k}+\omega_{\bf q} +\omega_{\bf k+q})^2} \ ,
\end{equation}
where we have transformed to the dimensionless vertex and frequencies
in the last expression.

The off-diagonal average $\langle b_{\bf k}b_{-\bf k}\rangle$ 
is determined by the lowest-order anomalous self-energies:
\begin{equation}
\delta G_{12}({\bf k},\omega) = G_0({\bf k},\omega) G_0(-{\bf k},-\omega)
\Sigma_{12}({\bf k},\omega) 
\end{equation}
with
$\Sigma_{12}({\bf k},\omega)=\Sigma_{HF}({\bf k})
+\Sigma_{12}^{(c)}({\bf k},\omega)
+\Sigma_{12}^{(d)}({\bf k},\omega)$.  
The first frequency-independent contribution from the Hartree-Fock 
self-energy (\ref{SEHF}) yields

\begin{eqnarray}
\delta S_{2,2}' & = & -\frac{9}{16S} \sum_{\bf k} 
\frac{\gamma_{\bf k} (1-\gamma_{\bf k}) }{\omega_{\bf k}^3}
\left(\frac{1}{2}c_1+c_2\gamma_{\bf k}\right)  \\
& = &
 -\frac{9}{32S}\, c_1c_2 + \frac{9}{32S}\,(c_2-c_1) \sum_{\bf k} 
\frac{\gamma_{\bf k} (1-\gamma_{\bf k}) }{\omega_{\bf k}^3} \ . \nonumber
\end{eqnarray}
Two other terms give identical contributions 
to $\langle b_{\bf k}b_{-\bf k}\rangle$ 
with the net result
\begin{equation}
\delta S_{2,2}'' = \frac{9}{8S} \sum_{\bf k}\,\frac{\gamma_{\bf k}}
{\omega_{\bf k}^2}\:\sum_{\bf q} 
\frac{\widetilde{\Gamma}_1({\bf q};{\bf k})
\widetilde{\Gamma}_2({\bf q};{\bf k})}
{\omega_{\bf k}+ \omega_{\bf q}+\omega_{\bf k-q}} \ .
\end{equation}
Combining all of the above terms together, we obtain:
\begin{eqnarray}
\label{BdeltaS2}
\delta S_2 & = & 
 -\frac{9}{16}\, c_1c_2 + \frac{9}{16}\,(c_2-c_1) \sum_{\bf k} 
\frac{\gamma_{\bf k} (1-\gamma_{\bf k}) }{\omega_{\bf k}^3} \nonumber   \\
 & + &
\frac{9}{4} \sum_{\bf k}\, \frac{\gamma_{\bf k}}{\omega_{\bf k}^2}\:
\sum_{\bf q} \frac{\widetilde{\Gamma}_1({\bf k},{\bf q})
\widetilde{\Gamma}_2(-{\bf k},{\bf q})}
{\omega_{\bf q}+ \omega_{\bf k-q}+\omega_{\bf k}} 
 \\
& + & 
\frac{3}{2} \sum_{\bf k}\, \frac{1+\frac{1}{2}\gamma_{\bf k}}{\omega_{\bf k}}\:
\sum_{\bf q} \frac{\widetilde{\Gamma}_2({\bf k},{\bf q})^2}
{(\omega_{\bf q}+ \omega_{\bf k+q}+\omega_{\bf k})^2}  \ . \nonumber
\end{eqnarray}

As discussed in Sec.~\ref{formalism}.B.2, one cannot use 
Eq.~(\ref{BdeltaS2}) for numerical evaluation of $\delta S_2$ directly 
as it leads to ambiguous results or simply does not converge. 
The way to regularize the above expression is to use an
analytical insight.\cite{Chubukov94} The integrand in 
the third term in (\ref{BdeltaS2}) 
can be reduced precisely to the divergent second  term for 
${\bf k}\rightarrow {\bf Q}$: 
\begin{eqnarray}
&& \frac{9}{4} \, \frac{\gamma_{\bf k}}{\omega_{\bf k}^2}\:
\sum_{\bf q}  \frac{\widetilde{\Gamma}_1({\bf k},{\bf q})
\widetilde{\Gamma}_2(-{\bf k},{\bf q})}
{\omega_{\bf q}+ \omega_{\bf k-q}+\omega_{\bf k}}
\Bigg|_{{\bf k}\rightarrow{\bf Q}} \nonumber \\
&& \quad \quad  = -\frac{9}{16}(c_2-c_1)  
\frac{\gamma_{\bf k} (1-\gamma_{\bf k}) }{\omega_{\bf k}^3}
+{\cal O}(\omega_{\bf k}^{-2}).
\label{Bcancel}
\end{eqnarray}
Then, the proper subtraction of the leading singularities is ensured
by re-expressing the {\it second} term in Eq.~(\ref{BdeltaS2}) 
as a double integral over ${\bf k}$ and ${\bf q}$ using Eq.~(\ref{Bcancel}).
An additional technical detail is that one should use projector-type
multipliers $P_1=\frac{2}{3}(1-\gamma_{\bf k})$ and 
$P_2=\frac{1}{3}(1+2\gamma_{\bf k})$ to avoid introducing extra 
singularities by the above conversion in the second term. 
These projectors obey $P_1+P_2=1$ and guarantee convergence of the 
integrals close to the ${\bf k}=0$ and 
${\bf k}={\bf Q}$ points, respectively. Altogether, 
this has led us to the analytically
identical, but numerically regular form of the $1/S^2$ correction to the
on-site magnetization:
\begin{eqnarray}
\delta S_2 =  -\frac{9}{16} c_1c_2 + \delta\widehat{S}^{3,1}
+ \delta\widehat{S}^{3,2} + \delta\widehat{S}^{3,3} \ ,
\label{BdeltaS2a}
\end{eqnarray}
where the second and third terms are regular from the start
due to projectors and the last term is the regularized 
combination of all singular terms. Specifically: 
$$
\delta\widehat{S}^{3,1}=
\sum_{\bf k} 
\frac{\left(1+\frac{1}{2}\gamma_{\bf k}\right)}{\omega_{\bf k}}
({\textstyle \frac{1}{2}}+\gamma_{\bf k})
\sum_{\bf q} \frac{\widetilde{\Gamma}_2({\bf k},{\bf q})^2}
{(\omega_{\bf q} + \omega_{\bf k+q} + \omega_{\bf k})^2},
$$
\vskip -5mm
$$
\delta\hat{S}^{3,2}=
\frac{3}{4} \sum_{\bf k}\, \frac{\gamma_{\bf k}}{\omega_{\bf k}^2} 
(1+2\gamma_{\bf k}) 
\sum_{\bf q} \frac{\widetilde{\Gamma}_1({\bf k},{\bf k-q})
\widetilde{\Gamma}_2(-{\bf k},{\bf q})}
{\omega_{\bf q}+ \omega_{\bf k-q}+\omega_{\bf k}} \ .
$$
The combination of the divergent terms is given by:
\begin{widetext}
$$ 
\delta\widehat{S}^{3,3} =
\frac{3}{2} \sum_{\bf k}\, 
\frac{\left(1-\gamma_{\bf k}\right)}{\omega_{\bf k}^2}
\,
\sum_{\bf q} \biggl[ \gamma_{\bf k} \, \biggl(
\frac{\widetilde{\Gamma}_1({\bf k},{\bf k-q})
\widetilde{\Gamma}_2(-{\bf k},{\bf q})}
{\omega_{\bf q}+ \omega_{\bf k-q}+\omega_{\bf k}}
-\frac{\widetilde{\Gamma}_1({\bf Q},{\bf Q-q})
\widetilde{\Gamma}_2(-{\bf Q},{\bf q})}
{\omega_{\bf q}+ \omega_{\bf Q-q}}\cdot\frac{\omega_{\bf Q}}{\omega_{\bf k}}\biggr)  
+ \frac{2}{3}\frac{\omega_{\bf k}\left(1+\frac{1}{2}\gamma_{\bf
    k}\right)\widetilde{\Gamma}_2({\bf k},{\bf q})^2} 
{(\omega_{\bf q}+ \omega_{\bf k+q}+\omega_{\bf k})^2} \biggr]\ .
$$ 
\end{widetext}
The results of the numerical integration of the individual terms are:
\begin{eqnarray*}
&& -\frac{9}{16}\,c_1 c_2   = 0.02019927 \ ,  \ \ 
\delta\widehat{S}^{3,1}= 0.017918(1)\ ,  \\ 
&& \delta\widehat{S}^{3,2}= 0.025496(2) \ , \ \ \ \
\delta\widehat{S}^{3,3}= -0.074660(5)\ .
\end{eqnarray*}
Altogether, they lead to the following value of the second-order
correction 
\begin{equation}
\delta S_2 = -0.011045(5) \ .
\label{Bnumbers1}
\end{equation}

\section{Higher-order singularities}
\label{app_parquet}

Singularities in the perturbative calculations of the bosonic excitation 
spectra are known since the early works by Pitaevskii\cite{Pitaevskii59,LL_IX} 
on the termination point in the phonon branch of\, $^4$He.
There is a renewed interest in the similar problems in the context of various
spin systems.\cite{Zhitomirsky06} 
On the other hand, the logarithmic singularities
in the fermionic spectra are also known to occur since the early works 
on the edge-singularities in metals,\cite{Mahan} in the context of numerous
aspects of 1D Luttinger liquids,\cite{Giamarchi} as well as in the newer 
physical systems such as graphene.\cite{Neto_RMP}

In all these problems, some physical process, often of a threshold nature,
leads to a non-analytic behavior of various quantities. 
Such a non-analyticity manifests itself in a breakdown of the perturbative 
expansion, {\it i.e.} as a singularity. The theoretical challenge is
to reconstruct  
the original non-analytic behavior from the singular terms in the 
perturbative expansion. In the Pitaevskii's case, such a 
reconstruction is rather straightforward  and consists of
resummation of the  leading divergent terms of the ``ladder'' (RPA) type.
In such a case, the original non-analyticity is straightforwardly related
to the singularity, {\it e.g.} $1/\ln|\omega|$ to $\ln|\omega|$,
respectively. 
In the fermionic systems, such a resummation is more complicated and involves 
the infamous ``parquet'' diagrams. However, often enough, the
reconstruction of the 
original non-analytic behavior is still possible from the analysis of a few 
most divergent terms of the expansion. In some well-known cases, the 
log-singularity resumes in a non-analyticity of a non-trivial fractional 
power-law type, $|\omega|^\alpha$.\cite{Mahan}

Since in our case we deal with bosonic excitations, one can expect  that the
Pitaevskii's consideration is the most relevant one. While this expectation,
with some minor corrections, turns out to be true, we would like to outline
some differences and similarities of the singularities in the spectra of 
noncollinear AFs in a somewhat broader context.

\begin{figure}[t]
\includegraphics[width=1.0\columnwidth]{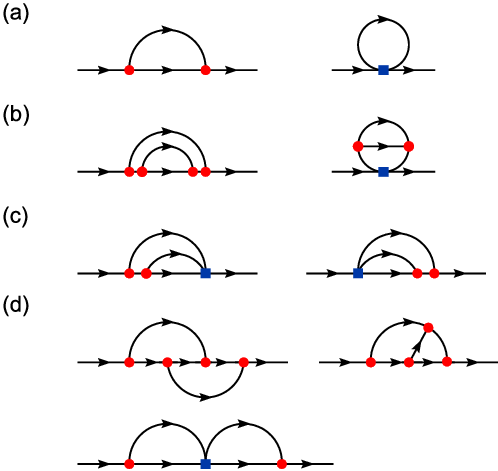}
\caption{(Color online) Schematic picture of all 
topologically inequivalent diagrams of order $1/S$, (a), and $1/S^2$, 
(b)-(d), respectively. The leading divergence in $1/S$ diagrams 
is $\ln|\Delta\omega|$ (first diagram in (a)), in $1/S^2$ order it is 
$(\ln|\Delta\omega|)^2$ (all diagrams in (d)). 
First two diagrams in (d) are identical.}
\label{parquet}
\end{figure}

One obvious difference of our problem from the fermionic case is the presence
of two coupling constants, three-particle, $V_3$, and  four-particle, $V_4$,
depicted in Fig.~\ref{vertices}. Because of that, the diagrammatic
expansion is more 
complicated in  our case. Since the singularity occurs for any value
of the spin, 
it is natural to group the diagrams by their order in $1/S$. By construction, 
the three-particle vertex is of order $1/\sqrt{S}$ relative to the magnon 
energy $\varepsilon_{\bf k}$ and it must occur in pairs in any
self-energy diagram, 
while the four-particle vertex is of order $1/S$. We depict schematically all 
topologically different diagrams of order $1/S$ and $1/S^2$ that occur
in such a  
theory in Fig.~\ref{parquet}(a) and Fig.~\ref{parquet}(b)-(d), respectively.

The singularity in our case is due to the ``bubble'' diagram 
[\#1 in Fig.~\ref{parquet}(a)], which leads to the logarithmic term for
certain ${\bf k}$-values of external lines of the diagram.
A simple analysis shows that out of the $1/S^2$-terms only diagrams in 
Fig.~\ref{parquet}(d) yield the higher power of the log,
$(\ln|\Delta\omega|)^2$. 
In fact, the first two diagrams in the (d) group are identical and can
be seen as  
vertex corrections  to the bubble in Fig.~\ref{parquet}(a).
The last diagram in the group (d) is the standard ``two-bubble'' member of the
ladder sequence considered by Pitaevskii. A closer inspection of the
second diagram  
in the (d) group in Fig.~\ref{parquet} makes it clear that it is also
the ``two-bubble'' 
diagram with the virtual magnon 
line and two 3-magnon vertices joining the ``bubbles''. This modifies
consideration by 
Pitaevskii, albeit only quantitatively, as it introduces extra interaction
together with some retardation into the ``ladder'' sequence. 
To rule out the presence of any ``non-ladder''-type diagrams with the higher
powers of the log and, hence, to eliminate the possibility of 
more complicated form of the non-analytic behavior, the following
observation is useful.  
In the parquet consideration,\cite{Mahan} it is both particle-particle and 
particle-hole diagrams that are singular. In our case of bosons with
non-conserved  
number of particles at $T=0$, the particle-hole diagrams are all
identically zero. 
Thus, the leading power of the log of any given diagram is simply the
number of  
non-equivalent regions of the diagram in which a vertical cut
intersects only two magnon 
lines. Therefore, the most-divergent 
diagram sequence is comprised of the ``ladder''-like diagrams only, 
with both $V_4$ and $(V_3)^2$ between the ``bubbles''. This can be
confirmed by an  
explicit analysis of the ``dangerous'' diagrams in the next, $1/S^3$
order (not shown 
in Fig.~\ref{parquet}).

Another qualitative difference of our problem from  the fermionic case 
in which the ``parquet'' summation is successful is that the latter 
consideration often exploits 
the long-wavelength character of the problem to obtain a universal  
answer. In our case, the singularity occurs at short wavelength. Therefore,
even in the hypothetical generalization of our problem to the bosons
with conserved  
number of particles and singular particle-hole ``bubbles'', 
the ``parquet''-like analysis is unlikely to be successful
as the coefficients of the leading log-powers will be independent from
each other. 

Altogether, presented analysis of higher-order divergences in the diagrammatic
sequence for the magnon propagator in the noncollinear AFs shows that the 
singularities are regularized by summation of the ``ladder'' diagrams
of the Pitaevskii 
type with the minor modification of the interaction vertex. The
regularized result 
can be written schematically as: 
\begin{eqnarray}
\Sigma\propto \frac{V_3^2\Pi}{1-\widetilde{V}\Pi}\propto A
+\frac{B}{\Pi}+\dots\, , 
\end{eqnarray}
where $\widetilde{V}=V_4+V_3 G V_3$, and $\Pi$ is the singular 
``bubble'' contribution ($\sim\ln|\Delta\omega|$).

\section{off-shell Dyson equation for complex energies}
\label{app_D}

Suppose that the Dyson equation
\begin{equation}
\label{DE_app}
\omega - \omega_{\bf k} -\Sigma({\bf k},\omega)= 0,
\end{equation}
has a solution at a complex energy
\begin{equation}
\omega = \widetilde\omega_{\bf k}-i\Gamma_{\bf k},\ \ \ \Gamma_{\bf k}>0\, ,
\end{equation}
where the sign of the imaginary part is dictated by causality.
The Dyson equation can be written for the real and imaginary parts as
\begin{eqnarray}
\label{DE_Re_app}
&&\widetilde\omega_{\bf k}=\omega_{\bf k} +
{\rm Re}[\Sigma({\bf k},\widetilde\omega_{\bf k}-i\Gamma_{\bf k})],\\
\label{DE_Im_app}
&&\Gamma_{\bf k}=-
{\rm Im}[\Sigma({\bf k},\widetilde\omega_{\bf k}-i\Gamma_{\bf k})]>0\ .
\end{eqnarray}
In the standard textbook approach\cite{Mahan} the damping is assumed
to be small, 
$\Gamma_{\bf k}\ll \widetilde\omega_{\bf k}$, and is neglected from
the self-energy 
in the right-hand side of Eqs.~(\ref{DE_Re_app}) and
(\ref{DE_Im_app}). In that case, 
only the real part of the Dyson equation needs to be solved self-consistently
while the damping is simply evaluated from the imaginary part of the
self-energy  
at the real $\omega=\widetilde\omega_{\bf k}$, which is the solution
of  (\ref{DE_Re_app}). 

Since we have encountered a problem in which the imaginary part of the
self-energy  
diverges at some ${\bf k}$-values for $\omega$ along the real axis, we
would like to  
deviate from the standard approach and keep finite  $\Gamma_{\bf k}$ in both
parts of the Dyson equation, solving them self-consistently for both 
$\widetilde\omega_{\bf k}$ and $\Gamma_{\bf k}$. However, on this path
one discovers 
a difficulty that is unrelated to the singularity and is much more generic. We
believe that it deserves a separate discussion.

Let us consider the self-energy of very
general form:
\begin{equation}
\label{S1_app}
\Sigma({\bf k},\omega) = \sum_{\bf q} \frac{|V_{{\bf k}, {\bf q}}|^2}
{\omega-\varepsilon_{\bf q}-\varepsilon_{{\bf k}-{\bf q}}+i0}\, .
\end{equation}
The self-energy of such type appears in the problem of phonon interaction
with electron-hole continuum, Cherenkov radiation, pair production in QED, 
to name a few.
All of these problems involve a (virtual) decay
of a particle into two other particles. For simplicity, the products 
of the decay are assumed to be free particles 
(no damping for $\varepsilon_{\bf q}$'s). 
The difficulty arises when substituting 
$\omega = \widetilde\omega_{\bf k}-i\Gamma_{\bf k}$ in
the denominator of (\ref{S1_app}). 
This shifts the pole in (\ref{DE_app}) into
the wrong (upper) $\omega$-half-plane making $\Gamma_{\bf k}<0$, which 
violates causality and renders the Dyson equation unsolvable. 

The resolution of this problem  within the formalism of 
causal diagram technique requires one to define properly
the energy conservation for particles with finite lifetime. 
The self-energy (\ref{S1_app}) is the result of
integrating the ``bubble''-like diagram over  the internal frequency. 
Specifically, assuming
for simplicity that the decay is into particles of the same species, 
one can write:
\begin{eqnarray}
\label{S2_app}
\Sigma({\bf k},\omega)\! =\! i\sum_{\bf q}\! \int_\varepsilon 
\ \frac{|V_{{\bf k}, {\bf q}}|^2}
{(\omega\! -\! \varepsilon\! -\! \varepsilon_{{\bf k}-{\bf q}}\! +\! i0)\,
(\varepsilon\! -\! \varepsilon_{\bf q}\! +\! i0)}
\end{eqnarray}
where $\int_\varepsilon\equiv\int d\varepsilon/2\pi$.
This step seems to make the problem even worse: substituting
$\omega = \widetilde\omega_{\bf k}-i\Gamma_{\bf k}$ in (\ref{S2_app}) shifts 
the pole of the first Green's function into the wrong half-plane, 
which renders the integral over internal frequency zero.

Making one more step back we recall that Eq.~(\ref{S2_app}) comes from 
the standard diagrammatic expression that  explicitly
reflects the conservation of energy within the decay process:
\begin{eqnarray}
\label{S3_app}
\Sigma({\bf k},\omega) \! = \! i\sum_{\bf q}\!\int_\varepsilon 
\!\int_{\varepsilon_1} \!  
\frac{|V_{{\bf k}, {\bf q}}|^2\, 
\delta(\omega - \varepsilon_1 - \varepsilon)}{
(\varepsilon_1\! -\! \varepsilon_{{\bf k}-{\bf q}}\! +\! i0)\,
(\varepsilon\! -\! \varepsilon_{\bf q}\! +\! i0)}\,
. 
\end{eqnarray}
Integrating (\ref{S3_app}) over $\varepsilon_1$ gives (\ref{S2_app}).
Now, the only place where ``external'' energy $\omega$ enters the self-energy 
is the $\delta$-function. Thus, one needs to generalize  the $\delta$-function
in Eq.~(\ref{S3_app}) for the case when $\omega$ turns complex. An
obvious trick is to 
represent the $\delta$-function by the Lorentzian of width $\Gamma_{\bf k}$.
Then one can rewrite $\Sigma({\bf k},\omega)$ for 
$\omega = \widetilde\omega_{\bf k}-i\Gamma_{\bf k}$ as:
\begin{eqnarray}
\label{S4_app}
\Sigma({\bf k},\omega)\! &=&\! i\sum_{\bf q}\! \int_\varepsilon 
\!\int_{\varepsilon_1} \! 
\frac{|V_{{\bf k}, {\bf q}}|^2 \, 2\Gamma_{\bf k}}
{(\widetilde\omega_{\bf k}\!-\!\varepsilon_1\!-\!\varepsilon)^2\!+\!
\Gamma_{\bf k}^2}\nonumber\\
&&\times\frac{1}{
(\varepsilon_1\!-\!\varepsilon_{{\bf k}\!-\!{\bf q}}\!+\!i0)\,
(\varepsilon\!-\!\varepsilon_{\bf q}\!+\!i0)}\,
. 
\end{eqnarray}
Note, that the above expression {\it does} respect the causality.
Integrating it by standard means we obtain the desired well-behaved result:
\begin{equation}
\label{S5_app}
\Sigma({\bf k},\omega) = \sum_{\bf q} \frac{|V_{{\bf k}, {\bf q}}|^2}
{\widetilde\omega_{\bf k}
-\varepsilon_{\bf q}-\varepsilon_{{\bf k}-{\bf q}}+i\Gamma_{\bf k}}\, .
\end{equation}
One can see that, starting from the basic diagrammatic rules we obtain the
result for the self-energy with the ``wrong'' sign of the imaginary
part of the energy  
of the decaying particle. This resolves the problem. In the form
(\ref{S5_app}) the  
self-energy has the quasiparticle pole in the correct (lower) half-plane
and yields the correct sign of $\Gamma_{\bf k}$ in the Dyson equation.
 Thus, the prescription is:
the energy of the decaying particle in the self-energy 
(\ref{S1_app}) should be taken from the advanced Green's function. The
``proper''  
Dyson equation then reads as 
\begin{equation}
\label{DE1_app}
\omega - \omega_{\bf k} -\Sigma({\bf k},\omega^*)= 0,
\end{equation}
for complex $\omega=\widetilde\omega_{\bf k}-i\Gamma_{\bf k}$.

Using the Matsubara technique, the
problem is resolved without any tricks with the 
$\delta$-function, but simply by forcing all the poles to be in the
correct half-plane. This does not give any additional insight, but 
means that the retarded self-energy simply corresponds to the
complex conjugation of $\omega$:
$\Sigma_{ret}({\bf k},\widetilde\omega_{\bf k}-i\Gamma_{\bf k})\equiv
\Sigma({\bf k},\widetilde\omega_{\bf k}+i\Gamma_{\bf k})$.
Thus, one can write the Dyson equation in an explicitly
self-consistent form as: 
\begin{equation}
\label{DEret_app}
\widetilde\omega_{\bf k}-i\Gamma_{\bf k} - \omega_{\bf k} -\Sigma({\bf
  k},\widetilde\omega_{\bf k}+i\Gamma_{\bf k})=0\,  
\end{equation}
with $\Sigma({\bf k},\omega)$ from (\ref{S1_app}). 
Changing $\Gamma_{\bf k}$ to an infinitesimal
$\delta$, this expression is nothing but the standard Dyson equation obtained
from the Matsubara approach, see Ref.~\onlinecite{Mahan},
Eqs.~(3.140)-(3.142).  

\section{Quasiparticle residue}
\label{app_residue}

Here we discuss an extension of the definition of the quasiparticle residue 
to the case of a particle with a finite lifetime.
The concept of the quasiparticle residue is introduced in the context 
of the problem of interacting particles where it is assumed that after 
``dressing'' the Green's function retains a well-defined pole 
at low energy. Close to that pole one can write the Green's function
in the main-pole approximation:
\begin{equation}
G_{\bf k}(\omega)\approx 
\frac{Z_{\bf k}}{\omega-\widetilde\omega_{\bf k}+i0}+G^{incoh}_{\bf k}(\omega)
\end{equation}
where the incoherent part of the Green's function
$G^{incoh}_{\bf k}(\omega)$ is regular at the ``new'' quasiparticle energy
$\omega=\widetilde\omega_{\bf k}$ and $Z_{\bf k}$ is the quasiparticle residue,
$Z_{\bf k}<1$. $Z_{\bf k}$  
is also the weight associated with the $\delta$-functional
peak of the Green's function at $\omega=\widetilde\omega_{\bf k}$. 
We wish to extend this to the case
when the quasiparticle has a finite lifetime. In other words, we 
hope to be able to write:
\begin{eqnarray}
G_{\bf k}(\omega)&=&\frac{1}{\omega-\omega_{\bf k}-\Sigma_{\bf k}(\omega)}\\
&&\approx 
\frac{Z_{\bf k}}{\omega-\widetilde\omega_{\bf k}+i\Gamma_{\bf k}}
+G^{incoh}_{\bf k}(\omega)\, .\nonumber
\end{eqnarray}
To get from the left-hand side to the right-hand side we need to
assume that there exist a solution of the Dyson equation
\begin{equation}
\omega-\omega_{\bf k}-\Sigma_{\bf k}(\omega)=0
\end{equation}
at some complex 
$\omega=\widetilde\omega_{\bf k}-i\Gamma_{\bf k}$, where
\begin{eqnarray}
&&\widetilde\omega_{\bf k}=\omega_{\bf k}+
{\rm Re}[\Sigma_{\bf k}(\widetilde\omega_{\bf k}-i\Gamma_{\bf k})]\\
&&\Gamma_{\bf k}=-{\rm Im}[\Sigma_{\bf k}(\widetilde\omega_{\bf
    k}-i\Gamma_{\bf k})]\, . 
\end{eqnarray}
Then, as in the standard approach, we  proceed by adding and subtracting 
$\Sigma_{\bf k}(\widetilde\omega_{\bf k}-i\Gamma_{\bf k})=
\widetilde\omega_{\bf k}-\omega_{\bf k}-i\Gamma_{\bf k}$ to the denominator of 
$G_{\bf k}(\omega)$:
\begin{eqnarray}
G_{\bf k}(\omega)&=&
\frac{1}{\omega-\omega_{\bf k}-\Sigma_{\bf k}(\omega)}\\
&=&
\frac{1}{\omega-\widetilde\omega_{\bf k}+i\Gamma_{\bf k}
  -\big(\Sigma_{\bf k}(\omega)-\Sigma_{\bf k}
(\widetilde\omega_{\bf k}-i\Gamma_{\bf k})\big)}\nonumber
\end{eqnarray}
This can be rewritten without approximation as:
\begin{eqnarray}
G_{\bf k}(\omega)&\equiv &
\frac{1}{\omega-\widetilde\omega_{\bf k}+i\Gamma_{\bf k}}\\
&&\times
\left[1-\left(\displaystyle\frac{\Sigma_{\bf k}(\omega)- 
\Sigma_{\bf k}(\widetilde\omega_{\bf k}-i\Gamma_{\bf k})}
{\omega-\widetilde\omega_{\bf k}+i\Gamma_{\bf k}}\right)\right]^{-1} .  
\nonumber
\end{eqnarray}
Near the pole, $\omega\rightarrow\widetilde\omega_{\bf k}-i\Gamma_{\bf k}$,
this expression finally yields:
\begin{eqnarray}
G_{\bf k}(\omega)\approx
\frac{Z_{\bf k}}{\omega-\widetilde\omega_{\bf k}+i\Gamma_{\bf k}}\, ,
\end{eqnarray}
where
\begin{eqnarray}
\label{Zk_app}
Z_{\bf k}\equiv \left[1-\left. \displaystyle
\frac{\partial\Sigma_{\bf k}(\omega)}{\partial\omega} 
\right|_{\omega=\widetilde\omega_{\bf k}-i\Gamma_{\bf k}}\right]^{-1}
\end{eqnarray}
which coincides with the standard definition of $Z_{\bf k}$ up to the 
change of $\widetilde\omega_{\bf k}$ to $\widetilde\omega_{\bf
  k}-i\Gamma_{\bf k}$. 
\cite{Mahan}
Therefore, generally speaking, the quasiparticle residue obtained this way 
is complex. However, one would prefer to have $Z_{\bf k}$ real in accord
with the expectation that the area under the Lorentzian (broadened
$\delta$-peak) should give the quasiparticle weight. For that, let 
us consider the spectral function:
\begin{eqnarray}
A_{\bf k}(\omega) &=& \frac{1}{\pi}\biggl(
\frac{\Gamma_{\bf k} {\rm Re}[Z_{\bf k}]}
{(\omega-\widetilde\omega_{\bf k})^2+\Gamma_{\bf k}^2} \nonumber \\
& - &
\frac{(\omega-\widetilde\omega_{\bf k}) {\rm Im}[Z_{\bf k}]}
{(\omega-\widetilde\omega_{\bf k})^2+\Gamma_{\bf k}^2}
\biggr)\ , 
\end{eqnarray}
where the first term has the Lorentzian form while the second one vanishes at 
$\omega_{\bf k}=\widetilde\omega_{\bf k}$.
Clearly, the second part of the expression is odd in $\omega$ and the
spectral weight of the quasiparticle peak, which is given by the integral of
$A_{\bf k}(\omega)$ along the real axis:
\begin{eqnarray}
\int_{-\infty}^\infty A_{\bf k}(\omega)d\omega \equiv  {\rm Re}[Z_{\bf k}]
\end{eqnarray}
is simply the real part of $Z_{\bf k}$ in (\ref{Zk_app}). 
This  gives a proper definition of the ``generalized'' quasiparticle residue:
\begin{equation}
\label{Zk1_app}
Z_{\bf k}\equiv {\rm Re}\biggl(\Bigl[
1 - \frac{\partial\Sigma_{\bf k}(\omega)}{\partial\omega} 
\Bigr|_{\omega=\widetilde\omega_{\bf k}-i\Gamma_{\bf k}}
\Bigr]^{-1}\biggr)
\end{equation}
Interestingly, this definition differs from the one
in Ref.~\onlinecite{Mahan}.

Our interest in this problem is twofold. First, we deal with 
quasiparticles that have finite damping due to decays.
Second, the decay part of the self-energy, Eq.~(\ref{SEa}), near 
the log-singularity has derivatives that are even more singular. 
In particular, an attempt to calculate the quasiparticle residue 
using the on-shell approach near the saddle-point singularity 
${\bf k}\rightarrow{\bf k^*}$ leads to: 
\begin{eqnarray}
\left.\frac{\partial\Sigma_{11}^{(a)}({\bf k},\omega)}{\partial\omega} 
\right|_{\omega_{\bf k}}\!=\! -\frac{1}{2} \sum_{\bf q} 
\frac{\widetilde{\Gamma}_1({\bf q};{\bf k})^2 }
{(\omega_{\bf k} - \omega_{\bf q} - \omega_{\bf k-q} + i0)^2}\, ,\nonumber
\end{eqnarray}
whose imaginary part is divergent as $\propto (k-k^*)^{-1}$
and the real part has a $\delta$-function singularity at $k=k^*$.
If, on the other hand,
one uses the solution of the Dyson equation together with our 
more general definition of $Z_{\bf k}$ (\ref{Zk1_app}), the result 
becomes regular. This is yet another way of saying that the 
$1/S$-expansion in the noncollinear AFs is singular and the usual on-shell
approach cannot be used.

Interestingly, the off-shell consideration gives, after some algebra, 
that in the large-$S$ limit, at the singular 
${\bf k}$-points $(\partial\Sigma/\partial\omega)\propto 1/\ln S$.
This means that the quasiparticle residue reaches the classical limit (=1)
very slowly as  
$$Z_{\bf k}\approx \left(1-\frac{A}{\ln S}\right)^{-1}$$ 
at these points.


\end{document}